\documentclass[fleqn,usenatbib]{mnras}

\pdfoutput=1

\usepackage[T1]{fontenc}
\usepackage{ae,aecompl}

\usepackage{graphicx}
\graphicspath{{img/}}
\usepackage{amsmath}
\usepackage{amssymb}
\usepackage{mathtools}
\usepackage{upgreek}
\usepackage{txfonts}
\usepackage{array}
\usepackage{multirow}

\newcommand{\sub}[1]{_\mathrm{#1}}

\newcommand{\dif}{\mathrm{d}}

\defcitealias{2012Sci...338.1196C}{CC12}

\title[Exoplanet recycling in WD debris discs]{Exoplanet recycling in massive white-dwarf debris discs}

\author[R. van Lieshout et al.]{%
R. van Lieshout,$^{1}$\thanks{E-mail: lieshout@ast.cam.ac.uk (RvL)}
Q. Kral$^{1}$,
S. Charnoz$^{2}$,
M. C. Wyatt$^{1}$ and
A. Shannon$^{3,4,1}$ \\
$^{1}$Institute of Astronomy, University of Cambridge, Madingley Road, Cambridge CB3 0HA, UK\\
$^{2}$Institut de Physique du Globe/Universit\'{e} Paris Diderot/CEA/CNRS, 75005 Paris, France\\
$^{3}$Department of Astronomy \& Astrophysics, The Pennsylvania State University, State College, PA 16801, USA\\
$^{4}$Center for Exoplanets and Habitable Worlds, The Pennsylvania State University, State College, PA 16802, USA}

\date{Accepted XXX. Received YYY; in original form ZZZ}

\pubyear{2018}

\begin{document}
\label{firstpage}
\pagerange{\pageref{firstpage}--\pageref{lastpage}}
\maketitle

\begin{abstract}
Several tens of white dwarfs are known to host circumstellar discs of dusty debris,
thought to arise from the tidal disruption of rocky bodies
originating in the star's remnant planetary system.
This paper investigates the evolution of such discs if they are very massive,
as may be the case if their progenitor was a terrestrial planet, moon, or dwarf planet.
Assuming the discs are physically thin and flat, like Saturn's rings,
their evolution is governed by Poynting--Robertson drag or viscous spreading,
where the disc's effective viscosity is due to self-gravity wakes.
For discs with masses $ \gtrsim $\,10$^{26} \, \mathrm{g} $,
located in the outer parts of the tidal disruption zone,
viscous spreading dominates the evolution,
and mass is transported both in- and outwards.
When outwards-spreading material flows beyond the Roche limit,
it coagulates into new (minor) planets in a process analogous to
the ongoing formation of moonlets at the outer edge of Saturn's rings.
The newly formed bodies migrate outwards by exchanging angular momentum with the disc
and coalesce into larger objects through mutual collisions.
Eventually, the disc's Roche-limit overflow recycles tens of percent of the original disc mass;
most ends up in a single large body near 2:1 mean-motion resonance with the disc's outer edge.
Hence, the recycling of a tidally disrupted super-Earth, for example,
could yield an Earth-mass planet on a \mbox{$\sim$10-h} orbit,
located in the habitable zone for 2-to-\mbox{10-Gyr}-old white dwarfs.
The recycling process also creates a population of smaller bodies just outside the Roche limit,
which may explain the minor planets recently postulated to orbit \mbox{WD~1145+017}.
\end{abstract}


\section{Introduction}
\label{s:intro}

Several tens of white dwarfs (WDs) are known to exhibit
excess near- and mid-IR emission,
revealing the presence of
circumstellar dust
(see \citealt{2016NewAR..71....9F} for a recent review).
A few of these systems also show double-peaked emission lines of metals,
which indicate that metallic gas is present as well
\citep[e.g.,][]{2006Sci...314.1908G}.
The estimated radial location of this circumstellar material
\citep[$r \lesssim 1\,\mathrm{R\sub{\odot}}$; e.g.,][]{2010ApJ...722.1078M}
is well inside the maximum stellar radius reached
by the host star during its asymptotic-giant-branch phase,
meaning that the material must have emerged during the WD stage.
More recently,
the presence of circumstellar dust
in the close vicinity of a WD
was confirmed by the discovery of periodic, irregularly shaped dips
in the light curve of \mbox{WD~1145+017}
that can be explained by transiting clouds of dust,
orbiting the star at a distance of
$ r \approx 1\,\mathrm{R\sub{\odot}}$
\citep{2015Natur.526..546V}.

WDs that exhibit IR excess also show evidence of
elements heavier than helium in their atmosphere.
This atmospheric metal pollution is a much more common phenomenon
than the IR excess,
present at detectable levels in 1/4 to 1/3 of WDs \citep{2010ApJ...722..725Z},
most of which do not exhibit detectable IR excess.
Given the short timescale on which metals sink out of
the observable part of a WD's atmosphere
(days to Myrs, depending on WD type and temperature; \citealt{2009A&A...498..517K}),
their presence must be the result of ongoing or recent accretion.
Spectroscopic analysis of polluted WDs can be used to
quantify the mass accretion rates required to maintain the quickly sinking pollution
and to obtain high-precision measurements of the elemental abundances of the accreted material.
The latter has revealed that the composition of the
accreted material
in most polluted WDs broadly
resembles that of the bulk Earth
\citep{2012MNRAS.424..333G,2014AREPS..42...45J},
although there are some interesting exceptions to this rule
\citep[e.g.,][]{2013Sci...342..218F,2017ApJ...836L...7X}.

\begin{figure}
  \includegraphics[width=\columnwidth]{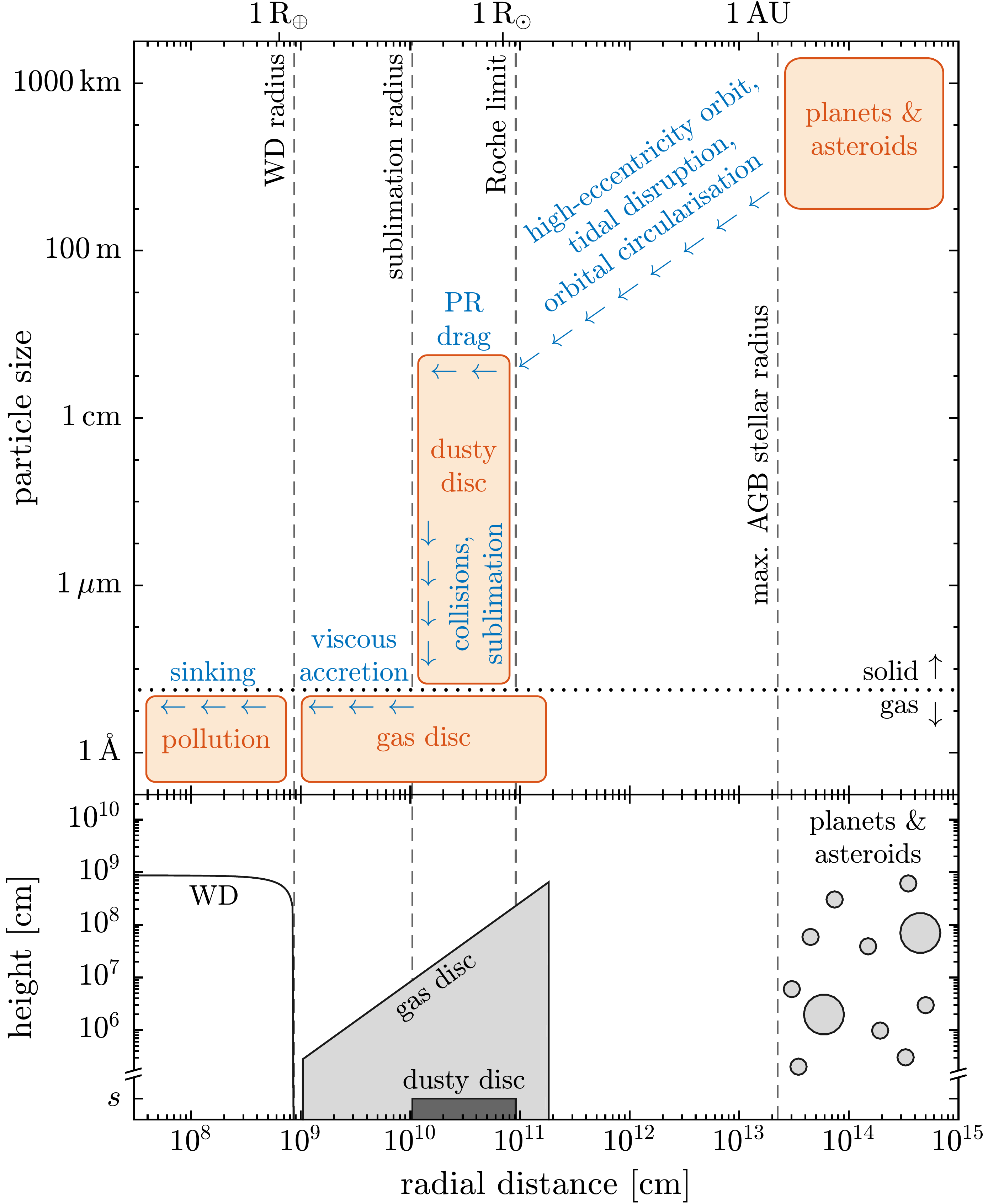}
  \caption{%
  Overview of
  a possible model for
  the circumstellar environment of polluted WDs
  and the processes leading to rocky material reaching the star.
  \textbf{Top panel:} Main
  components of a polluted WD system (red boxes)
  and the dominant mass flow through the system (blue arrows).
  \textbf{Bottom panel:} Geometry corresponding to this model.
  Note the break in the vertical axis;
  $ s $ denotes the typical radius of debris particles in the dusty disc.}
  \label{fig:overview}
\end{figure}

Given these observational clues,
the circumstellar material and atmospheric metal pollution are thought to
originate from rocky bodies in the WD's remnant planetary system,
which survives beyond
the maximum asymptotic-giant-branch stellar radius
($r \gtrsim 1\,\mathrm{AU}$).
Figure~\ref{fig:overview} (top panel) summarises
some of the current ideas on how mass may flow
from this outer reservoir of large bodies
to the WD's surface,
traversing 5 orders of magnitude in radial distance and at least 12 in particle size.
In the depicted scenario,
rocky bodies are placed onto highly eccentric, star-grazing orbits,
leading to their tidal disruption as they pass within the WD's Roche limit
\citep{2003ApJ...584L..91J}.
The debris produced in the tidal disruption is suggested to
then spread along the orbit of the progenitor \citep{2014MNRAS.445.2244V},
migrate towards the WD \citep{2015MNRAS.451.3453V},
and eventually form
a compact circumstellar disc \citep{2003ApJ...584L..91J},
from which material accretes onto the WD.
There are, however, still many unanswered questions regarding
the exact nature of the bodies that undergo this process,
the mechanism that places them on their extreme orbits,
and the physics governing the subsequent steps of disc formation and accretion.
In order to understand, for example,
what the composition of WD pollution can tell us about exoplanetary materials,
it is important to gain a better understanding of
the processes governing the circumstellar environment of polluted WDs.

In the multistage process
of transporting rocky material from a WD's remnant planetary system
to its stellar surface,
the debris-disc phase offers
a useful `checkpoint'
where theory can be compared with several sets of observations.
Specifically, models of WD debris discs must be able to simultaneously explain
the typical shape of
the IR-excess spectral energy distribution
and the mass accretion rates derived from the atmospheric metal pollution
\citep[see also][]{2017MNRAS.468..154B}.
A popular disc model is that of
\citet{2003ApJ...584L..91J}, who suggested that the discs are
compact, opaque, and flat
(much like Saturn's rings;
see also Fig.~\ref{fig:overview}, bottom panel, and Sect.~\ref{s:disc_basics}),
and showed that this model can reproduce
the properties of the IR excess.
Subsequent theoretical work on the behaviour of such discs
has shown that
the accretion of disc material due to Poynting--Robertson (PR) drag
can explain the minimum mass accretion rates found for disc-bearing WDs
\citep{2011ApJ...732L...3R,2011ApJ...741...36B}.

The global evolution of a WD debris disc depends critically on its total mass,
but this quantity cannot be fully constrained from the observed IR excess,
because the disc may be optically thick (as in the model of \citealt{2003ApJ...584L..91J}).
The discs are usually
assumed to contain an amount of mass
comparable to that of a Solar-System asteroid
($\lesssim$\,10$^{23}\,$g),
based on the measured masses of atmospheric pollution,
which reflect the amount of mass that was accreted within the last sinking timescale
\citep[see, e.g., Fig.~6 of][and references therein]{2016RSOS....350571V}.
Nonetheless,
WD debris discs that are much more massive than 10$^{23}\,$g may exist.
For one,
the distribution of pollution masses has a tail towards higher values,
with the most polluted WD known, \mbox{HE~0446-2531}, possibly containing as much as
$ 8 \times 10^{24}\,$g
(almost the mass of Pluto)
of metals in its atmosphere
\citep{2000A&A...363.1040F,2012ApJ...749..154G}.
Also,
pollution masses only provide a lower limit on the potential disc mass,
because (1)~the accretion of metals onto the WD may have been going on for much longer than one sinking timescale
and (2)~additional mass may remain locked up in the circumstellar debris disc.

The evolution of very massive WD debris disc
has not yet been investigated in detail
and
it is the subject of this paper.
Specifically, we consider discs
that have the flat geometry proposed by \citet{2003ApJ...584L..91J},
but contain the mass equivalent of
a dwarf planet, major moon, or terrestrial planet.
A possible formation scenario for such discs
may be the tidal disruption of a massive rocky body
(see also Sect.~\ref{s:occur_disc}),
but our study is ambivalent to the processes that led to the formation of the disc,
taking an already-formed disc as its starting point.

A physical process that we pay special attention to is viscous spreading.
This process is insignificant for the evolution of low-mass WD debris discs
\citep[their Appendix~A]{2008ApJ...674..431F,2012MNRAS.423..505M},
but
should be taken into account
for the high-mass discs that we consider here,
because its effects become stronger at higher surface densities.
From planetary-ring studies,
it is known that the viscous spreading of a tidal disc can result in
the formation of gravitationally bound objects at the disc's outer edge,
specifically explaining the presence and properties of
the small moons located close to the outer edge of Saturn's rings
\citep[hereafter \citetalias{2012Sci...338.1196C}]
{2010Natur.465..752C,2012Sci...338.1196C}.
We find that a similar process may
occur in massive WD debris discs.
Thus,
a large rocky body (such as a planet) that
undergoes a tidal disruption around a WD,
forming a massive debris disc,
can be partially `recycled' into
a set of second-generation (minor) planets
\citep[cf.][]{2015MNRAS.450.4233B}.
This recycling mechanism
may be the origin of the close-in, massive bodies
that have been suggested to orbit \mbox{WD~1145+017} to explain
some aspects of this WD's peculiar transits
\citep{2015Natur.526..546V,2016ApJ...818L...7G,2016MNRAS.458.3904R}.

\begin{table}
  \centering
  \caption{Definition of symbols.}
  \label{tbl:symb_defs}
 \begin{tabular}{ >{\raggedright\arraybackslash}p{0.88cm} >{\raggedright\arraybackslash}p{5.11cm} >{\centering\arraybackslash}p{1.16cm} }
    \hline
  Symbol &                              Meaning & See also \\
    \hline
  $ A $ &                               Shearing rate of adjacent annuli in the disc & Eq.~\eqref{eq:torque_visc} \\
  $ c $ &                               Speed of light & \\
  $ C\sub{R} $ &                        Prefactor in Roche-limit equation & Eq.~\eqref{eq:r_roche} \\
  $ C\sub{visc} $ &                     Self-gravitational viscosity correction factor & Eq.~\eqref{eq:visc_sg} \\
  $ D\sub{tr} $ &                       Full transit duration & Eq.~\eqref{eq:tr_dur} \\
  $ f_{\varphi\mathrm{,PR}} $ &         Azimuthal PR-drag force per unit area & Eq.~\eqref{eq:force_per_area_pr} \\
  $ G $ &                               Gravitational constant & \\
  $ L_\star $ &                         WD luminosity & Sect.~\ref{s:disc_evol_pr} \\
  $ m $ &                               Grain mass & Sect.~\ref{s:disc_basics} \\
  $ M_\star $ &                         WD mass & \\
  $ M\sub{crit} $ &                     Disc mass above which viscous spreading dominates PR drag & Eq.~\eqref{eq:mdisk_crit} \\
  $ M\sub{disc} $ &                     Initial disc mass & Eq.~\eqref{eq:sigma} \\ 
  $ M\sub{form} $ &                     Disc mass above which viscosity spreads disc to beyond Roche limit & Eq.~\eqref{eq:mdisk_form} \\
  $ M\sub{p} $ &                        Mass of (minor) planet & Sect.~\ref{s:delta_q} \\
  $ M\sub{recyc} $ &                    Total mass in newly formed (minor) planets & Eq.~\eqref{eq:f_recyc_simple} \\
  $ M\sub{T} $ &                        Toomre mass, i.e., mass enclosed in a circu- lar portion of the disc with diameter $ \lambda\sub{T} $ & Eq.~\eqref{eq:mass_toomre} \\
  $ \dot{M}\sub{out} $ &                Outward mass flow rate at the Roche limit caused by viscous spreading & Eq.~\eqref{eq:mdot_roche} \\
  $ \dot{M}\sub{PR} $ &                 PR-drag-induced inward mass flow rate & Eq.~\eqref{eq:mdot_pr_thick} \\
  $ n $ &                               Number of (minor) planets & Sect.~\ref{s:delta_q} \\
  $ N\sub{in} $ &                       Number of in-transit photometry points & Eq.~\eqref{eq:tr_snr} \\
  $ p $ &                               Integer identifying mean-motion resonance & Sect.~\ref{s:largest_planet} \\
  $ p\sub{tr} $ &                       Geometric transit probability & Eq.~\eqref{eq:tr_prob} \\
  $ P_{2:1} $ &                         Keplerian orbital period at $ r_{2:1} $ & Eq.~\eqref{eq:period_2to1} \\
  $ P\sub{R} $ &                        Keplerian orbital period at Roche limit & Eq.~\eqref{eq:period_roche} \\ 
  $ q $ &                               Planet-to-WD mass ratio & Eq.~\eqref{eq:delta_q_defs} \\
  $ q\sub{0} $ &                        $ q $ at start of pyramidal regime & Eq.~\eqref{eq:q_0} \\
  $ q\sub{c} $ &                        $ q $ at end of continuous regime & Eq.~\eqref{eq:delta_c_q_c} \\
  $ q\sub{d} $ &                        $ q $ at end of discrete regime & Eq.~\eqref{eq:delta_d_q_d} \\
  $ Q\sub{T} $ &                        Toomre's gravitational stability parameter & Eq.~\eqref{eq:q_toomre} \\
  $ r $ &                               Radial distance & \\
  $ r_0 $ &                             Initial (central) radius of the disc & \\ 
  $ r_{2:1} $ &                         Radius at which the 2:1 MMR of an orbiting body coincides with the Roche limit & Eq.~\eqref{eq:r_2to1} \\
  $ r\sub{subl} $ &                     Sublimation radius & Sect.~\ref{s:disc_basics} \\
  $ r\sub{H} $ &                        Mutual Hill radius of disc particles & Eq.~\eqref{eq:r_hill_star} \\
  $ r\sub{H}^* $ &                      $ r\sub{H} $ scaled to the sum of the particles' radii & Eq.~\eqref{eq:r_hill_star} \\
  $ r\sub{R} $ &                        Roche limit & Eq.~\eqref{eq:r_roche} \\
  $ R_\star $ &                         WD radius & \\
  $ R\sub{p} $ &                        Radius of (minor) planet & Sect.~\ref{s:delta_q} \\
  $ s $ &                               Grain radius & Sect.~\ref{s:disc_basics} \\
  $ t $ &                               Time & \\ 
  $ t\sub{c} $ &                        Time interval between formation of two bodies at $ \left\{ \varDelta\sub{c}, q\sub{c} \right\} $ & Eq.~\eqref{eq:t_c} \\
  $ t\sub{crit} $ &                     Disc evolution time at transition from viscous spreading to PR drag & Eq.~\eqref{eq:t_crit} \\
  $ t\sub{cool} $ &                     WD cooling age & Sect.~\ref{s:num_setup} \\
  $ t\sub{d} $ &                        Time interval between formation of two bodies at $ \left\{ \varDelta\sub{d}, q\sub{d} \right\} $ & Eq.~\eqref{eq:t_d} \\
  $ t\sub{disc} $ &                     Disc lifetime, i.e., until no disc mass remains & Sect.~\ref{s:disc_lifetime} \\
  $ t\sub{evol} $ &                     Disc evolution timescale, i.e., on which changes in local surface density occur & Fig.~\ref{fig:time_visc_pr} \\
  $ t\sub{PR} $ &                       PR-drag disc-evolution timescale & Eq.~\eqref{eq:t_pr_thick} \\
  $ t\sub{start} $ &                    Time when Roche-limit overflow commences & Sect.~\ref{s:sigma_roche} \\
  $ t\sub{T} $ &                        Time interval between formation of two bodies of mass $ M\sub{T}( r\sub{R} ) $ & Eq.~\eqref{eq:t_toomre} \\
  $ t\sub{visc} $ &                     Viscous disc-evolution timescale & Eq.~\eqref{eq:t_visc} \\
  $ T_\star $ &                         WD effective temperature & \\
  $ \mathcal{T} $ &                     Disc's viscous timescale at Roche limit scaled to local orbital period & Eq.~\eqref{eq:dimless_time} \\
  $ Z $ &                               Auxiliary variable for discrete regime & Eq.~\eqref{eq:delta_d_q_d} \\
  $ Z_0 $ &                             Numerical constant, $ Z_0 \approx 0.682 $ & Eq.~\eqref{eq:delta_d_q_d} \\
    \hline
  \end{tabular}
\end{table}
\begin{table}
  \centering
  \contcaption{\vphantom{Definition of symbols.}}
  \label{tbl:symb_defs2}
 \begin{tabular}{ >{\raggedright\arraybackslash}p{0.88cm} >{\raggedright\arraybackslash}p{5.11cm} >{\centering\arraybackslash}p{1.16cm} }
    \hline
  Symbol &                              Meaning & See also \\
    \hline
  $ \alpha $, $ \beta $ &               Power-law indices for distance and sur- face-density dependence of viscosity & Sect.~\ref{s:disc_evol_visc} \\
  $ \varGamma\sub{PR} $ &               PR drag torque & Eq.~\eqref{eq:torque_dens_pr} \\
  $ \varGamma\sub{visc} $ &             Viscous torque & Eq.~\eqref{eq:torque_visc} \\
  $ \delta $ &                          Initial scaled disc width & Eq.~\eqref{eq:sigma} \\
  $ \delta\sub{tr} $ &                  Transit depth & Eq.~\eqref{eq:tr_depth} \\
  $ \varDelta $ &                       Scaled distance beyond Roche limit & Eq.~\eqref{eq:delta_q_defs} \\
  $ \varDelta_0 $ &                     $ \varDelta $ at start of pyramidal regime & Eq.~\eqref{eq:delta_0} \\
  $ \varDelta_{2:1} $ &                 $ \varDelta $ at which the 2:1 MMR of an orbiting body coincides with the Roche limit & Eq.~\eqref{eq:delta_2to1} \\
  $ \varDelta\sub{c} $ &                $ \varDelta $ at end of continuous regime & Eq.~\eqref{eq:delta_c_q_c} \\
  $ \varDelta\sub{d} $ &                $ \varDelta $ at end of discrete regime & Eq.~\eqref{eq:delta_d_q_d} \\
  $ \zeta $ &                           Impinging angle of stellar radiation & Eq.~\eqref{eq:zeta} \\
  $ \eta $ &                            Numerical coefficient in Toomre's gravi- tational stability criterion, $ \eta \approx 3.36 $ & Eq.~\eqref{eq:q_toomre} \\
  $ \lambda\sub{T} $ &                  Toomre's critical wavelength & Eq.~\eqref{eq:lambda_toomre} \\
  $ \nu $ &                             Effective kinetic viscosity & Eq.~\eqref{eq:visc_sg} \\
  $ \rho_\star $ &                      WD mean density & Eq.~\eqref{eq:r_roche} \\
  $ \rho\sub{d} $ &                     Material density of disc particles & Sect.~\ref{s:disc_basics} \\
  $ \sigma_r $ &                        Radial velocity dispersion & Eq.~\eqref{eq:vel_disp_r} \\
  $ \sigma\sub{SB} $ &                  Stefan--Boltzmann constant & \\
  $ \sigma\sub{phot} $ &                Photometric precision on a single measurement in a light curve & Eq.~\eqref{eq:tr_snr} \\
  $ \varSigma $ &                       Disc's surface mass density & Eq.~\eqref{eq:sigma} \\
  $ \varSigma\sub{crit} $ &             Disc's surface density above which viscous spreading dominates PR drag & Eq.~\eqref{eq:sigma_crit_time} \\
  $ \varSigma\sub{form} $ &             Disc's surface density above which viscosity spreads out disc to beyond Roche limit & Fig.~\ref{fig:snapshots} \\
  $ \varSigma\sub{R} $ &                Disc's surface density at the Roche limit & Sect.~\ref{s:sigma_roche} \\ 
  $ \varSigma\sub{visc} $ &             surface-density profile with constant $ t\sub{visc} $ & Sect.~\ref{s:disc_evol_visc} \\ 
  $ \tau_\bot $ &                       Disc's vertical optical depth & Eq.~\eqref{eq:tau_vert} \\
  $ \tau_\parallel $ &                  Disc's optical depth to radial stellar radiation & Eq.~\eqref{eq:tau_par} \\
  $ \phi_r $ &                          Momentum-transfer efficiency for PR drag & Eq.~\eqref{eq:force_per_area_pr} \\
  $ \varOmega\sub{K} $ &                Keplerian angular frequency & Sect.~\ref{s:grav_stab} \\ 
    \hline
  \end{tabular}
\end{table}

The rest of this paper is organised as follows.
In Sect.~\ref{s:analytical}, we use analytical arguments
to investigate the properties and evolution of massive WD debris discs,
and of the second-generation (minor) planets they may produce.
In Sect.~\ref{s:numerical}, these findings are checked and expanded on
using a numerical model adapted from studies of planetary ring systems.
In Sect.~\ref{s:discussion}, we discuss
various aspects of the exoplanet-recycling scenario, including
the prospects of testing its predictions observationally;
whether the scenario can help explain
some of the unresolved properties of the transits of \mbox{WD~1145+017};
and the potential of forming second-generation exoplanets in this manner
around other astrophysical objects, such as pulsars, hot subdwarfs, and brown dwarfs.
We finish with a set of conclusions in Sect.~\ref{s:conclusions}.
For reference, a list of symbol definitions used in this paper is given
in Table~\ref{tbl:symb_defs}.

\section{Analytical considerations}
\label{s:analytical}

In this section, we analytically investigate
the evolution of massive WD debris discs
and the second-generation planets that they may produce.
We start by
introducing key concepts and assumptions, and
specifying a prescription for the disc (Sect.~\ref{s:prelim}).
Next, we assess the gravitational stability of WD debris discs,
which is important because it affects the disc's effective viscosity
(Sect.~\ref{s:grav_stab}).
We then investigate the evolution of these discs under the effects of
viscous spreading and Poynting--Robertson drag,
and determine what discs are able
to spawn a new generation of (minor) planets (Sect.~\ref{s:disc_evol}).
Lastly, we explore what the basic properties of these newly formed bodies might be
(Sect.~\ref{s:planets}).
A brief summary of our analytical findings is included at the end of the section
(Sect.~\ref{s:analytical_summary}).

\subsection{Preliminaries}
\label{s:prelim}

\subsubsection{The Roche limit}
\label{s:r_roche}

A central concept in this study is the Roche limit:
the distance within which self-gravity cannot hold together a low-mass body
against the tidal forces and Keplerian shear caused by a large central mass.\footnote{%
Sometimes, the term `Roche limit' is used to refer strictly to
the incompressible-fluid case
with $ C\sub{R} \approx 2.456 $ in Eq.~\eqref{eq:r_roche}.
We use the term more generally to mean the radial distance inside of which
tides and shear cause an approaching secondary body to break up,
or prevent a ring of particles from accumulating into a larger body
\citep{2006Icar..183..331H}.}
For material with an internal density of $ \rho\sub{d} $, orbiting a WD of
radius $ R_\star $,
mass $ M_\star $,
and mean density $ \rho_\star $,
it is given by \citep[e.g.,][]{Roche1849,1969efe..book.....C}
\begin{align}
  \label{eq:r_roche}
  r\sub{R}
    & = C\sub{R} \left( \frac{ \rho_\star }{ \rho\sub{d} } \right)^{1/3} R_\star
      = C\sub{R} \left( \frac{ 3 }{ 4 \uppi } \frac{ M_\star }{ \rho\sub{d} } \right)^{1/3}
      \nonumber \\
    & \approx 1.3 \, \mathrm{R_\odot} \;
        \Biggl( \frac{ C\sub{R} }{ 2.0 } \Biggr) \,
        \Biggl( \frac{ M_\star }{ \mathrm{0.6\,M\sub{\odot}} } \Biggr)^{1/3}
        \Biggl( \frac{ \rho\sub{d} }{ \mathrm{3\,g\,cm^{-3}} } \Biggr)^{-1/3}.
\end{align}
Here, the prefactor $ C\sub{R} $ is a constant between about
1.3 and 2.9,
which depends on the strength, shape, orientation, and spin state
of the low-mass body
\citep[see][]{1996EM&P...72..113H,2006Icar..183..331H,2008Icar..193..283H}.
The classical value found by \citet{Roche1849} is $ C\sub{R} \approx 2.456 $,
valid for homogeneous, synchronously rotating, incompressible fluid bodies.
We keep $ C\sub{R} $ as a free parameter for most of our analysis,
but when needed we take $ C\sub{R} = 2.0 $ as a fiducial value.
This value is valid for `rubble piles'
(i.e., particle aggregates held together solely by gravity)
whose shape is such that they fill their Roche lobe
(\citealt{2007Sci...318.1602P}; $ \gamma \approx 1.6 $ in their notation).
It can explain the observed densities of Saturn's innermost moonlets
\citep{2007Sci...318.1602P} and is roughly consistent with
the observed architectures of the Saturnian and Uranian systems,
where the Roche limit marks the boundary between
quasi-continuous particle rings and discrete
moons
\citep{2013ApJ...765L..28T}.

Combining Eq.~\eqref{eq:r_roche} with Kepler's third law
gives the orbital period
of a body located at the Roche limit
\begin{equation}
  \label{eq:period_roche}
  P\sub{R}
    = \sqrt{ \frac{ 3 \uppi C\sub{R}^3 }{ G \rho\sub{d} } }
    \approx 5.4 \, \mathrm{h} \;
      \Biggl( \frac{ C\sub{R} }{ 2.0 } \Biggr)^{3/2}
      \Biggl( \frac{ \rho\sub{d} }{ \mathrm{3\,g\,cm^{-3}} } \Biggr)^{-1/2}.
\end{equation}
It is worth noting that
the dependence on the central mass has been eliminated here.

\subsubsection{Basic properties of the particle disc}
\label{s:disc_basics}

We consider geometrically thin, flat, compact, axisymmetric discs,
made up of solid debris particles,
similar in many respects to Saturn's rings.
With `geometrically thin' we mean that the disc's thickness
is considerably smaller than the WD's diameter,
while `flat' means that the thickness does not vary with radius.
The disc may converge to this state through
successive energy-dissipating collisions between the constituent particles
(e.g., \citealt{1977A&A....54..895B,1979Icar...38...54C},
but see \citealt{2017ApJ...844..116K} for a contrasting view).
With `compact' we mean that the disc is located within the Roche limit $ r\sub{R} $,
as expected for a disc resulting from
a tidal disruption followed by orbital shrinking and circularisation of the debris.
Our study will concentrate mostly (but not exclusively)
on discs located in the outer parts of the tidal disruption zone
($ 0.5 \lesssim r / r\sub{R} < 1 $).
Debris discs may preferentially be found here,
because the outer parts of the tidal disruption zone
have a larger cross-section to incoming bodies
than the inner parts.

For simplicity, we assume that the particles making up the disc are all
identical spheres with radius $ s $, internal density $ \rho\sub{d} $,
and mass $ m = 4 \uppi \rho\sub{d} s^3 / 3 $.
We will use $ \rho\sub{d} = \mathrm{3\,g\,cm^{-3}} $
as a fiducial value for the density,
in line with the inferred rocky composition of
WD atmospheric pollution \citep{2014AREPS..42...45J}.
Regarding the typical sizes of particles in WD debris discs,
there are only a few (relatively weak) observational constraints:
Firstly, several WDs exhibit 10\,$\upmu$m silicate emission features,
indicating the presence of (at least some) $ s \sim 1\,\upmu$m dust
\citep[e.g.][]{2007AJ....133.1927J,2009AJ....137.3191J}.
Secondly, \citet{1990ApJ...357..216G}
put a rough upper limit of $ s \lesssim 1\,$m
on the size of the particles orbiting \mbox{G29-38}
based on their detection of IR pulsations
(interpreted as the WD's optical pulsations
that are reprocessed into the IR by the circumstellar dust,
putting an upper limit on the thermal inertia of the dust grains).
These two results bracket grain sizes of $ s \sim 1\,$cm,
which we will use as a fiducial value when one is needed.
Additional motivation for using this value
comes from the analogy with Saturn's rings,
in which particles of a few cm are the most abundant \citep[e.g.,][]{2009sfch.book..459C}.
As we will see later, however,
most of the conclusions made in this study
ultimately turn out to be independent of grain size,
as long as the particles are not too large
(see Sects.~\ref{s:visc_vs_pr} and \ref{s:num_setup}).

Our study is focussed specifically on massive discs, with masses $ M\sub{disc} $
roughly ranging from that of Ceres ($\sim$\,10$^{24}\,$g)
to that of a super-Earth (10$^{28}$ to 10$^{29}\,$g).
Assuming that the disc initially constitutes
a relatively narrow annulus at radius $ r $,
its surface mass density can be estimated as
\begin{equation}
  \label{eq:sigma}
  \varSigma
    = \frac{ M\sub{disc} }{ 2 \uppi \delta r^2 }
    \approx 3 \times 10^4 \, \mathrm{g\,cm^{-2}} \;
      \Biggl( \frac{ \delta }{ 0.1 } \Biggr)^{-1}
      \Biggl( \frac{  M\sub{disc} }{ 10^{26}\,\mathrm{g} } \Biggr) \,
      \Biggl( \frac{ r }{ 1\,\mathrm{R_\odot} } \Biggr)^{-2},
\end{equation}
where $ \delta $ is a dimensionless factor that characterises the disc's width
(i.e., the disc width is roughly \mbox{$ \delta \times r $}).
Using the assumption of equal-sized, spherical grains,
we can write the vertical optical depth of the disc as
\begin{equation}
  \label{eq:tau_vert}
  \tau_\bot
    = \frac{ 3 }{ 4 } \frac{ \varSigma }{ \rho\sub{d} s }
    \approx 3 \times 10^3 \;
      \Biggl( \frac{ \rho\sub{d} }{ \mathrm{3\,g\,cm^{-3}} } \Biggr)^{-1}
      \Biggl( \frac{ s }{ \mathrm{1\,cm} } \Biggr)^{-1}
      \Biggl( \frac{ \varSigma }{ \mathrm{10^4\,g\,cm^{-2}} } \Biggr).
\end{equation}
This also assumes that the grains can be treated
using the geometrical-optics approximation, which is the case
if the particles are much larger than
the wavelength of the radiation under consideration.
Since the disc is assumed to be vertically much smaller than the WD,
but horizontally much larger,
stellar radiation impinges on the disc at a shallow angle $ \zeta $.
Following the `lamp-post illumination model' of
\citeauthor{2011ApJ...732L...3R} (\citeyear{2011ApJ...732L...3R}; see also \citealt{1985A&A...146..366F}),
this impinging angle is
\begin{equation}
  \label{eq:zeta}
  \zeta
    \simeq \frac{ 4 }{ 3 \uppi } \frac{ R_\star }{ r }
    \approx 0.3\degr \;
      \Biggl( \frac{ R\sub{\star} }{ \mathrm{0.0125\,R\sub{\odot}} } \Biggr) \,
      \Biggl( \frac{ r }{ 1\,\mathrm{R_\odot} } \Biggr)^{-1},
\end{equation}
and the disc's optical depth to the impinging stellar radiation
is
\begin{align}
  \label{eq:tau_par}
  \tau_\parallel
      = \frac{ \tau_\bot }{ \zeta }
      \simeq \frac{ 9 \uppi }{ 16 } \frac{ \varSigma }{ s \rho\sub{d} } \frac{ r }{ R_\star }
    & \approx 5 \times 10^5 \;
        \Biggl( \frac{ R\sub{\star} }{ \mathrm{0.0125\,R\sub{\odot}} } \Biggr) \,
        \Biggl( \frac{ \rho\sub{d} }{ \mathrm{3\,g\,cm^{-3}} } \Biggr)^{-1}
    \nonumber \\
    & \quad \times
        \Biggl( \frac{ s }{ \mathrm{1\,cm} } \Biggr)^{-1}
        \Biggl( \frac{ \varSigma }{ \mathrm{10^4\,g\,cm^{-2}} } \Biggr) \,
        \Biggl( \frac{ r }{ 1\,\mathrm{R_\odot} } \Biggr)^{-1}.
\end{align}
This shows that the massive discs considered in this study
can be assumed to be opaque to the stellar radiation.

\subsubsection{The sublimation radius and the gaseous disc}
\label{s:r_subl}

Very close to the WD,
disc temperatures will become high enough for the dust material to sublimate.
The distance $ r\sub{subl} $ at which this happens
depends on the WD temperature, disc composition, and ambient gas pressure.
For simplicity, we use a fixed value of
$ r\sub{subl} = 0.15\,\mathrm{R_\odot} $,
roughly consistent with theoretical expectations and observational estimates
\citep[e.g.,][]{2007ApJ...663.1285J,2012ApJ...760..123R}.
Inside this sublimation radius, the solid debris quickly turns to gas,
which is assumed to then rapidly accrete onto the WD
via a gaseous viscous accretion disc
\citep[e.g.,][]{2012MNRAS.423..505M}.

In this work we mostly ignore the possible back-reaction of the gas
that spreads outwards from $ r\sub{subl} $ on the particle disc.
\citet{2011MNRAS.416L..55R} and \citet{2012MNRAS.423..505M}
have found that
the drag due to this gas
may cause a run-away accretion of the particle disc onto the WD.
Whether this process happens, however, depends critically on
the strength of the positive feedback of the gas on the solid disc
(called $ \mathcal{F} $ by \citealt{2011MNRAS.416L..55R} and \citealt{2012MNRAS.423..505M}),
which is highly uncertain because it is controlled by
ill-constrained parameters of the gas disc such as its viscosity ($ \alpha $)
and the critical Reynolds number above which the gas flow becomes fully turbulent
($\mathit{R\kern-.04em e}_*$).
Ignoring gas drag simply corresponds to assuming a low feedback
($ \mathcal{F} < 1 $).
Furthermore, gas
outside the sublimation radius
is expected to rapidly condense onto the solid particle disc,
which may hinder the run-away process.
Nevertheless, we briefly discuss the possibility of disc destruction
due to gas-drag-induced run-away accretion in Sect.~\ref{s:destroy_disc}.

\subsection{Gravitational stability of the disc}
\label{s:grav_stab}

To evaluate the gravitational stability of massive WD debris discs,
we follow the analyses of \citet{2001Icar..154..296D} and \citet{2009sfch.book..413S}
for planetary rings.
Given the (presumed) similarity between WD debris discs and planetary rings,
their reasoning can be directly applied here.

The gravitational stability of a differentially rotating disc
against axisymmetric perturbations
can be assessed using Toomre's parameter $ Q\sub{T} $ \citep{1964ApJ...139.1217T},
which for a Keplerian particle disc is
\begin{equation}
  \label{eq:q_toomre}
  Q\sub{T} = \frac{ \sigma_r \varOmega\sub{K} }{ \eta G \varSigma },
  \qquad
  \eta \approx 3.36.
\end{equation}
Here,
$ \sigma_r $ is the radial velocity dispersion of the disc particles,
$ \varOmega\sub{K} = \sqrt{ G M_\star / r^3 } $ is the Keplerian angular frequency,
and $ G $ is the gravitational constant.
Gravitational instabilities occur for $ Q_\mathrm{T} \lesssim 2 $
and have a typical length scale of the order of
Toomre's critical wavelength
\citep{1964ApJ...139.1217T,1966ApJ...146..810J}
\begin{align}
  \label{eq:lambda_toomre}
  \lambda\sub{T}
    & = \frac{ 4 \uppi^2 G \varSigma }{ \varOmega\sub{K}^2 }
      = 3 \uppi \left( \frac{ C\sub{R} r }{ r\sub{R} } \right)^3
          \frac{ \varSigma }{ \rho\sub{d} } \nonumber \\
    & \approx 2.5 \, \mathrm{km} \;
      \Biggl( \frac{ C\sub{R} }{ 2.0 } \Biggr)^{3}
      \Biggl( \frac{ \rho\sub{d} }{ \mathrm{3\,g\,cm^{-3}} } \Biggr)^{-1}
      \Biggl( \frac{ \varSigma }{ \mathrm{10^4\,g\,cm^{-2}} } \Biggr) \,
      \Biggl( \frac{ r / r\sub{R} }{ 1 } \Biggr)^{3}.
\end{align}

To determine the correct expression for the radial velocity dispersion $ \sigma_r $,
it is important to determine what processes govern the motions of the particles.
In the discs considered here,
low-velocity, bouncing collisions are frequent
and they set the base level for the particle motions.
These physical impacts regulate the velocity dispersion to be of the order of
the difference in orbital velocities between radially adjacent particles
$ 2 s \varOmega\sub{K} $ \citep{1977A&A....54..895B,1988AJ.....95..925W,1995Icar..117..287S}.
The velocity dispersion can be enhanced above this base level
by close gravitational encounters between the disc particles,
which yield motions of the order of
their mutual escape speed
$ \sqrt{ 2 G m / s } $ \citep[e.g.,][]{1995Icar..117..287S,1999Icar..137..152O}.
If the velocity dispersion due to physical impacts is already higher than this,
however, the effect of gravitational encounters will be negligible.
Thus, the radial velocity dispersion can be expressed as
\begin{equation}
  \label{eq:vel_disp_r}
  \sigma_r \sim \max \left[
    2 s \varOmega\sub{K},
    \sqrt{ 2 G m / s } \;\!
  \right].
\end{equation}
The transition between the two regimes in this equation occurs at
$ r / r\sub{R} = 2^{1/3} / C\sub{R} \approx 0.63 \left( C\sub{R} / 2.0 \right)^{-1} $.

Inserting the expression for $ \sigma_r $ into Eq.~\eqref{eq:q_toomre},
together with $ Q_\mathrm{T} \lesssim 2 $,
and rewriting the result using Eqs.~\eqref{eq:r_roche} and \eqref{eq:tau_vert}
gives the relatively simple gravitational stability criterion
\begin{equation}
  \label{eq:tau_grav_stab}
  \tau_\bot \gtrsim \max \left[
    \frac{ \uppi }{ \eta } \left( \frac{ r\sub{R} }{ C\sub{R} r } \right)^3,
    \sqrt{ \frac{ 1 }{ 2 } } \frac{ \uppi }{ \eta }
      \left( \frac{ r\sub{R} }{ C\sub{R} r } \right)^{3/2}
  \right],
\end{equation}
or equivalently \citep[cf. Eq.~(13) of][]{2001Icar..154..296D}
\begin{equation}
  \label{eq:tau_grav_stab2}
  \tau_\bot \gtrsim \max \left(
    0.08 {r\sub{H}^*}^{-3}, \,
    0.2 {r\sub{H}^*}^{-3/2}
  \right),
\end{equation}
with the transition between the two regimes
at $ r\sub{H}^* = 6^{-1/3} \approx 0.55 $.
Here, the auxiliary variable $ r\sub{H}^* $
(closely related to the Roche limit)
is defined as the ratio of
the mutual Hill radius of two disc particles
$ r\sub{H} = [ 2 m / ( 3 M_\star ) ]^{1/3} r $
to the sum of their physical radii:
\begin{align}
  \label{eq:r_hill_star}
  r\sub{H}^*
    = \frac{ r\sub{H} }{ 2 s }
    & = \left( \frac{ \uppi }{ 9 } \frac{ \rho\sub{d} }{ M_\star } \right)^{1/3} r
      \approx 0.7 \;
      \Biggl( \frac{ M\sub{\star} }{ \mathrm{0.6\,M\sub{\odot}} } \Biggr)^{-1/3}
      \Biggl( \frac{ \rho\sub{d} }{ \mathrm{3\,g\,cm^{-3}} } \Biggr)^{1/3}
      \Biggl( \frac{ r }{ 1\,\mathrm{R_\odot} } \Biggr) \nonumber \\
    & = \frac{ C\sub{R} }{ 12^{1/3} } \frac{ r }{ r\sub{R} }
      \approx 0.87 \;
        \Biggl( \frac{ C\sub{R} }{ 2.0 } \Biggr) \,
        \Biggl( \frac{ r / r\sub{R} }{ 1 } \Biggr).
\end{align}

\begin{figure}
  \includegraphics[width=\columnwidth]{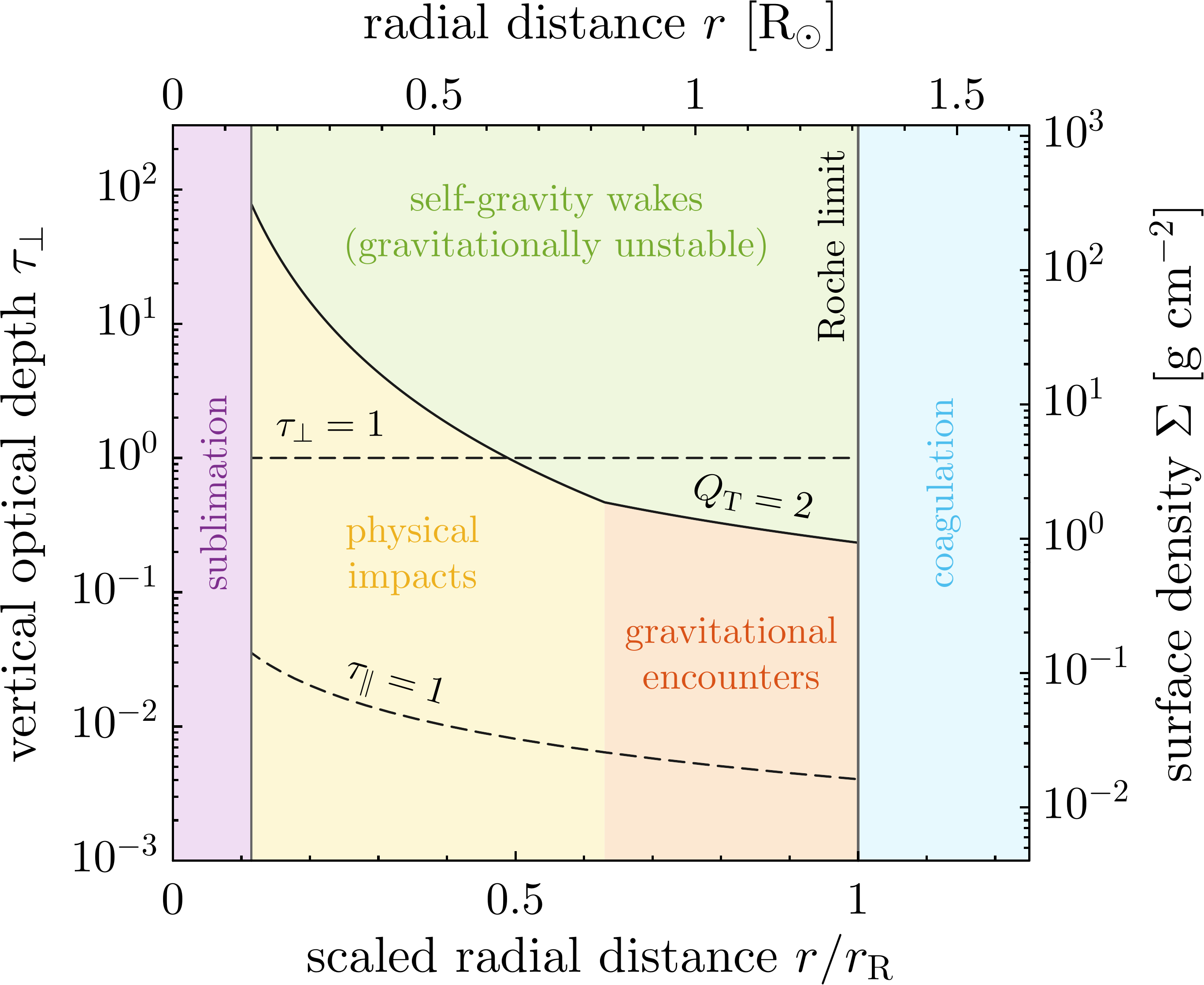}
  \caption{%
  Gravitational stability criterion for a particle disc
  (solid line; Eq.~\eqref{eq:tau_grav_stab}),
  as a function of radial distance $ r $ scaled to the Roche limit $ r\sub{R} $,
  assuming a Roche prefactor of $ C\sub{R} = 2.0 $.
  Coloured shadings demarcate parts of parameter space
  in which different processes dominate
  \citep[cf. Fig.~14.7 of][]{2009sfch.book..413S}.
  Dashed lines indicate where the disc becomes optically thick
  vertically ($ \tau_\bot = 1 $) and optically thick to the impinging stellar radiation
  ($ \tau_\parallel = 1 $; Eq.~\eqref{eq:tau_par}; valid for the top axis
  and a stellar radius of $ R_\star = 0.0125\,\mathrm{R\sub{\odot}} $).
  Vertical lines mark the dust sublimation radius
  (at $ r\sub{subl} = 0.15\,\mathrm{R_\odot} $)
  and the Roche limit.
  The absolute distances indicated along the top axis
  assume a stellar mass of $ M_\star = 0.6\,\mathrm{M\sub{\odot}} $,
  a material density of $ \rho\sub{d} = 3\,\mathrm{g\,cm}^{-3} $,
  and $ C\sub{R} = 2.0 $ (see Eq.~\eqref{eq:r_roche}).
  The surface densities indicated along the right-hand-side axis
  assume $ \rho\sub{d} = 3\,\mathrm{g\,cm}^{-3} $
  and a particle size of $ s = 1\,\mathrm{cm} $
  (see Eq.~\eqref{eq:tau_vert}).}
  \label{fig:grav_stab}
\end{figure}

Figure~\ref{fig:grav_stab} shows the gravitational stability criterion.
Comparing the typical vertical optical depth of
the discs we consider in this study
($ \tau_\bot \gtrsim 10^2 $; see Eq.~\eqref{eq:tau_vert}) to this criterion
demonstrates that the discs are clearly gravitationally unstable.
In fact, gravitational instabilities will occur in
any vertically optically thick disc
located in the outer half of the tidal disruption zone.
In such discs, the interplay between
gravitational instability, tidal forces, and Keplerian shear
will lead to the continuous creation and destruction of
transient density enhancements known as self-gravity wakes
(\citealt{1992Natur.359..619S}; see also \citealt{1966ApJ...146..810J}).
The wakes are typically spaced by
distances of around $ \lambda\sub{T} $
and their presence has been demonstrated
observationally for Saturn's dense A and B rings \citep{2009sfch.book..375C}.
In a gravitationally unstable disc,
the radial velocity dispersion no longer follows Eq.~\eqref{eq:vel_disp_r}
but instead adjusts to a higher value
such that $ Q\sub{T} \simeq 2 $
\citep{1995Icar..117..287S,2012PThPS.195...48S}.
Importantly,
the self-gravity wakes strongly affect the disc's effective viscosity:
the wake structures lead to systematic motions of the disc particles
and they exert gravitational torques,
thereby enhancing the angular-momentum transport in the disc \citep{2001Icar..154..296D}.

\subsection{Disc evolution}
\label{s:disc_evol}

Two processes are likely to be important for
the evolution of massive debris discs around WDs:
viscous spreading \citep[see, e.g.,][]{1981ARA&A..19..137P}
and Poynting--Robertson (PR) drag \citep[see, e.g.,][]{1979Icar...40....1B}.
The former is known to govern the long-term evolution of massive planetary rings
\citep{2010Icar..209..771S},
while the latter is key in the evolution of (asteroid-mass) WD debris discs
\citep{2011ApJ...732L...3R,2011ApJ...741...36B}.
In this section, we review how both processes operate
and determine which of the two dominates under what conditions.
A more rigorous analysis of their interplay using a torque-density balance
is presented in Appendix~\ref{app:visc_pr_torque}.

\begin{figure*}
  \includegraphics[height=6.9cm]{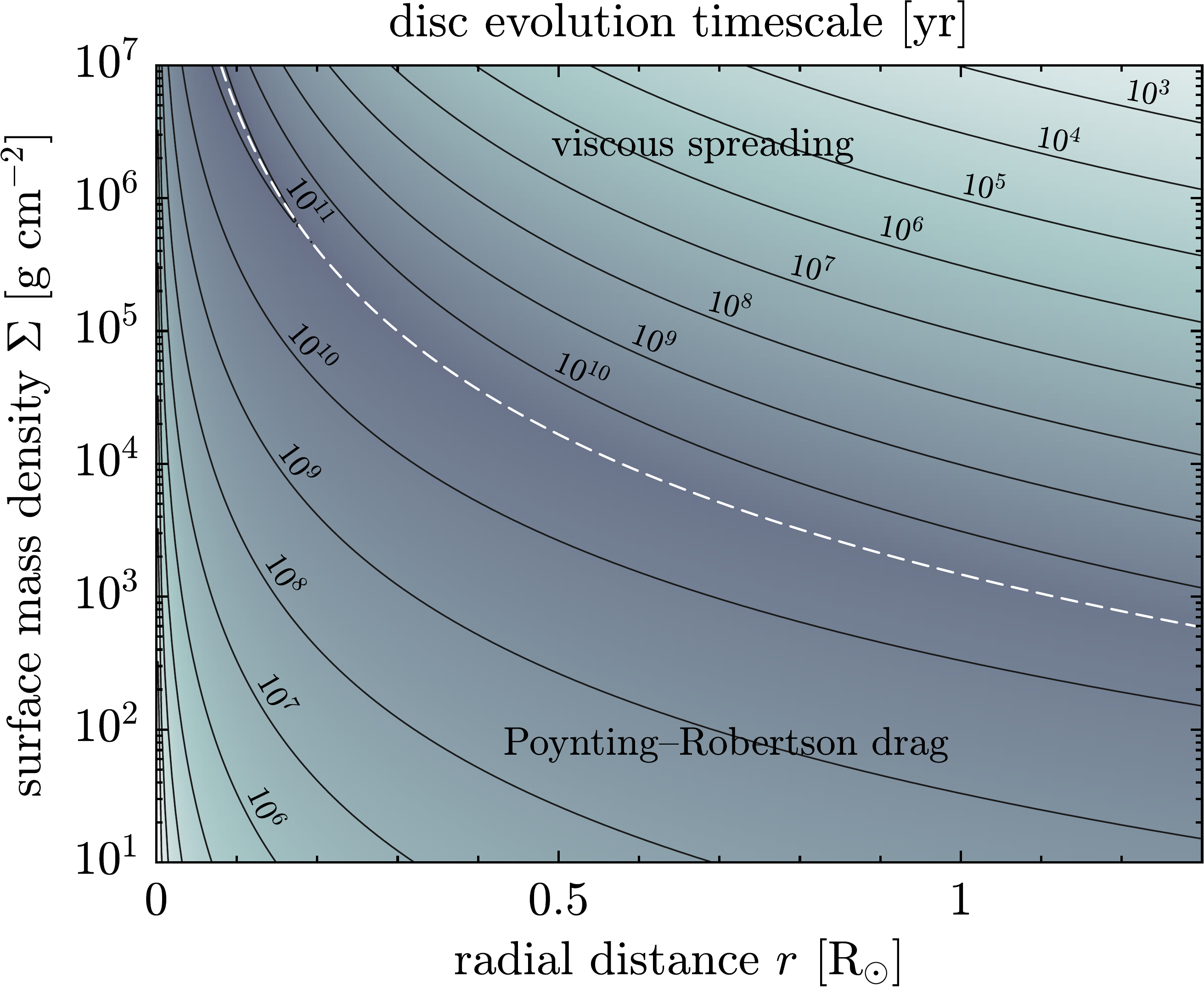}
  \hfill
  \includegraphics[height=6.9cm]{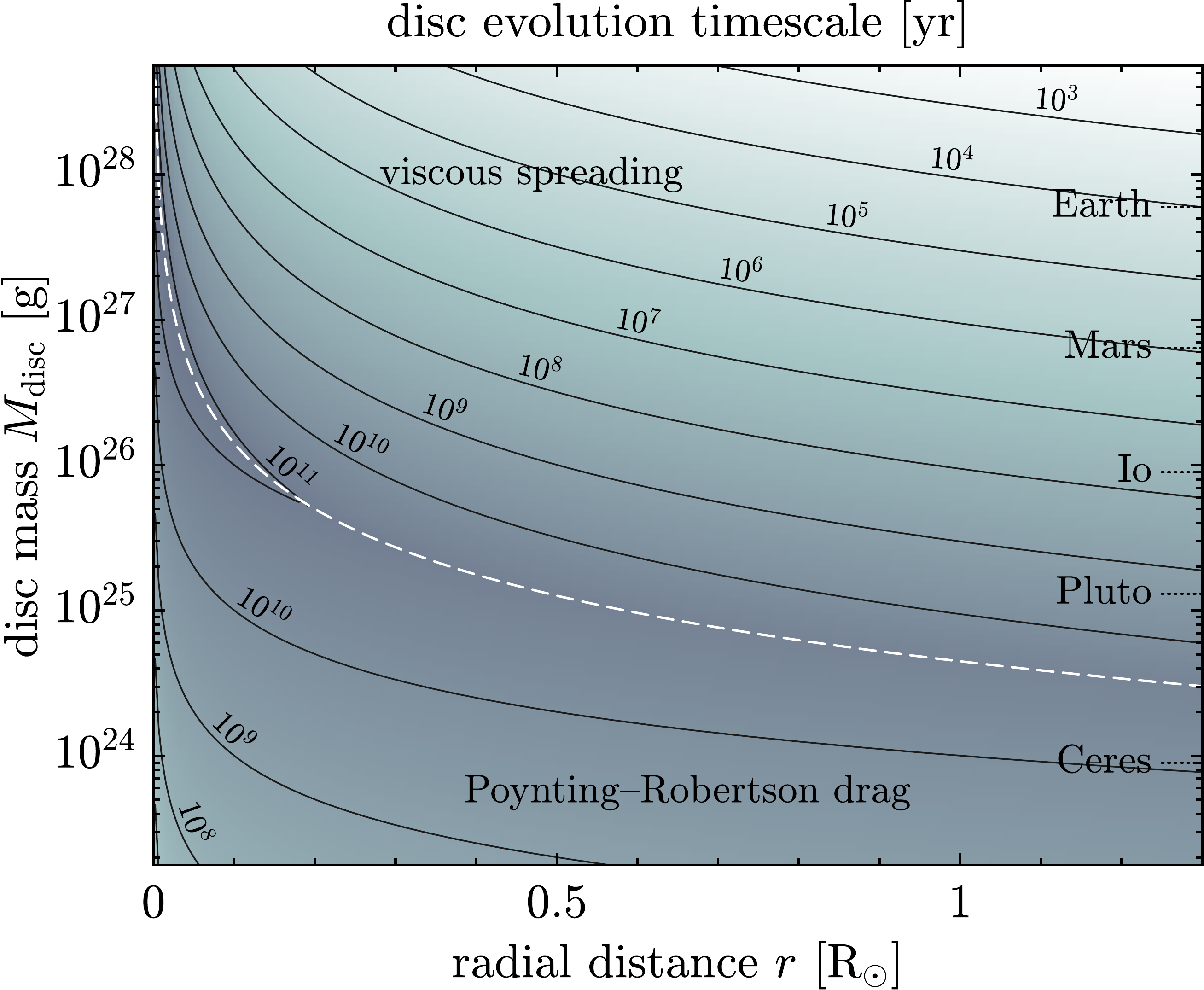}
  \caption{%
  Contours of equal disc evolution timescale
  $ t\sub{evol} = \min \left( t\sub{visc}, t\sub{PR} \right) $
  as a function of radial distance $ r $
  and surface mass density $ \varSigma $ (\textbf{left-hand panel})
  or disc mass $ M\sub{disc} $ (\textbf{right-hand panel}).
  The dashed white lines indicate conditions
  for which the two timescales are equal
  (corresponding to Eqs.~\eqref{eq:sigma_crit_time} and \eqref{eq:mdisk_crit}).
  We have assumed stellar parameters
  $ T_\star = 10{,}000\,\mathrm{K} $,
  $ M_\star = 0.6\,\mathrm{M\sub{\odot}} $,
  $ R_\star = 0.0125\,\mathrm{R\sub{\odot}} $,
  and a material density of $ \rho\sub{d} = 3\,\mathrm{g\,cm}^{-3} $.
  Disc masses were computed from surface densities
  using a disc-width parameter of $ \delta = 0.1 $.
  For reference, the masses of several Solar-System objects are indicated.}
  \label{fig:time_visc_pr}
\end{figure*}

\subsubsection{Viscous evolution}
\label{s:disc_evol_visc}

Viscous spreading of an astrophysical disc
transports angular momentum outwards, along with some of the disc's mass,
while moving most of the mass inwards.
The disc's viscosity depends on whether it is gravitationally stable.
As determined in Sect.~\ref{s:grav_stab},
the discs considered in this study are gravitationally unstable
and exhibit self-gravity wakes that
enhance the effective viscosity.
Dimensional analysis can be used to show
that the effective kinetic viscosity
of a gravitationally unstable (fluid) disc should be of the form
 $ \nu \sim \lambda\sub{T}^2 \varOmega\sub{K} \sim G^2 \varSigma^2 / \varOmega\sub{K}^3 $
\citep[e.g.,][]{1987MNRAS.225..607L}.
For particle discs, \citeauthor{2001Icar..154..296D}
(\citeyear{2001Icar..154..296D}; see also \citealt{2012AJ....143..110Y}) find
\begin{equation}
  \label{eq:visc_sg}
  \nu
    \simeq C\sub{visc}( r\sub{H}^{*} )
      \frac{ G^2 \varSigma^2 }{ \varOmega\sub{K}^3 },
  \qquad C\sub{visc}( r\sub{H}^{*} ) \simeq 26 {r\sub{H}^*}^5.
\end{equation}
Here, $ C\sub{visc} $ is an empirical correction factor, determined from
a fit to viscosities derived from numerical simulations,
with values $ 0.41 \lesssim C\sub{visc} \lesssim 13 $
in the range $ 0.5 \leq r / r\sub{R} \leq 1 $ (assuming $ C\sub{R} = 2.0 $).
The strong additional radial dependence
in $ C\sub{visc} $
is thought to be a result of the incompressibility of the particles
that make up the disc.

The evolution of discs whose viscosity is given by
a double power-law function of radius and surface density
(i.e., $ \nu \propto \varSigma^{\;\!\alpha} r^{\,\beta} $)
can usually be described analytically in terms of similarity solutions
\citep{1991MNRAS.248..754P}.
Unfortunately,
the power-law exponents in the present case
($ \alpha = 2 $, $ \beta = 19/2 $)
fall outside the regime in which these known solutions are valid.
The approximate evolution of the disc, however,
can still be understood
by considering its viscous timescale,
i.e., the characteristic timescale on which viscous spreading causes
changes in the local surface density of the disc.
For an initially narrow disc at radius $ r $,
the viscous timescale is given by
\citep[e.g.,][]{1981ARA&A..19..137P,Hahn2009}
\begin{equation}
  \label{eq:t_visc}
  t\sub{visc}
    \sim \frac{ r^2 }{ 12 \nu }
    \simeq
      \frac{ 3^{7/3} }{ 104 \uppi^{5/3} } 
      \frac{ 1 }{ \sqrt{ G } }
      \frac{ M_\star^{19/6} }{ \rho\sub{d}^{5/3} }
      \frac{ 1 }{ \varSigma^2 r^{15/2} }.
\end{equation}
Because the effective viscosity caused by gravitational instabilities increases rapidly with distance,
the outer parts of the disc evolve much faster than the inner parts.
For an initially narrow disc,
the outer edge will therefore move out faster than the inner edge moves in.
To conserve angular momentum, however, most of the disc's mass moves inwards.
The combined effect of these different factors is that
the disc develops a pile-up of material close to its inner edge.
After some time, when the disc's initial radial distribution has faded,
the surface density will roughly follow a profile of equal viscous timescale,
with a radial dependence $ \varSigma\sub{visc} \propto r^{-15/4} $.

\subsubsection[Poynting-Robertson-drag disc evolution]{Poynting--Robertson-drag disc evolution}
\label{s:disc_evol_pr}

Stellar radiation that hits the disc exerts PR drag on the disc
and as a result disc material will spiral inwards towards the WD.
For a disc that is optically thick to the impinging stellar radiation ($ \tau_\parallel \gg 1 $),
the PR-drag-induced mass flow is given by \citep{2011ApJ...732L...3R}
\begin{equation}
  \label{eq:mdot_pr_thick}
  \dot{M}\sub{PR}(r)
    \simeq \zeta \frac{ L_\star }{ c^2 }
    \simeq \frac{ 16 }{ 3 } \frac{ \sigma\sub{SB} }{ c^2 } \frac{ T_\star^4 R_\star^3 }{ r }.
\end{equation}
Here,
$ c $ is the speed of light,
$ L_\star = 4 \uppi R_\star^2 \sigma\sub{SB} T_\star^4 $ is the WD's luminosity,
$ \sigma\sub{SB} $ is the Stefan--Boltzmann constant,
and $ T_\star $ is the WD's effective temperature.
Note that this mass flow (for an optically thick disc) is independent of
any disc properties such as surface density $ \varSigma $ or grain size $ s $.

The global evolution of PR-drag-dominated WD debris discs was studied
in further detail by \citet{2011ApJ...741...36B}.
They find that 
the typical timescale for the evolution of optically thick discs is given by\footnote{%
This `PR-drag disc-evolution timescale' is
not to be confused with the more commonly used `PR-drag timescale',
i.e., the time it takes for an individual particle to migrate
from some radius onto its host star.
These two are only the same (to within a factor of order unity)
for optically thin discs \citep{2011ApJ...741...36B}.}
\begin{equation}
  \label{eq:t_pr_thick}
  t\sub{PR}
    \sim \frac{ 2 \uppi \varSigma r^2 }{ \dot{M}\sub{PR} }
    \simeq \frac{ 3 \uppi }{ 8 } \frac{ c^2 }{ \sigma\sub{SB} }
        \frac{ 1 }{ T_\star^4 R_\star^3 } \varSigma r^3.
\end{equation}
Specifically for narrow discs
located in outer parts of the tidal disruption zone
(i.e., the type of discs we focus on here),
\citet[see their Sect.~4.1.1]{2011ApJ...741...36B} find the following behaviour:
(1)~surface densities in the bulk of the disc gradually go down;
(2)~the disc develops a sharp outer edge that moves inwards with time; and
(3)~an optically thin ($ \tau_\parallel \lesssim 1 $) tail appears,
extending inwards from the inner edge of the optically thick region,
through which the disc material drains into the sublimation zone.

\subsubsection[Viscous spreading vs. Poynting-Robertson drag]{Viscous spreading vs. Poynting--Robertson drag}
\label{s:visc_vs_pr}

The scaling of the viscous and PR-drag disc-evolution timescales
with surface density and radius
($ t\sub{visc} \propto \varSigma^{-2} r^{-15/2} $;
$ t\sub{PR} \propto \varSigma r^3 $)
shows that
viscous spreading plays a larger role in the evolution of
dense discs at large distances,
while PR drag is more important close to the star
and for lower surface densities
(see Fig.~\ref{fig:time_visc_pr}).
By setting the two timescales equal to each other
($ t\sub{visc} = t\sub{PR} $)
and solving for $ \varSigma $, we find the critical surface mass density
$ \varSigma\sub{crit} $ above which
viscous spreading dominates the disc's evolution rather than PR drag
(dashed white line in Fig.~\ref{fig:time_visc_pr}, left-hand panel)
\begin{equation}
  \label{eq:sigma_crit_time}
  \varSigma\sub{crit}
    \simeq
        \frac{ 3^{4/9} }{ 13^{1/3} \uppi^{8/9} }
        \frac{ \sigma\sub{SB}^{1/3} }{ c^{2/3} G^{1/6} }
        \frac{ T_\star^{4/3} M_\star^{19/18} R_\star }{ \rho\sub{d}^{5/9} }
        \frac{ 1 }{ r^{7/2} }.
\end{equation}
For an assumed value for the disc-width parameter $ \delta $,
this critical surface density translates (via Eq.~\eqref{eq:sigma}) into
a critical disc mass separating viscous-spreading and PR-drag dominated discs
(dashed white line in Fig.~\ref{fig:time_visc_pr}, right-hand panel)
\begin{align}
  \label{eq:mdisk_crit}
  M\sub{crit}
    & \simeq
        \frac{ 2 \times 3^{4/9} \uppi^{1/9} }{ 13^{1/3} } 
        \frac{ \sigma\sub{SB}^{1/3} }{ c^{2/3} G^{1/6} }
        \frac{ T_\star^{4/3} M_\star^{19/18} R_\star }{ \rho\sub{d}^{5/9} }
        \frac{ \delta }{ r^{3/2} } \nonumber \\
    & \approx
        4.5 \times 10^{24} \, \mathrm{g} \; 
        \Biggl( \frac{ T\sub{\star} }{ \mathrm{10{,}000\,K} } \Biggr)^{4/3}
        \Biggl( \frac{ M\sub{\star} }{ \mathrm{0.6\,M\sub{\odot}} } \Biggr)^{19/18}
        \Biggl( \frac{ R\sub{\star} }{ \mathrm{0.0125\,R\sub{\odot}} } \Biggr)
    \nonumber \\
    & \quad \times
        \Biggl( \frac{ \rho\sub{d} }{ \mathrm{3\,g\,cm^{-3}} } \Biggr)^{-5/9}
        \Biggl( \frac{ \delta }{ 0.1 } \Biggr) \,
        \Biggl( \frac{ r }{ \mathrm{1\,R}\sub{\odot} } \Biggr)^{-3/2}.
\end{align}

Noting that the timescale prescriptions used to derive
Eq.~\eqref{eq:sigma_crit_time}
are order-of-magnitude estimates,
we rederive $ \varSigma\sub{crit} $ in Appendix~\ref{app:visc_pr_torque}
using a torque-density balance between viscous spreading and PR drag.
The expression for $ \varSigma\sub{crit} $ found from
this exercise is the same as Eq.~\eqref{eq:sigma_crit_time}
except for a small difference in the numerical coefficient, which is a factor
$ (5/2)^{1/3} \approx 1.36 $ 
lower in the torque-density-balance result.
For reasons of internal consistency,
we continue our analysis using
Eq.~\eqref{eq:sigma_crit_time} as prescription for $ \varSigma\sub{crit} $.

We stress that the boundary between
viscous and PR-drag disc evolution
is independent of grain size.
This is the case as long as the particles are small enough to ensure
that the disc is gravitationally unstable,
so that using Eq.~\eqref{eq:visc_sg} for the effective viscosity is valid.
Comparing the typical values of $ \varSigma\sub{crit} $
(Fig.~\ref{fig:time_visc_pr})
with the gravitational-stability criterion
(Fig.~\ref{fig:grav_stab})
shows that a marginally viscous-spreading-dominated disc
will be gravitationally unstable for particle sizes of $ s \lesssim 1 $\,m
(through Eq.~\eqref{eq:tau_vert}).
For more massive discs this maximum grain size goes up,
making the validity requirement less stringent.
Furthermore, Fig.~\ref{fig:grav_stab} shows that
any disc that is gravitationally unstable is also
optically thick to the stellar irradiation ($ \tau_\parallel \gg 1 $),
so our treatment of PR drag in the derivation of $ \varSigma\sub{crit} $
is justified as well.

\subsubsection{Planet-producing discs}
\label{s:mdisc_form}

Discs whose initial mass
is substantially above the critical level $ M\sub{crit} $
may viscously spread to beyond the Roche limit,
where disc material can
coagulate
to ultimately form a new generation of (minor) planets
(see Sect.~\ref{s:planets}).
More specifically,
there is a disc mass $ M\sub{form} $
above which an initially narrow disc
located at radius $ r_0 $
will radially spread enough
for some of its mass to end up beyond the Roche limit.
We now proceed to estimate this minimum disc mass for planet formation.

An analytical approximation to $ M\sub{form} $ can be derived by assuming that
half of the disc's initial mass spreads outwards from $ r_0 $
and that this material follows
a radial surface-density profile $ \varSigma\sub{visc} \propto r^{-15/4} $
(appropriate for viscously spreading discs; see Sect.~\ref{s:disc_evol_visc}).
For the disc's outer edge to reach the Roche limit,
the outwards-spreading material needs to maintain a surface density above
the critical level $ \varSigma\sub{crit} $
at all distances between the initial radius and the Roche limit
(otherwise PR drag will halt further outward spreading).
Because $ \varSigma\sub{visc} $ decreases (slightly) faster with radius than
$ \varSigma\sub{crit} $,
one only needs to assess this
requirement
at the most distant point, the Roche limit.
The least massive disc to reach the Roche limit will thus have
a surface-density profile that
has a radial dependence of $ \varSigma\sub{visc} \propto r^{-15/4} $
and passes through the point
$ \left\{ r\sub{R}, \varSigma\sub{crit}( r\sub{R} ) \right\} $.
The profile that meets these two conditions
is given by
\begin{equation}
  \label{eq:sigma_visc_t_trans_roche}
    \varSigma\sub{visc}[ t\sub{crit}( r\sub{R} ) ]
    \simeq
      \frac{ 3^{19/36} C\sub{R}^{1/4} }{ 2^{1/6} 13^{1/3} \uppi^{35/36} } 
      \frac{ \sigma\sub{SB}^{1/3} }{ c^{2/3} G^{1/6} }
      \frac{ T_\star^{4/3} M_\star^{41/36} R_\star }{ \rho\sub{d}^{23/36} }
      \frac{ 1 }{ r^{15/4} },
\end{equation}
where $ t\sub{crit} $ is to be defined later
(see Sect.~\ref{s:disc_lifetime}).\footnote{%
Equation~\eqref{eq:sigma_visc_t_trans_roche} can be derived from
Eq.~\eqref{eq:t_visc} by setting $ t\sub{visc} = t\sub{crit}( r\sub{R} ) $,
as given by Eq.~\eqref{eq:t_crit}, and solving for $ \varSigma $.}

Following our assumption that half the disc's mass spreads outwards,
a viscously spreading disc will reach the Roche limit if
half its mass $ M\sub{disc} / 2 $ exceeds the mass
contained between $ r_0 $ and $ r\sub{R} $
in the surface-density distribution
given by Eq.~\eqref{eq:sigma_visc_t_trans_roche}.
Thus, the minimum disc mass for planet formation
can be calculated as
\begin{align}
  \label{eq:mdisk_form}
  M\sub{form}
    & \simeq
        \int\limits_{ r_0 }^{ r\sub{R} }
          4 \uppi r \varSigma\sub{visc}[ t\sub{crit}( r\sub{R} ) ] \dif r
    \nonumber \\
    & \simeq
        \frac{ 32 \uppi^{11/18} }{ 7 \times 3^{1/18} 13^{1/3} C\sub{R}^{3/2} } 
        \frac{ \sigma\sub{SB}^{1/3} }{ c^{2/3} G^{1/6} }
        \frac{ T_\star^{4/3} M_\star^{5/9} R_\star }{ \rho\sub{d}^{1/18} }
        \left[ \left( \frac{ r\sub{R} }{ r_0 } \right)^{7/4} - 1 \right]
    \nonumber \\
    & \approx
        3.5 \times 10^{25} \, \mathrm{g} \; 
        \Biggl( \frac{ T\sub{\star} }{ \mathrm{10{,}000\,K} } \Biggr)^{4/3}
        \Biggl( \frac{ M\sub{\star} }{ \mathrm{0.6\,M\sub{\odot}} } \Biggr)^{5/9}
        \Biggl( \frac{ R\sub{\star} }{ \mathrm{0.0125\,R\sub{\odot}} } \Biggr)
    \nonumber \\
    & \quad \times
        \Biggl( \frac{ C\sub{R} }{ 2.0 } \Biggr)^{-3/2}
        \Biggl( \frac{ \rho\sub{d} }{ \mathrm{3\,g\,cm^{-3}} } \Biggr)^{-1/18}
        \, \Biggl[ \left( \frac{ r\sub{R} }{ r_0 } \right)^{7/4} - 1 \Biggr].
\end{align}
The distance-dependent factor in this equation (in square brackets)
equals unity at $ r_0 / r\sub{R} = 2^{-4/7} \approx 0.67 $
and goes to zero for $ r_0 / r\sub{R} \rightarrow 1 $,
owing to the fact that a disc
that is initially already very close to the Roche limit
need not be very massive to spread beyond it.

\begin{figure}
  \includegraphics[width=\columnwidth]{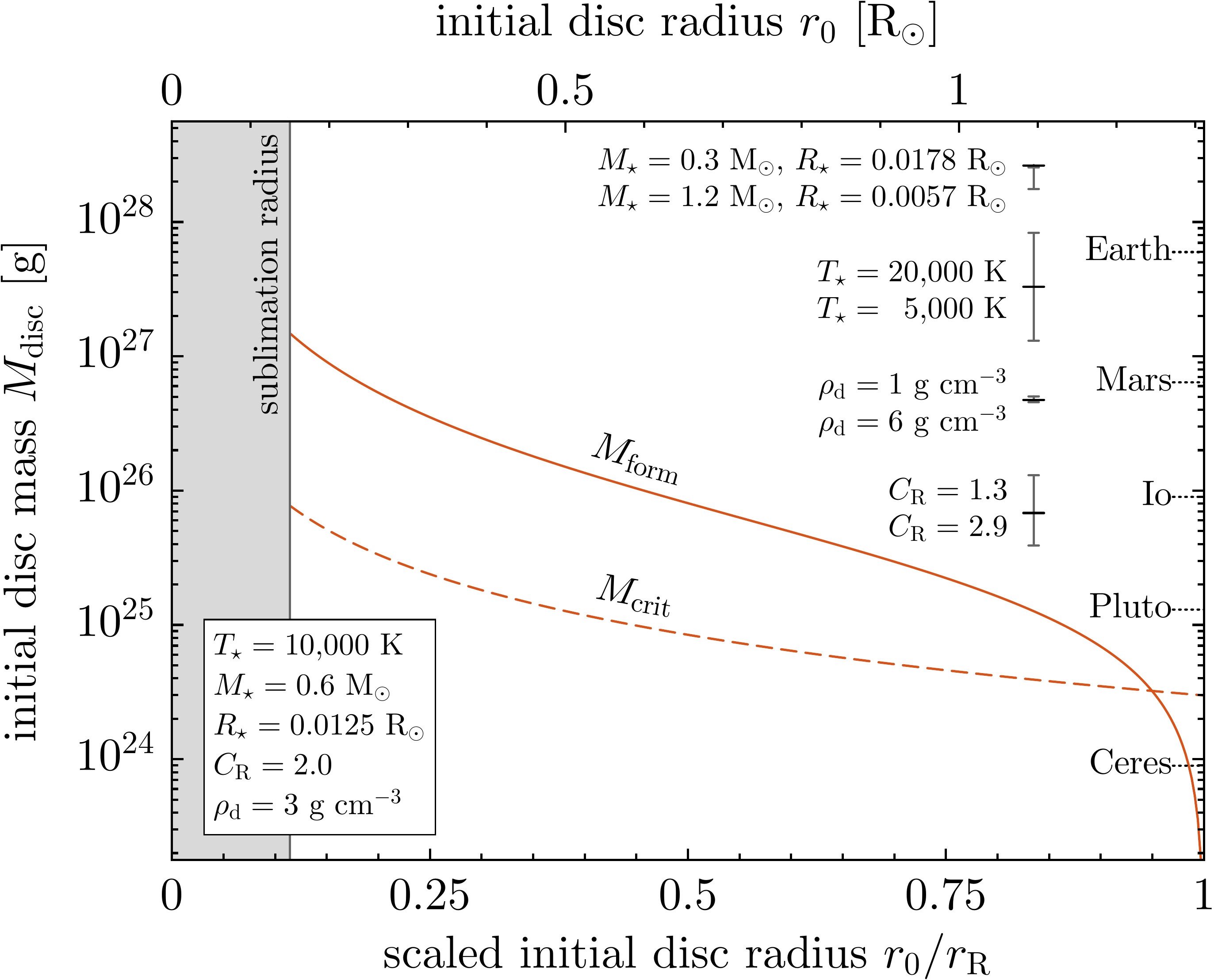}
  \caption{%
  Minimum disc mass for planet formation $ M\sub{form} $ (above which
  viscosity spreads an initially narrow disc to beyond the Roche limit;
  Eq.~\eqref{eq:mdisk_form}; solid line)
  as a function of the disc's initial radius $ r_0 $
  scaled to the Roche limit $ r\sub{R} $.
  Also shown is the critical disc mass $ M\sub{crit} $
  (above which viscous spreading dominates the disc's evolution over PR drag;
  Eq.~\eqref{eq:mdisk_crit}; dashed line)
  for a disc-width parameter of $ \delta = 0.1 $.
  Both curves are computed using the fiducial parameter values
  listed in the bottom-left corner (box).
  Error bars in the top right indicate how $ M\sub{form} $ would
  change relative to the fiducial level (dark horizontal bars)
  by varying the stellar mass $ M_\star $, stellar temperature $ T_\star $,
  material density $ \rho\sub{d} $, or Roche prefactor $ C\sub{R} $
  within reasonable ranges.
  We used the WD mass-radius relation
  to change the stellar radius $ R_\star $ along with $ M_\star $.
  Note that the absolute-distance values given along the top axis change
  if $ M_\star $, $ \rho\sub{d} $, or $ C\sub{R} $
  is altered from its fiducial value.}
  \label{fig:mdisk_form}
\end{figure}

Figure~\ref{fig:mdisk_form} shows the minimum disc mass for planet formation
as a function of the disc's initial radius $ r_0 $.
This demonstrates that
massive debris discs located in the outer parts of the tidal disruption zone
(i.e., Io-mass discs with $ r_0 / r\sub{R} \gtrsim 0.5 $)
will spread out to beyond the Roche limit.\footnote{%
For initial radii very close to the Roche limit,
much less massive discs can also reach the Roche limit,
and in Fig.~\ref{fig:mdisk_form}
the $ M\sub{form} $ curve even drops below the $ M\sub{crit} $ one.
This is because a well-defined $ M\sub{crit} $
requires a non-zero scaled disc width $ \delta $,
while the derivation of $ M\sub{form} $
assumes that the disc is initially infinitely thin.}
How $ M\sub{form} $ is affected by the uncertainties in various parameters
is illustrated by the error bars in the figure.
When the WD mass-radius relation is taken into account,
the stellar-mass and radius dependencies
roughly cancel each other,
except for the most massive WDs.
Specifically, using Eq.~(2.4) of
\citeauthor{2016RSOS....350571V} (\citeyear{2016RSOS....350571V};
originally from \citealt{1972ApJ...175..417N})
as mass-radius relation,
the combined effect of $ M\sub{\star} $ and $ R\sub{\star} $
on $ M\sub{form} $ varies by less than
10\% for WD masses
in the range
$ 0.1\,\mathrm{M_\odot} \lesssim M_\star \lesssim 1\,\mathrm{M_\odot} $.

\subsubsection{Disc lifetimes}
\label{s:disc_lifetime}

The lifetime of a disc $ t\sub{disc} $ is defined as the time after which
all the material that originally resides in the disc has either
flowed beyond the Roche limit or disappeared inside the sublimation radius.
For PR-drag-dominated discs, the evolution timescale decreases as material is removed.
Their lifetime
is therefore dominated by the initial stages of disc evolution,
and can be approximated by
\citep[cf. Eq.~(49) of][]{2012MNRAS.423..505M}
\begin{align}
  \label{eq:t_disc_pr}
  t\sub{disc}( M\sub{disc} \ll M\sub{crit} )
    & \simeq
        \frac{ M\sub{disc} }{ \dot{M}\sub{PR}( r_0 ) }
      \simeq
        \frac{ 3 }{ 16 } \frac{ c^2 }{ \sigma\sub{SB} }
        \frac{ M\sub{disc} r_0 }{ T_\star^4 R_\star^3 }
    \nonumber \\
    & \approx
        1.0 \, \mathrm{Gyr} \;
        \Biggl( \frac{ T\sub{\star} }{ \mathrm{10{,}000\,K} } \Biggr)^{-4}
        \Biggl( \frac{ R\sub{\star} }{ \mathrm{0.0125\,R\sub{\odot}} } \Biggr)^{-3}
    \nonumber \\
    & \quad \times
        \Biggl( \frac{ M\sub{disc} }{ \mathrm{10^{24}\,g} } \Biggr) \;
        \Biggl( \frac{ r_0 }{ 1\,\mathrm{R_\odot} } \Biggr).
\end{align}
This is equivalent to the PR-drag disc-evolution timescale $ t\sub{PR} $
for the initial state of the disc
multiplied by a factor $ \delta $
(see Eq.~\eqref{eq:t_pr_thick}).

For discs dominated by viscous spreading,
the evolution timescale instead increases as the surface density goes down.
A viscosity-dominated disc will spread out slower and slower,
until its surface density reaches the critical surface density $ \varSigma\sub{crit} $.
Once this happens, PR drag will take over the evolution of the disc's outer parts,
developing a sharp outer edge that moves inwards with time \citep{2011ApJ...741...36B}.
The disc's evolution timescale when the transition occurs $ t\sub{crit} $
can be found from Eq.~\eqref{eq:t_visc} or \eqref{eq:t_pr_thick}
with $ \varSigma = \varSigma\sub{crit} $, which both give
\begin{equation}
  \label{eq:t_crit}
  t\sub{crit}
    \simeq
      \frac{ 3^{13/9} \uppi^{1/9} }{ 8 \times 13^{1/3} } 
      \frac{ c^{4/3} }{ \sigma\sub{SB}^{2/3} G^{1/6} }
      \frac{ M_\star^{19/18} }{ T_\star^{8/3} R_\star^2 \rho\sub{d}^{5/9} }
      \frac{ 1 }{ \sqrt{r} }.
\end{equation}
Of particular interest is the value of this timescale
at the Roche limit $ t\sub{crit}( r\sub{R} ) $.
Since Eq.~\eqref{eq:t_crit} is a decreasing function of $ r $,
and the disc is confined to $ r < r\sub{R} $,
this is the lowest possible value of $ t\sub{crit} $.

Around the time of the transition
from viscosity-dominated to PR-drag-dominated disc evolution,
a disc that is massive enough to reach the Roche limit
will undergo viscous evolution for a timespan of about $ t\sub{crit}( r\sub{R} ) $,
followed by PR-drag evolution for another leg of time of roughly the same duration.
The total lifetime of such discs is dominated by
the duration of this episode around the transition.
It is independent of disc mass and can be approximated by
\begin{align}
  \label{eq:t_disc_plan}
  t\sub{disc}( M\sub{disc} \gg M\sub{form} )
    & \simeq 2 t\sub{crit}( r\sub{R} )
    \nonumber \\
    & \simeq
      \frac{ 3^{23/18} \uppi^{5/18} }{ 2^{5/3} 13^{1/3} \sqrt{ C\sub{R} } } 
      \frac{ c^{4/3} }{ \sigma\sub{SB}^{2/3} G^{1/6} }
      \frac{ M_\star^{8/9} }{ T_\star^{8/3} R_\star^2 \rho\sub{d}^{7/18} }
    \nonumber \\
    & \approx
        78 \, \mathrm{Gyr} \; 
        \Biggl( \frac{ C\sub{R} }{ 2.0 } \Biggr)^{-1/2}
        \Biggl( \frac{ T\sub{\star} }{ \mathrm{10{,}000\,K} } \Biggr)^{-8/3}
        \Biggl( \frac{ M\sub{\star} }{ \mathrm{0.6\,M\sub{\odot}} } \Biggr)^{8/9}
    \nonumber \\
    & \quad \times
        \Biggl( \frac{ R\sub{\star} }{ \mathrm{0.0125\,R\sub{\odot}} } \Biggr)^{-2}
        \Biggl( \frac{ \rho\sub{d} }{ \mathrm{3\,g\,cm^{-3}} } \Biggr)^{-5/9}.
\end{align}
This shows that massive WD debris discs
can survive against viscous spreading and PR drag
for longer than the age of the universe.
They may, of course, still be destroyed
by mechanisms not accounted for in our analysis
(see Sect.~\ref{s:destroy_disc}).

\subsection{Second-generation (minor) planets}
\label{s:planets}

When material from a massive disc ($ M\sub{disc} \gtrsim M\sub{form} $)
flows beyond the Roche limit,
its constituent particles are no longer prevented from coagulating by tidal forces
and can stick together in gravitationally bound aggregates.
These aggregates can then merge to form larger bodies,
eventually producing a set of second-generation (minor) planets.
In this section, we investigate these processes
and infer what the basic properties of the newly formed bodies may be.

\subsubsection{Disc properties at the Roche limit}
\label{s:sigma_roche}

The planet-formation process
depends directly on the state of the disc at its outer edge -- i.e., at the Roche limit.
Specifically, the properties of the newly formed (minor) planets
are determined by the outward mass flow rate through the Roche limit $ \dot{M}\sub{out} $,
which, in turn, is controlled by the disc's local surface density.
For this reason, we often express formulae in this section in terms of
the disc's Roche-limit surface density
$ \varSigma\sub{R} \equiv \varSigma( r\sub{R} ) $.
A rough estimate of this quantity
at the time when planet formation starts $ t\sub{start} $
can be obtained by assuming that half the disc's mass spreads out
between $ r_0 $ and $ r\sub{R} $,
following the viscous-spreading-dictated
surface-density profile $ \varSigma\sub{visc} \propto r^{-15/4} $
(see Sect.~\ref{s:disc_evol_visc}).
This gives
\begin{align}
  \label{eq:sigma_roche}
  \varSigma\sub{R}( t\sub{start} )
    & \simeq
        \frac{ 7 }{ 16 }
        \frac{ M\sub{disc} }{ \uppi r\sub{R}^2 }
        \left[ \left( \frac{ r\sub{R} }{ r_0 } \right)^{7/4} - 1 \right]^{-1}
    \nonumber \\
    & \approx
        1.7 \times 10^3 \, \mathrm{g\,cm^{-2}} \;
        \Biggl( \frac{ C\sub{R} }{ 2.0 } \Biggr) \,
        \Biggl( \frac{ M_\star }{ \mathrm{0.6\,M\sub{\odot}} } \Biggr)^{1/3}
        \Biggl( \frac{ \rho\sub{d} }{ \mathrm{3\,g\,cm^{-3}} } \Biggr)^{-1/3}
    \nonumber \\
    & \quad \times
        \Biggl( \frac{  M\sub{disc} }{ 10^{26}\,\mathrm{g} } \Biggr) \,
        \Biggl[ \left( \frac{ r\sub{R} }{ r_0 } \right)^{7/4} - 1 \Biggr]^{-1}.
\end{align}
Over time, the Roche-limit surface density will slowly go down.
When it approaches the local critical level $ \varSigma\sub{crit}( r\sub{R} ) $,
the outward mass flow through the Roche limit is gradually quenched by PR drag,
eventually terminating the production of new planets.
Therefore,
an approximate lower limit on the value of $ \varSigma\sub{R} $
relevant to second-generation planet formation
is
(see Fig.~\ref{fig:time_visc_pr}, left-hand panel)
\begin{align}
  \label{eq:sigma_crit_roche}
  \varSigma\sub{crit}( r\sub{R} )
    & \simeq
        \frac{ 2^{7/3} \uppi^{5/18} }{ 3^{13/18} 13^{1/3} C\sub{R}^{7/2} }
        \frac{ \sigma\sub{SB}^{1/3} }{ c^{2/3} G^{1/6} }
        \frac{ T_\star^{4/3} R_\star \rho\sub{d}^{11/18} }{ M_\star^{1/9} }
    \nonumber \\
    & \approx
        570 \, \mathrm{g\,cm^{-2}} \; 
        \Biggl( \frac{ T\sub{\star} }{ \mathrm{10{,}000\,K} } \Biggr)^{4/3}
        \Biggl( \frac{ M\sub{\star} }{ \mathrm{0.6\,M\sub{\odot}} } \Biggr)^{-1/9}
        \Biggl( \frac{ R\sub{\star} }{ \mathrm{0.0125\,R\sub{\odot}} } \Biggr)
    \nonumber \\
    & \quad \times
        \Biggl( \frac{ C\sub{R} }{ 2.0 } \Biggr)^{-7/2}
        \Biggl( \frac{ \rho\sub{d} }{ \mathrm{3\,g\,cm^{-3}} } \Biggr)^{11/18}.
\end{align}
From
Eqs.~\eqref{eq:sigma_roche} and \eqref{eq:sigma_crit_roche}
we find that
the value of $ \varSigma\sub{R} $
typically lies in the range
$ 10^2\,\mathrm{g\,cm^{-2}} \lesssim \varSigma\sub{R} \lesssim 10^5\,\mathrm{g\,cm^{-2}} $
for the discs we consider
while they are overflowing the Roche limit.

Given that
changes in the disc's local surface density happen on the local viscous timescale,
the disc's mass outflow rate at the Roche limit can be approximated by
\citepalias[cf. Eq.~(S4) of][]{2012Sci...338.1196C}
\begin{align}
  \label{eq:mdot_roche}
  \dot{M}\sub{out}
    & \simeq
        \frac{ \uppi r\sub{R}^2 \varSigma\sub{R} }
          { t\sub{visc}( \varSigma\sub{R}, r\sub{R} ) }
      \simeq
        \frac{ 3^{5/6} 13 C\sub{R}^{19/2} }{ 2^{10/3} \sqrt{ \uppi } }
        \sqrt{ G } \,
        \frac{ \varSigma\sub{R}^3 }{ \rho\sub{d}^{3/2} }
   \nonumber \\
   & \approx
       6.5 \times 10^{7} \, \mathrm{g\,s^{-1}} \;
       \Biggl( \frac{ C\sub{R} }{ 2.0 } \Biggr)^{19/2}
       \Biggl( \frac{ \rho\sub{d} }{ \mathrm{3\,g\,cm^{-3}} } \Biggr)^{-3/2}
       \Biggl( \frac{ \varSigma\sub{R} }{ \mathrm{10^3\,g\,cm^{-2}} } \Biggr)^{3}.
\end{align}
As expected,
for a Roche-limit surface mass density at the critical level
(i.e., $ \varSigma\sub{R} = \varSigma\sub{crit}( r\sub{R} ) $),
the outward mass flow rate due to viscous spreading
roughly matches
the inward mass flow rate from PR drag:
$ \dot{M}\sub{out}[ \varSigma\sub{crit}( r\sub{R} ) ] \simeq \dot{M}\sub{PR}( r\sub{R} ) / 2 $.
The discrepancy of a factor 2 can be explained by the approximate nature of
Eqs.~\eqref{eq:sigma_crit_time} and \eqref{eq:mdot_roche}.

\subsubsection{Planetesimal formation through gravitational instabilities}
\label{s:m_toomre}

As the disc expands beyond the Roche limit,
gravitational instabilities cause portions of the material beyond $ r\sub{R} $
to collapse into gravitationally bound aggregates.
Since the instabilities typically have sizes around Toomre's critical wavelength
$ \lambda\sub{T} $ (see Eq.~\eqref{eq:lambda_toomre}),
these newly formed, rubble-pile-type planetesimals
will have masses of the order of the local Toomre mass $ M\sub{T} $,
defined as the mass enclosed in a circular portion of the disc
with diameter $ \lambda\sub{T} $
\citep[cf.][]{1973ApJ...183.1051G,2006MNRAS.373.1039N,2007ApJ...657..521M,2009ApJ...703.1363M}.\footnote{%
Stricktly speaking, the relevant length scale is the wavelength of the fastest growing unstable mode
$ \lambda\sub{fgm} = 2 \sigma_r / ( G \varSigma ) $,
rather than $ \lambda\sub{T} $,
which is the wavelength of the longest instability.
For $ Q\sub{T} \simeq 2 $ (to which the gravitationally unstable disc material is expected to converge;
see Sect.~\ref{s:grav_stab}), however,
these two length scales are approximately the same.}
At the Roche limit, the Toomre mass is
\begin{align}
 \label{eq:mass_toomre}
 M\sub{T}( r\sub{R} )
   & =
       \uppi \left[ \frac{ \lambda\sub{T}( r\sub{R} ) }{ 2 } \right]^2
       \varSigma\sub{R}
     =
       \frac{ 9 \uppi^3 C\sub{R}^6 }{ 4 }
       \frac{ \varSigma\sub{R}^3 }{ \rho\sub{d}^2 }
   \nonumber \\
   & \approx
       5.0 \times 10^{11} \, \mathrm{g} \;
       \Biggl( \frac{ C\sub{R} }{ 2.0 } \Biggr)^6
       \Biggl( \frac{ \rho\sub{d} }{ \mathrm{3\,g\,cm^{-3}} } \Biggr)^{-2}
       \Biggl( \frac{ \varSigma\sub{R} }{ \mathrm{10^3\,g\,cm^{-2}} } \Biggr)^{3}.
\end{align}
The typical value of $ 5.0 \times 10^{11} $\,g
corresponds to a body with a radius of about $ 34 $\,m,
assuming its density is the same as the disc material.
As $ \varSigma\sub{R} $ goes down with time,
the mass of the planetesimals decreases.
Since the disc's mass outflow rate $ \dot{M}\sub{out} $ goes down at the same rate,
however, Toomre-mass bodies are created at a constant rate,
namely one planetesimal per time interval
\begin{equation}
  \label{eq:t_toomre}
  t\sub{T}
    = \frac{ M\sub{T}( r\sub{R} ) }{ \dot{M}\sub{out} }
    \simeq \frac{ 2^{4/3} 3^{2/3} \uppi^3 }{ 13 C\sub{R}^5 } P\sub{R}
    \approx 2.1 \, \mathrm{h} \;
      \Biggl( \frac{ C\sub{R} }{ 2.0 } \Biggr)^{-7/2}
      \Biggl( \frac{ \rho\sub{d} }{ \mathrm{3\,g\,cm^{-3}} } \Biggr)^{-1/2},
\end{equation}
i.e., a few bodies per local orbital period.

The region just beyond the Roche limit may be populated with
many co-orbital planetesimals with masses of around $ M\sub{T}( r\sub{R} ) $ each.
What initially happens to these bodies
will depend on their orbital evolution
and the outcomes of mutual collisions.
Here, we ignore the details of the initial coagulation process.
Instead, we follow \citet{2010Natur.465..752C} and \citetalias{2012Sci...338.1196C} in assuming that
the planetesimals will accumulate into a single large body
orbiting just beyond the Roche limit.
Once this body is formed,
it will absorb any additional material flowing out of the disc,
as long as it is close enough to the Roche limit
(see Sect.~\ref{s:delta_q_cont_disc}).

\subsubsection{Minor-planet growth and migration}
\label{s:delta_q}

A body orbiting close to the disc's outer edge
will grow by absorbing any disc material that flows beyond the Roche limit.
At the same time, this body
will experience torques from the disc
associated with mean-motion resonances (MMRs),
causing it to migrate outwards
\citep[e.g.,][]{1980ApJ...241..425G}.
The interaction between this simultaneous growth and outward migration
was studied in detail by
\citetalias{2012Sci...338.1196C},
who developed a one-dimensional analytical model for
the evolution of bodies formed by a viscously spreading particle disc
that overflows the Roche limit.
While their model was designed to study
moons produced by planetary rings
\citep[see also][]{2010Natur.465..752C,2011Icar..216..535C},
it is formulated very generally, so
we can readily apply many of its predictions to the present case of
WD debris discs spawning second-generation (minor) planets.

Assuming a constant mass outflow rate
from the disc's outer edge at $ r\sub{R} $
(prescribed by Eq.~\eqref{eq:mdot_roche}),
\citetalias{2012Sci...338.1196C} find that
the interplay of
growth and outward migration
leads to specific relations between
a body's mass $ M\sub{p} $ and
its separation from the disc's outer edge $ r - r\sub{R} $.
These are best expressed in terms of the dimensionless
quantities
\begin{equation}
  \label{eq:delta_q_defs}
  q \equiv \frac{ M\sub{p} }{ M_\star },
  \qquad
  \varDelta \equiv \frac{ r - r\sub{R} }{ r\sub{R} },
\end{equation}
(i.e., the minor-planet mass ratio and the body's scaled distance from the Roche limit)
and depend on just one free parameter:
the disc's viscous evolution timescale at the Roche limit, scaled to the local orbital period\footnote{%
When expressed in terms of absolute physical quantities
rather than the dimensionless $ \varDelta $, $ q $, and $ \mathcal{T} $
(called $ \tau\sub{disk} $ in the notation of \citetalias{2012Sci...338.1196C}),
there are small differences between our formulae and those of \citetalias{2012Sci...338.1196C}.
This is due to a difference in prescription for the viscous timescale
(\citetalias{2012Sci...338.1196C} omit the factor $ 1 / 12 $)
and a different treatment of the Roche prefactor
(fixed to $ C\sub{R} = 2.456 $ by \citetalias{2012Sci...338.1196C}).}
\begin{align}
  \label{eq:dimless_time}
  \mathcal{T}
    & \equiv \frac{ t\sub{visc}( \varSigma\sub{R}, r\sub{R} ) }{ P\sub{R} }
      \simeq \frac{ 4 \uppi^{1/3} }{ 3^{2/3} 13 C\sub{R}^9 } 
        \frac{ M_\star^{2/3} \rho\sub{d}^{4/3} }{ \varSigma\sub{R}^2 }
    \nonumber \\
    & \approx 2.1 \times 10^{13} \; 
        \Biggl( \frac{ C\sub{R} }{ 2.0 } \Biggr)^{-9}
        \Biggl( \frac{ M\sub{\star} }{ \mathrm{0.6\,M\sub{\odot}} } \Biggr)^{2/3}
        \Biggl( \frac{ \rho\sub{d} }{ \mathrm{3\,g\,cm^{-3}} } \Biggr)^{4/3}
        \Biggl( \frac{ \varSigma\sub{R} }{ \mathrm{10^3\,g\,cm^{-2}} } \Biggr)^{-2}.
\end{align}
Given the values for $ \varSigma\sub{R} $ found in Sect.~\ref{s:sigma_roche},
we expect $ \mathcal{T} $ to typically fall in the range $ 10^9 \lesssim \mathcal{T} \lesssim 10^{15} $
for the discs we consider.

\begin{figure}
  \includegraphics[width=\columnwidth]{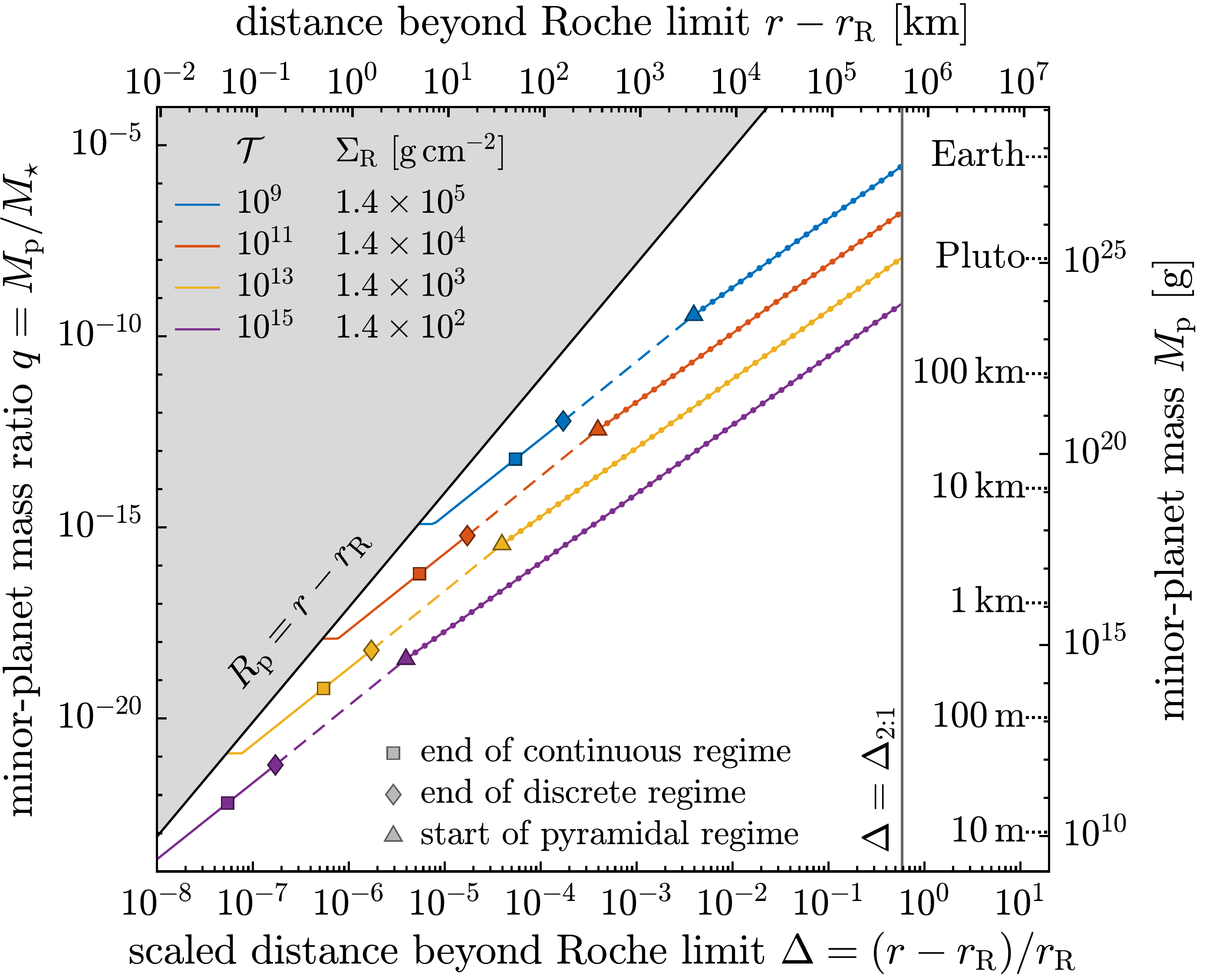}
  \caption{%
  Minor-planet mass ratio $ q $ as a function of
  the scaled distance beyond the Roche limit $ \varDelta $,
  for four different values
  of the dimensionless Roche-limit disc-evolution timescale $ \mathcal{T} $.
  The mass-distance relations are composed of several parts:
  an absolute minimum set by the Roche-limit Toomre mass
  (Eq.~\eqref{eq:mass_toomre};
  small horizontal line segments at low $ \varDelta $),
  Eq.~\eqref{eq:q_delta_cont_disc} for the continuous and discrete regimes
  ($ q \propto \varDelta^2 $, $ \varDelta < \varDelta\sub{d} $),
  Eq.~\eqref{eq:q_delta_pyra} for the pyramidal regime
  ($ q \propto \varDelta^{9/5} $, $ \varDelta > \varDelta\sub{0} $),
  and a hypothetical part connecting the discrete and pyramidal regimes
  (dashed lines).
  The knots in the pyramidal-regime line illustrate
  the average spacing of individual objects
  according to the number densities given by Eq.~\eqref{eq:dn_ddelta_pyra}.
  Square, diamond, and triangle symbols
  mark the boundaries of the different regimes
  (Eqs.~\eqref{eq:delta_c_q_c}, \eqref{eq:delta_d_q_d},
  \eqref{eq:delta_0}, and \eqref{eq:q_0}).
  A vertical grey line shows $ \varDelta_{2:1} $,
  the outermost scaled distance for which
  the migration rates underpinning the mass-distance relations are applicable.
  In terms of dimensional quantities,
  the Roche-limit surface mass densities $ \varSigma\sub{R} $
  corresponding to the four lines are indicated in the legend,
  and absolute minor-planet masses and distances are given by
  the right-hand-side and top axes, respectively.
  For these dimensional versions (and for the Toomre masses),
  we use the stellar parameters
  $ T_\star = 10{,}000\,\mathrm{K} $,
  $ M_\star = 0.6\,\mathrm{M\sub{\odot}} $,
  $ R_\star = 0.0125\,\mathrm{R\sub{\odot}} $,
  a material density of $ \rho\sub{d} = 3\,\mathrm{g\,cm}^{-3} $,
  and a Roche prefactor of $ C\sub{R} = 2.0 $.
  The top-left corner of parameter space (grey) is excluded,
  as this corresponds to bodies that are
  still in contact with the Roche limit
  (assuming their density equals that of the disc material).}
  \label{fig:delta_q}
\end{figure}

In the following two subsections, we review
the model of \citetalias{2012Sci...338.1196C}
and work out a few additional details.
For reference, Fig.~\ref{fig:delta_q} already shows
an overview of
the resulting minor-planet mass-distance relations
for a few representative values of $ \mathcal{T} $.
These different \mbox{$ \mathcal{T} $-values} can correspond to, e.g.,
the state of a given system at subsequent points in time,
or to the states of systems with different initial disc masses
and/or initial disc radii at the same point in time
(see Sect.~\ref{s:sigma_roche}).
In Fig.~\ref{fig:delta_q} and also in the rest of this paper,
we assume that the mean density of the (minor) planets
is the same as that of the disc material,
so the radius of these bodies can be written as 
$ R\sub{p} = [ 3 M\sub{p} / ( 4 \uppi \rho\sub{d} ) ]^{1/3} $.

\subsubsection{The continuous and discrete regimes}
\label{s:delta_q_cont_disc}

In the initial stage of planet formation (the `continuous regime'),
a single growing body
immediately absorbs any disc material that flows beyond the Roche limit.
At the same time, this body moves away from the disc
(through angular-momentum exchange via MMRs)
at a rate that depends on its mass and distance to the disc's outer edge.
As the body grows, its outward migration rate increases,
but as it moves away from the disc, the migration rate rapidly decreases
($ \dif \varDelta / \dif t \propto q \varDelta^{-3} $; \citetalias{2012Sci...338.1196C}).
The mass-distance relation for the continuous regime is found by integrating
this migration rate
under the assumption of a constant mass-absorption rate,
which yields \citepalias{2012Sci...338.1196C}
\begin{equation}
  \label{eq:q_delta_cont_disc}
  q \simeq \frac{ 3^{3/2} }{ 2^3 \sqrt{ \mathcal{T} } } \, \varDelta^2.
\end{equation}
When the growing, outwards-migrating body
moves beyond about two Hill radii from the edge of the disc,
it cannot absorb the disc material directly anymore,
ending the continuous regime.
This occurs at the scaled distance $ \varDelta\sub{c} $
and mass ratio $ q\sub{c} $, given by \citepalias{2012Sci...338.1196C}
\begin{equation}
  \label{eq:delta_c_q_c}
  \varDelta\sub{c}
    \simeq \sqrt{ \frac{ 3 }{ \mathcal{T} } }
    \approx \frac{ 1.73 }{ \sqrt{ \mathcal{T} } },
  \qquad
  q\sub{c}
    \simeq \frac{ 3^{5/2} }{ 2^3 \mathcal{T}^{3/2} }
    \approx \frac{ 1.95 }{ \mathcal{T}^{3/2} }.
\end{equation}

Once a body migrates beyond $ \varDelta\sub{c} $,
the planet-formation process enters what is called the `discrete regime'.
Here, the further growth of the initial body happens via the transfer
of small bodies that themselves accrete according to the continuous regime.
On average, the mass-distance relation still follows Eq.~\eqref{eq:q_delta_cont_disc},
but the growth of the first-formed body will be discontinuous.
The discrete regime ends when this object moves too far away
to absorb additional newly formed continuous-regime bodies
(two Hill radii from $ \varDelta\sub{c} $).
The scaled distance $ \varDelta\sub{d} $ and mass-ratio $ q\sub{d} $
at this point are given by \citepalias{2012Sci...338.1196C}
\begin{equation}
  \label{eq:delta_d_q_d}
  \varDelta\sub{d}
    \simeq \frac{ 1 }{ Z_0^3 } \sqrt{ \frac{ 3 }{ \mathcal{T} } }
    \approx \frac{ 5.45 }{ \sqrt{ \mathcal{T} } },
  \qquad
  q\sub{d}
    \simeq \frac{ 3^{5/2} }{ 2^3 Z_0^6 \mathcal{T}^{3/2} }
    \approx \frac{ 19.3 }{ \mathcal{T}^{3/2} },
\end{equation}
where $ Z_0 \approx 0.682 $ is a numerical constant
(specifically, the real-valued root of $ Z^3 + Z - 1 $).

Given that the dimensionless disc-evolution timescale $ \mathcal{T} $
is very large for the WD debris discs we consider,
the continuous and discrete regimes end very close to the Roche limit,
and correspond to relatively small bodies
(see Fig.~\ref{fig:delta_q}).
In a WD debris disc that is forming planets,
mass continuously flows
through the region between the Roche limit and $ \varDelta\sub{d} $,
transported initially by continuous-regime bodies and subsequently by discrete-regime bodies.
While the mass-ratios of these bodies and the extents of the regimes all
change with time as the disc's Roche-limit surface density $ \varSigma\sub{R} $ goes down,
the rates at which bodies are created at $ \left\{ \varDelta\sub{c}, q\sub{c} \right\} $
and $ \left\{ \varDelta\sub{d}, q\sub{d} \right\} $
turn out to remain constant, independent of the properties of the disc.
The corresponding time intervals
between the formation of two subsequent bodies are given by
\citepalias[cf.][]{2012Sci...338.1196C}
\begin{equation}
  \label{eq:t_c}
  t\sub{c}
    = \frac{ q\sub{c} M_\star }{ \dot{M}\sub{out} }
    = \frac{ 3^{13/6} \sqrt{13} \, C\sub{R}^{5/2} }{ 2^{8/3} \sqrt{\uppi} } P\sub{R}
    \approx 4.4 \, \mathrm{d} \;
      \Biggl( \frac{ C\sub{R} }{ 2.0 } \Biggr)^4
      \Biggl( \frac{ \rho\sub{d} }{ \mathrm{3\,g\,cm^{-3}} } \Biggr)^{-1/2} \! ,
\end{equation}
for bodies with mass ratio $ q\sub{c} $ located at scaled distance $ \varDelta\sub{c} $,
and
\begin{equation}
  \label{eq:t_d}
  t\sub{d}
    = \frac{ q\sub{d} M_\star }{ \dot{M}\sub{out} }
    = \frac{ 3^{13/6} \sqrt{13} \, C\sub{R}^{5/2} }{ 2^{8/3} \sqrt{\uppi} \, Z_0^6 } P\sub{R}
    \approx 44 \, \mathrm{d} \;
      \Biggl( \frac{ C\sub{R} }{ 2.0 } \Biggr)^4
      \Biggl( \frac{ \rho\sub{d} }{ \mathrm{3\,g\,cm^{-3}} } \Biggr)^{-1/2},
\end{equation}
for bodies with mass ratio $ q\sub{d} $ at $ \varDelta\sub{d} $.

\subsubsection{The pyramidal regime}
\label{s:delta_q_pyra}

Beyond scaled distance $ \varDelta\sub{d} $,
many minor planets can exist simultaneously,
each one migrating outwards at a rate that depends on
its mass and its distance to the disc's edge.
The approximate minor-planet mass-distance relation in this domain can be found
by considering an idealised scenario in which bodies of equal mass
are formed at a given distance at regular intervals in time.
In this case,
later-formed bodies will gradually catch up with earlier-formed ones,
because the rate of outward migration decreases with distance.
When two bodies come to within two mutual Hill radii of one another,
they are assumed to undergo a perfect-merger collision,
creating a body of double the mass and angular momentum.
This process of catch-up then repeats itself,
but with bodies that are twice the mass of the original building blocks,
and that are formed half as frequently,
at the location of the first merger.
In turn, this also repeats itself, and so on
(see Sect.~S6 of \citetalias{2012Sci...338.1196C} for a more detailed explanation).
The resulting hierarchical pattern of mergers
(giving this stage the name `pyramidal regime')
yields the mass-distance relation\footnote{%
This can be derived from Eq.~(S18) of \citetalias{2012Sci...338.1196C}
(which gives the scaled distance of pyramidal-regime bodies
when they are about to merge)
by solving for $ q $ and multiplying the result by $ 3 / 2 $.
The multiplication accounts for the fact that
pyramidal-regime bodies on average have a mass that is
a factor $ 3 / 2 $ higher than the mass used in \citetalias{2012Sci...338.1196C}'s Eq.~(S18),
because mergers result in bodies of twice the mass at the same distance.}
\begin{equation}
  \label{eq:q_delta_pyra}
  q \simeq \frac{ 3^{13/5} }{ 2^{16/5} \mathcal{T}^{3/5} } \, \varDelta^{9/5}.
\end{equation}

Another prediction for the pyramidal regime is that the
number of minor planets $ n $ follows
the number-density-distance relation
$ \dif n / \dif \varDelta \propto 1 / \varDelta $
\citepalias{2012Sci...338.1196C},
meaning that for higher $ \varDelta $ the bodies are farther apart
(specifically, they are equidistant in $ \log( \varDelta ) $;
see knotted lines in Fig.~\ref{fig:delta_q}).
To find the normalisation for this relation,
we note that, at any given time, there are either one or two bodies
between two consecutive locations where mergers happen,
or an average of $ 3 / 2 $ bodies per interval between merger locations.
Because a merging event corresponds to a doubling of the planet mass,
the \mbox{$ \varDelta $-values} corresponding to two consecutive merger locations
are a factor $ 2^{5/9} $ apart (from Eq.~\eqref{eq:q_delta_pyra}).
Thus, the average number of pyramidal-regime bodies per unit of scaled distance is
\begin{equation}
  \label{eq:dn_ddelta_pyra}
  \frac{ \dif n }{ \dif \varDelta }
      \simeq \frac{ 3 }{ 2 \left( 2^{5/9} - 1 \right) \varDelta }
      \approx \frac{ 3.19 }{ \varDelta },
\end{equation}
and, on average, the bodies are a factor $ ( 3 / 2 )^{5/9} \approx 1.25 $
apart in $ \varDelta $.

Considering Eqs.~\eqref{eq:q_delta_pyra} and \eqref{eq:dn_ddelta_pyra}
in the limit $ \varDelta \rightarrow 0 $ reveals that
there is a scaled distance $ \varDelta\sub{0} $
inside of which the minor-planet number density is so high
that bodies are spaced less than two mutual Hill radii $ r\sub{H} $ from one another.
Inside of this distance, the pyramidal-regime relations cannot hold,
because bodies would already have merged earlier.
Therefore,
$ \varDelta\sub{0} $ and the corresponding mass-ratio $ q_0 $
mark the formal start of the pyramidal regime.
These can be computed by solving $ \dif n / \dif r = ( 2 r\sub{H} )^{-1} $,
using Eqs.~\eqref{eq:q_delta_pyra} and \eqref{eq:dn_ddelta_pyra},
as well as the approximation $ r \simeq r\sub{R} $ (i.e., $ \varDelta_0 \ll 1 $),
which gives
\begin{align}
  \label{eq:delta_0}
  \varDelta\sub{0}
    & \simeq \frac{ 3^{23/6} }{ 2^{11/6} \left( 2^{5/9} - 1 \right)^{5/2} \sqrt{ \mathcal{T} } }
      \approx \frac{ 125 }{ \sqrt{ \mathcal{T} } },
  \\
  \label{eq:q_0}
  q_0
    & \simeq \frac{ 3^{19/2} }{ 2^{13/2} \left( 2^{5/9} - 1 \right)^{9/2} \mathcal{T}^{3/2} }
      \approx \frac{ 1.13 \times 10^4 }{ \mathcal{T}^{3/2} }.
\end{align}
In the region between $ \varDelta\sub{d} $ and $ \varDelta_0 $,
where formally the mass-distance relation is undefined,
we expect the relation to follow a trend similar to
Eqs.~\eqref{eq:q_delta_cont_disc} and \eqref{eq:q_delta_pyra}.
For this reason, the discrete and pyramidal regimes
are connected with a dashed line in Fig.~\ref{fig:delta_q}.

Migration via disc torques operates as long as a body has first-order MMRs
overlapping with the disc
(i.e., resonances for which the mean motions obey a ratio
$ p + 1 : p $, where $ p $ is a positive integer).
The farthest-reaching first-order interior MMR is the 2:1 resonance.
Hence, disc torques can at most move a planet out to the radius
where the planet's 2:1 MMR coincides with the edge of the disc
at the Roche limit.
This radius and the corresponding scaled distance from the Roche limit are given by
\begin{align}
  \label{eq:r_2to1}
  r_{2:1}
    & = 2^{2/3} r\sub{R}
      \approx 2.1 \, \mathrm{R_\odot} \;
        \Biggl( \frac{ C\sub{R} }{ 2.0 } \Biggr) \,
        \Biggl( \frac{ M_\star }{ \mathrm{0.6\,M\sub{\odot}} } \Biggr)^{1/3}
        \Biggl( \frac{ \rho\sub{d} }{ \mathrm{3\,g\,cm^{-3}} } \Biggr)^{-1/3},
  \\
  \label{eq:delta_2to1}
  \varDelta_{2:1}
    & = 2^{2/3} - 1 \approx 0.587.
\end{align}
Beyond $ r_{2:1} $, a planet would only have
higher-order MMRs (e.g., 5:2, 3:1, and 4:1) overlapping with the disc
and (for low-eccentricity orbits) these are
inefficient at transferring angular momentum
\citep[e.g.,][]{1984prin.conf..589D}.
Hence, the pyramidal-regime relations cease to hold beyond $ \varDelta_{2:1} $.\footnote{%
Close to $ \varDelta_{2:1} $, the relations will already become less accurate,
because the derivations of \citetalias{2012Sci...338.1196C} use the approximation
$ r \simeq r\sub{R} $ (i.e., $ \varDelta \ll 1 $).}

We can now calculate
the global properties of the pyramidal-regime minor-planet population
when it is filled from $ \varDelta_0 $ to $ \varDelta_{2:1} $.
The total number of bodies in the pyramidal regime
is given by
\begin{align}
  \label{eq:n_pyra}
  n
    & =
        \int\limits_{ \varDelta\sub{0} }^{ \varDelta_{2:1} }
          \frac{ \dif n }{ \dif \varDelta } \dif \varDelta
      \simeq
        \frac{ 3 }{ 2 \left( 2^{5/9} - 1 \right) }
        \ln \left[ \frac{ 2^{11/6} }{ 3^{23/6} } \left( 2^{2/3} - 1 \right)
          \left( 2^{5/9} - 1 \right)^{5/2} \sqrt{ \mathcal{T} } \right]
    \nonumber \\
    & \!\begin{multlined}[b][0.9\displaywidth]
      \approx 32 + \log_{10} \left[
        \Biggl( \frac{ C\sub{R} }{ 2.0 } \Biggr)^{-33.09}
        \Biggl( \frac{ M\sub{\star} }{ \mathrm{0.6\,M\sub{\odot}} } \Biggr)^{2.45}
        \Biggl( \frac{ \rho\sub{d} }{ \mathrm{3\,g\,cm^{-3}} } \Biggr)^{4.90} \right.
      \\
        \times \left.
          \Biggl( \frac{ \varSigma\sub{R} }{ \mathrm{10^4\,g\,cm^{-2}} } \Biggr)^{-7.35}
        \right].
     \end{multlined}
\end{align}
The majority of these will be small bodies close to the Roche limit.
We note that this does not take into account the fact that
a tightly packed system may not be dynamically stable
on the timescale on which the bodies are renewed.
Dynamical instabilities may lead to additional collisions
and hence a lower total number of bodies with a wider spacing
(see also Sect.~\ref{s:wd1145_model_longer}).
To investigate this issue in more detail
would require
a more thorough treatment of
the gravitational interactions between the bodies
\citep[see, e.g.,][]{2015ApJ...799...40H,2017ApJ...836..109S}.

The mass ratio of all the bodies in the pyramidal regime combined
can be computed as
\begin{equation}
  \label{eq:q_pyra}
  \frac{ M\sub{pyra} }{ M_\star }
      =
        \int\limits_{ \varDelta\sub{0} }^{ \varDelta_{2:1} }
          q( \varDelta ) \frac{ \dif n }{ \dif \varDelta } \dif \varDelta
      \simeq
        \frac{ 3^{8/5} 5 \, \bigl( 2^{2/3} - 1 \bigr)^{9/5} }
          { 2^{21/5} \left( 2^{5/9} - 1 \right) \mathcal{T}^{3/5} }
      \approx \frac{ 1.29 }{ \mathcal{T}^{3/5} },
\end{equation}
where we have used $ \varDelta_0 \ll \varDelta_{2:1} $ in evaluating the integral.
This is equivalent to about $ 1.7 \, q( \varDelta_{2:1} ) $,
using Eq.~\eqref{eq:q_delta_pyra},
demonstrating that the largest body in the pyramidal regime, farthest from the WD,
generally holds more than half the total pyramidal-regime mass.
In terms of absolute physical quantities,
the total mass that can be accommodated in the pyramidal regime is
\begin{align}
  \label{eq:mass_pyra}
  M\sub{pyra}
    & \simeq
        \frac{ 5 \times 3^2 13^{3/5} \bigl( 2^{2/3} - 1 \bigr)^{9/5} C\sub{R}^{27/5} }
          { 2^{27/5} \left( 2^{5/9} - 1 \right) \uppi^{1/5} }
        \frac{ M_\star^{3/5} \varSigma\sub{R}^{6/5} }{ \rho\sub{d}^{4/5} }
    \nonumber \\
    & \approx 1.6 \times 10^{25} \, \mathrm{g} \; 
        \Biggl( \frac{ C\sub{R} }{ 2.0 } \Biggr)^{27/5}
        \Biggl( \frac{ M\sub{\star} }{ \mathrm{0.6\,M\sub{\odot}} } \Biggr)^{3/5}
        \Biggl( \frac{ \rho\sub{d} }{ \mathrm{3\,g\,cm^{-3}} } \Biggr)^{-4/5}
    \nonumber \\
    & \quad \times
        \Biggl( \frac{ \varSigma\sub{R} }{ \mathrm{10^3\,g\,cm^{-2}} } \Biggr)^{6/5}.
\end{align}

\subsubsection{A large planet at $ r_{2:1} $}
\label{s:largest_planet}

The pyramidal regime can only accommodate a limited amount of mass in minor planets,
and this amount decreases with time as $ \varSigma\sub{R} $ goes down.
Conversely,
the total mass in newly formed planets $ M\sub{recyc} $ increases with time
as more disc material flows beyond the Roche limit.
This accumulated mass in planets is dominated by
the early mass flow through the Roche limit
and therefore $ M\sub{recyc} $ approaches a constant value at late times,
which is of the order of
$ \dot{M}\sub{out}( t\sub{start} ) t\sub{visc}[ \varSigma\sub{R}( t\sub{start} ), r\sub{R} ] \simeq \uppi r\sub{R}^2 \varSigma\sub{R}( t\sub{start} ) $.
In most planet-forming disc systems, there will thus be a point in time
when $ M\sub{recyc} $ exceeds $ M\sub{pyra} $.
Once this happens,
the excess material will naturally end up at $ r_{2:1} $.

Planets cannot migrate efficiently beyond $ r_{2:1} $ via disc torques,
because they would have no more first-order MMRs overlapping with the disc.
Other mechanisms may be considered
that might transport mass beyond $ r_{2:1} $.
For planetary satellites,
the planet's tides can push moons further outwards
(if they are located outside the planet's corotation radius).
However, because WDs are so compact,
the newly formed planets that we consider here
are not massive enough to generate significant tidal bulges on the WD.
Another possible mechanism to consider is resonance capture between two planets,
which might push the outer body beyond $ r_{2:1} $,
while the inner one is still inside $ r_{2:1} $,
migrating outwards due to disc torques.
Studies looking into this effect in satellite systems, however, find that
the additional outward migration that can be achieved in this way is modest
at most \citep{2015ApJ...799...40H,2017ApJ...836..109S}.

We conclude that
the outward migration of second-generation planets
beyond $ r_{2:1} $ is likely to be very limited.
Therefore,
if a planet-forming disc is massive enough
that the bodies forming out of it reach $ r_{2:1} $,
most of the material flowing beyond the Roche limit
will accumulate into a single large body stalled at $ r_{2:1} $.
A rough estimate of
the total mass in newly formed planets $ M\sub{recyc} $
can now be made
by considering
angular-momentum conservation between the initial state and late times
(ignoring the small angular-momentum losses due to PR drag).
Assuming that all of the disc's initial angular momentum
(carried by mass $ M\sub{disc} $ at $ r_0 $)
ends up at $ r_{2:1} $,
the disc's mass-recycling efficiency
is given by
\begin{equation}
  \label{eq:f_recyc_simple}
    \frac{ M\sub{recyc} }{ M\sub{disc} }
    \simeq \frac{ 1 }{ 2^{1/3} } \sqrt{ \frac{ r_0 }{ r\sub{R} } }
    \approx 0.56 \;
      \Biggl( \frac{ r_0 / r\sub{R} }{ 0.5 } \Biggr)^{1/2}.
\end{equation}
This shows that a significant portion
(in many cases more than half)
of the disc's original mass
can be recycled into new planets.
For instance, given a debris disc of initial mass
$ M\sub{disc} \sim 10^{28} $\,g (i.e., the mass of a super-Earth),
the largest planet could easily have the mass of the Earth.

Finally,
a clear prediction can be made regarding the orbital period of
the largest planet.
Since
this body
is located around $ r_{2:1} $,
where its 2:1 MMR coincides with the Roche limit,
its orbital period will approximately be
\begin{equation}
  \label{eq:period_2to1}
  P_{2:1}
    = 2 P\sub{R}
    \approx 11 \, \mathrm{h} \;
      \Biggl( \frac{ C\sub{R} }{ 2.0 } \Biggr)^{3/2}
      \Biggl( \frac{ \rho\sub{d} }{ \mathrm{3\,g\,cm^{-3}} } \Biggr)^{-1/2}.
\end{equation}

\subsection{Summary of analytical findings}
\label{s:analytical_summary}

Based on the analytical considerations presented in the preceding sections,
we make the following
tentative (to be tested numerically in the following section)
conclusions about massive WD debris discs
(with assumed properties as described in Sect.~\ref{s:disc_basics})
located in the outer parts of the tidal disruption zone
\mbox{($ 0.5 \lesssim r / r\sub{R} < 1 $):}
\begin{enumerate}
  \item Geometrically thin but vertically optically thick discs
        are gravitationally unstable
        (see Eqs.~\eqref{eq:tau_grav_stab} and \eqref{eq:tau_grav_stab2};
        Fig.~\ref{fig:grav_stab}),
        leading to self-gravity wakes that
        enhance the disc's effective viscosity (see Eq.~\eqref{eq:visc_sg}).
  \item For surface mass densities higher than
        $ \varSigma \sim 10^{3} $ to $ 10^{4} $\,g\,cm$^{-2}$
        (corresponding to the initial state of discs more massive than Pluto;
        $ M\sub{disc} \gtrsim 10^{25}\,\mathrm{g} $),
        the disc's global evolution is dominated by viscous spreading
        rather than PR drag
        (see Eqs.~\eqref{eq:sigma_crit_time} and \eqref{eq:mdisk_crit};
        Fig.~\ref{fig:time_visc_pr}).
  \item Discs more massive than Io
        ($ M\sub{disc} \gtrsim 10^{26}\,\mathrm{g} $)
        will viscously spread
        enough for some of their mass to end up beyond the Roche limit
        (see Eq.~\eqref{eq:mdisk_form}; Fig.~\ref{fig:mdisk_form}).
  \item These massive discs can survive 
        against viscous spreading and PR drag
        for longer than the age of the universe
        (see Eq.~\eqref{eq:t_disc_plan}).
\end{enumerate}
Disc material flowing out beyond the Roche limit will coagulate
to form a new generation of (minor) planets.
These newly formed bodies will migrate outwards
(due to angular momentum transfer with the disc via MMRs)
and merge with one another to form larger bodies.
Based on the model of \citetalias{2012Sci...338.1196C},
we expect the new (minor) planets to have the following properties:
\begin{enumerate}
  \item Tens of per cent of the initial disc mass can be recycled into
        second-generation bodies (see Eq.~\eqref{eq:f_recyc_simple}).
  \item Most of the recycled mass ends up in a single object
        orbiting such that its 2:1 MMR coincides with the Roche limit,
        which means its orbital period will be of the order of $\sim$\,10\,h
        (see Eq.~\eqref{eq:period_2to1}).
  \item In addition, a population of some tens
        (see Eq.~\eqref{eq:n_pyra})
        of smaller bodies is predicted in the region
        extending inwards from this largest planet down to the Roche limit,
        most of which are small bodies huddled in the innermost regions
        (see Eq.~\eqref{eq:dn_ddelta_pyra}).
  \item Beyond a narrow region close to the Roche limit
        where the active formation of new bodies takes place,
        the distribution of masses of these bodies as a function of
        distance from the Roche limit roughly follows a power law
        with an exponent close to 2
        (see Eqs.~\eqref{eq:q_delta_cont_disc} and \eqref{eq:q_delta_pyra};
        Fig.~\ref{fig:delta_q}).
  \item The ongoing formation of new bodies
        close to the Roche limit happens on timescales of
        hours, days, and months,
        independent on the properties of the disc
        (three different timescales for three different types of objects;
        see Eqs.~\eqref{eq:t_toomre}, \eqref{eq:t_c}, and \eqref{eq:t_d}).
\end{enumerate}

\section{Numerical simulations}
\label{s:numerical}

We now proceed to
investigate the evolution of WD-debris-disc-planet systems
using numerical techniques.
Numerical simulations
make it possible to
obtain detailed, quantitative estimates of the state of these systems
as a function of time,
as opposed to the predictions for their behaviour in certain limiting cases
that can be found analytically.
Furthermore, the numerical simulations
allow us to let go of several simplifying assumptions
that were necessarily made in the previous section,
thereby testing the validity of the analytical predictions.
Some specific improvements that were made with respect to the analytical treatment are
(1)~using a more complete prescription for the disc's effective viscosity,
(2)~computing disc-planet torques associated with individual MMRs,
rather than using an average torque due to many MMRs,
(3)~including the back-reaction of the planets on the disc, and
(4)~following the evolution of individual minor planets,
as opposed to starting with equal-mass building blocks.
We briefly describe our numerical model and how we use it in Sect.~\ref{s:num_methods}.
The results of the simulations
and a comparison with the analytical predictions
are presented in Sect.~\ref{s:num_results}.

\subsection{Methods}
\label{s:num_methods}

\subsubsection{Model description}
\label{s:num_model}

We employ a modified version of
\textsc{hydrorings},
a hybrid viscous-disc/discrete-body model
originally developed to study the Saturnian ring-satellite system
\citep{2010Natur.465..752C,2011Icar..216..535C,2010Icar..209..771S}.
It is a one-dimensional model,
which is necessary to achieve the computational speeds
needed
to follow the evolution
of disc-planet systems over multi-Gyr timescales.
The code
consists of two main
components:
(1)~a viscous-evolution solver that describes the evolution of
the particle disc inside the Roche limit as a continuous medium
and
(2)~an orbital-elements tracker
that models the migration and growth of
discrete bodies outside the Roche limit.
These two modules are coupled by prescriptions for
(a)~how new bodies are spawned when disc material overflows the Roche limit and
(b)~the subsequent interaction of these bodies with the disc
through angular momentum exchange at MMRs.
The model
can successfully reproduce the masses and locations of
many of Saturn's regular satellites
and has also been tested on the formation of Earth's Moon
\citep{2010Natur.465..752C}.

The model's
disc module solves
the viscous diffusion equation with added external torques
\citep[e.g., Eq.~(3) of][]{1996ApJ...460..832T}
on a discretised radial grid
using a second-order time-stepping method.
The prescription used for the disc's effective kinetic viscosity
is detailed in Sect.~2.2 of \citet{2010Icar..209..771S}.
This includes the translational and collisional components of viscosity
\citep{1978Icar...34..240G,1984SvA....28..574S},
as well as the contribution from self-gravity wakes for $ Q\sub{T} < 2 $
\citep{2001Icar..154..296D}.
For gravitationally unstable discs in the outer part of the tidal disruption zone,
the total effective kinetic viscosity equals the one given by Eq.~\eqref{eq:visc_sg}.
Torques on the disc from first-order MMRs with external bodies
are computed using the formalism of \citet{1987Icar...69..157M}.
For this study, PR drag was added as an additional torque on the disc,
computed by integrating Eq.~\eqref{eq:torque_dens_pr} over each radial bin.
The implementation of PR drag in the code was tested
by reproducing the results of \citet[specifically, their Fig.~7]{2011ApJ...741...36B}.
Any inwards-flowing disc material that passes through
the sublimation radius $ r\sub{subl} $
is removed from the simulation and its angular momentum is lost.
This material is assumed to be converted to gas,
which is rapidly accreted
onto the WD via a gaseous viscous accretion disc
\citep[not modelled here, but see][]{2012MNRAS.423..505M}.

At each time step,
any disc material that has spread beyond the Roche limit $ r\sub{R} $
is converted into a new minor planet,
which is subsequently tracked as a discrete body
(fully described by its mass, semi-major axis, and eccentricity).
The semi-major axes of the bodies are modelled to evolve
under the action of disc torques associated with first-order MMRs
(already computed for the back-reaction of the bodies on the disc; see above).
Changes in the eccentricities of the bodies
due to mutual perturbations are modelled
using an instantaneous-close-encounter formalism \citep{2011Icar..216..535C}.
When two bodies
come to within 2.2 mutual Hill radii from one another
\citep{2007Icar..189..523K},
they are assumed to merge
into a new body whose mass is the combined mass of the two progenitors
and whose orbital elements are dictated by the conservation of angular momentum.
The code
does not take into account
resonant interactions between orbiting bodies
\citep[see discussion in Sect. S3.3 of ][]{2010Natur.465..752C}
and
the treatment of tidal interactions between the planets and the central object
was switched off for the present study (see Sect. \ref{s:largest_planet}).

\subsubsection{Setup of the model runs}
\label{s:num_setup}

One of the main goals of this work is to investigate
which WD debris discs are able to form second-generation (minor) planets
and what the properties of the newly formed bodies may be.
To answer these questions using our numerical model,
we ran a set of simulations
with a different initial disc mass $ M\sub{disc} $ and radius $ r_0 $
for each run,
keeping all other parameters fixed.
Table~\ref{tbl:sim_pars} lists the parameter values we adopted.
Most of our choices are motivated in Sect.~\ref{s:prelim},
but some additional details are provided below.

\begin{table}
  \centering
  \caption{Parameter values used in the numerical simulations.}
  \label{tbl:sim_pars}
  \begin{tabular}{lcr@{\,}l} 
    \hline
    Parameter & Symbol & \multicolumn{2}{c}{Value(s)} \\
    \hline
  WD effective temperature & $ T_\star $ & 10{,}000 & K \\
  WD mass & $ M_\star $ & 0.6 & $ \mathrm{M\sub{\odot}} $ \\
  WD radius & $ R_\star $ & 0.0125 & $ \mathrm{R\sub{\odot}} $ \\
  sublimation radius & $ r\sub{subl} $ & 0.15 & $ \mathrm{R\sub{\odot}} $ \\
  Roche limit & $ r\sub{R} $ & 1.3 & $ \mathrm{R\sub{\odot}} $ \\
  grain radius & $ s $ & 1 & cm \\
  material density & $ \rho\sub{d} $ & 3 & $ \mathrm{g\,cm}^{-3} $ \\
  initial scaled disc width & $ \delta $ & 0.1 & \\
  initial disc radius & $ r_0 $ & 0.2 to 1.2 & $ \mathrm{R\sub{\odot}} $ \\
  initial disc mass & $ M\sub{disc} $ & $ 10^{23.5} $ to $ 10^{28.5} $ & g \\
    \hline
  \end{tabular}
\end{table}

\begin{figure*}
  \includegraphics[height=6.645cm]{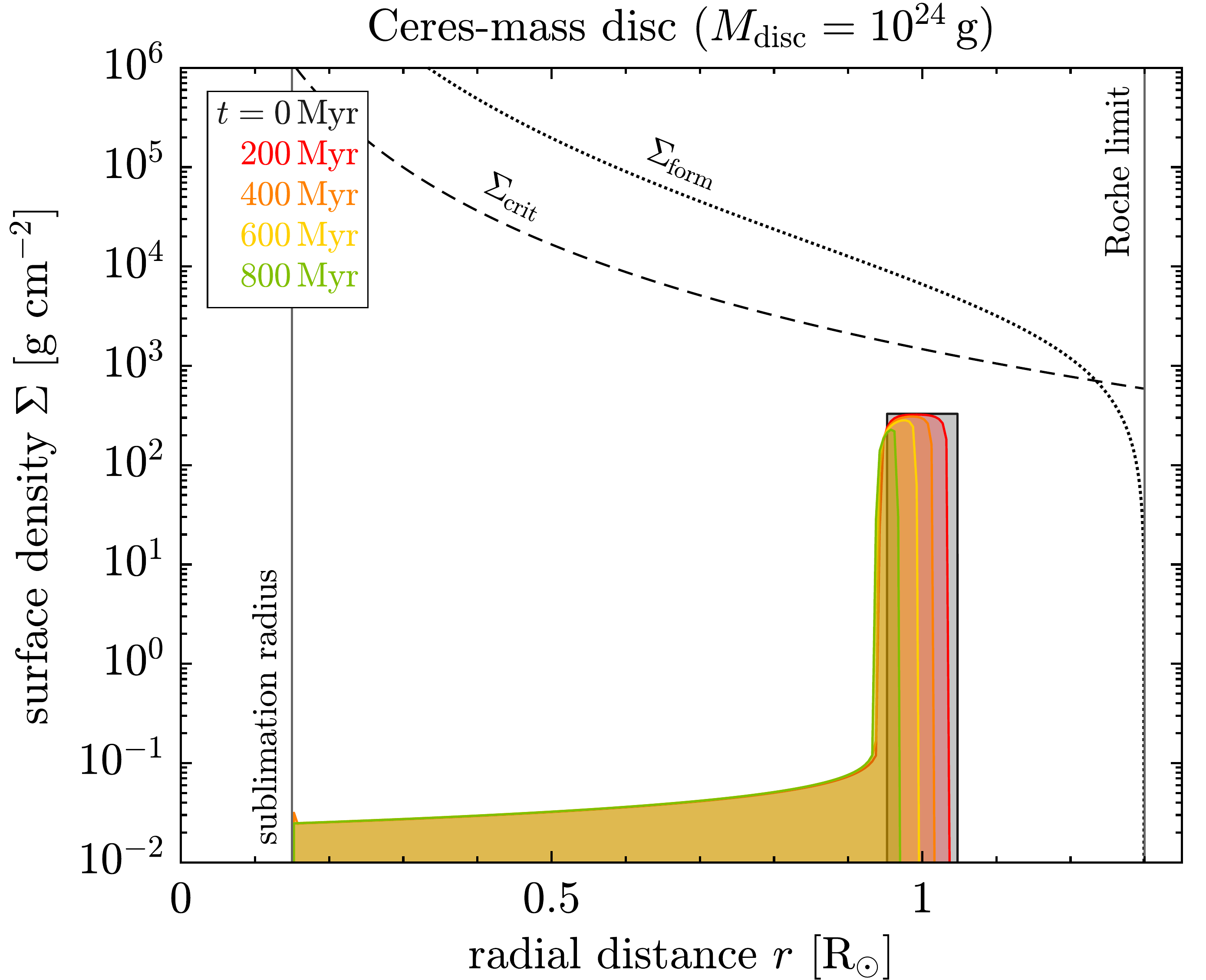}
  \hfill
  \includegraphics[height=6.645cm]{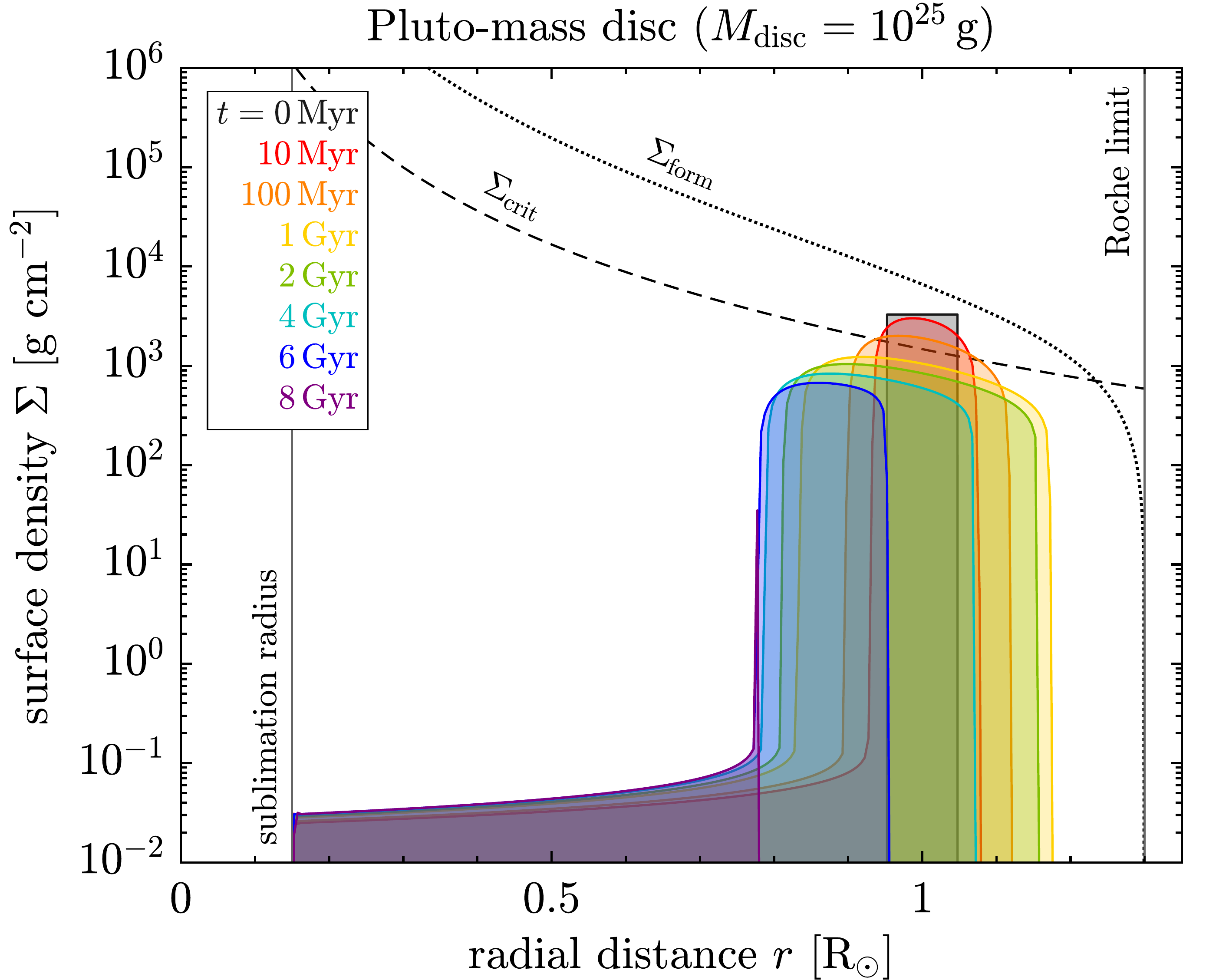}\vspace{9pt}
  \includegraphics[height=6.645cm]{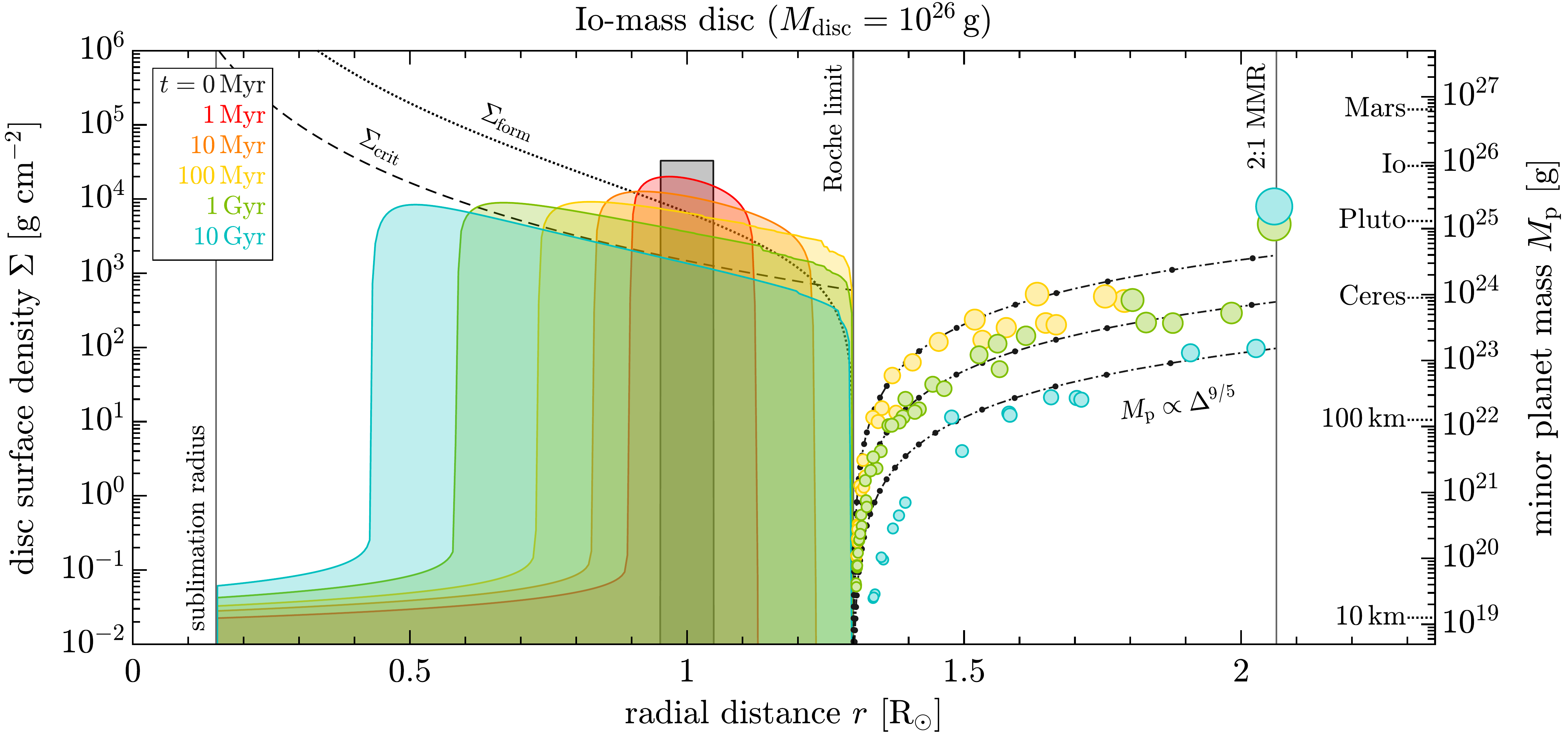}
  \caption{%
  Three examples of the simulated evolution of a WD debris disc, using different initial masses,
  leading to evolution governed by different processes.
  The coloured lines show the surface-density profile of the discs at different points in time.
  The dashed black curve indicates the critical surface density
  $ \varSigma\sub{crit} $ (Eq.~\eqref{eq:sigma_crit_time}),
  below which disc evolution is dominated by PR drag.
  The dotted black curve shows the minimum initial surface density
  for planet formation $ \varSigma\sub{form} $
  (i.e., the equivalent of $ M\sub{form} $,
  converted to a surface density using Eq.~\eqref{eq:sigma} with $ \delta = 0.1 $).
  In the bottom panel,
  the coloured circles indicate the semi-major axes and masses
  of the (minor) planets that spawned from the disc,
  with larger symbols for more massive objects.
  The dash-dotted black curves show the mass-distance relation
  Eq.~\eqref{eq:q_delta_pyra},
  scaled to roughly match the minor-planet population at different times
  (using,
  from top to bottom,
  $ \mathcal{T} = 8 \times 10^{13} $, $ 1.2 \times 10^{15} $, and $ 1.8 \times 10^{16} $).
  Thick black dots
  in these curves indicate the typical spacing between individual objects
  predicted by Eq.~\eqref{eq:dn_ddelta_pyra}.
  Vertical grey lines mark the sublimation radius $ r\sub{subl} $,
  the Roche limit $ r\sub{R} $, and the radial distance $ r_{2:1} $
  at which point a planet's 2:1 MMR coincides with the Roche limit.}
  \label{fig:snapshots}
\end{figure*}

For the stellar parameters, we chose values that reflect a typical WD
\citep[e.g.,][]{2013ApJS..204....5K}.
By adopting a fixed effective temperature for the WD, we ignore its cooling,
which will decrease the critical surface density over time.
WD effective temperature roughly scales as $ T_\star \propto t\sub{cool}^{-7/20} $ \citep{2010A&ARv..18..471A},
where $ t\sub{cool} $ is the WD's cooling age.
Given that a 10{,}000\,K WD has a cooling age of $\sim$0.6\,Gyr,
the timescale on which
$ \varSigma\sub{crit} \propto T_\star^{4/3} \propto t\sub{cool}^{-7/15} $
will halve is about $\sim$2.6\,Gyr.

While particle size needs to be specified in the model,
and the particle size for WD debris discs is highly uncertain,
the value of this parameter does not play a major role in
the global evolution of the massive discs we consider
(see Sect.~\ref{s:visc_vs_pr}).
For the inwards-extending tail of material caused by PR drag
\citep[see][]{2011ApJ...741...36B},
grain size determines the surface density but not the inward mass flow,
which is the parameter that controls the disc's global evolution.

The radial grid used in the model contains 230 bins
between $ r\sub{subl} $ and $ r\sub{R} $
of width $ 0.005\,\mathrm{R\sub{\odot}} \approx 3480\,\mathrm{km} $.
This resolution
also sets the distance beyond the Roche limit at which new bodies appear.
Comparing the resolution with the distances marking the domains of the various
regimes of planet formation (Fig.~\ref{fig:delta_q}) shows that
only the pyramidal regime will be simulated in the vast majority of runs.

As initial radial surface-density distribution,
we used a top-hat function centred at $ r_0 $
with a width given by $ \delta $ (see Eq.~\eqref{eq:sigma}).
We ran a small grid of models to investigate the effects of disc mass and radius.
The initial central radius of the disc $ r_0 $
was varied between 0.2\,R$_\odot$ and 1.2\,R$_\odot$ in steps of 0.1\,R$_\odot$.
The initial disc mass $ M\sub{disc} $ was varied between
about $ 3 \times 10^{23} $\,g and $ 3 \times 10^{28} $\,g
in logarithmic steps of a factor $ \sqrt{10} $.

Each simulation is continued long enough
for the particle disc to fully disappear
-- i.e., until all material is either locked in (minor) planets
or lost at the sublimation radius.
While in many cases this means simulating times
that are beyond the current age of the universe,
we generally only present results from the first 10\,Gyr of the simulations.
Results beyond $ t = 10 $\,Gyr
are used only
to assess the validity of the analytical disc-lifetime predictions
made in Sect.~\ref{s:disc_lifetime}.

\subsection{Results}
\label{s:num_results}

\subsubsection{Three illustrative examples}
\label{s:num_res_examples}

Before discussing the properties of
the set of simulations as a whole,
we examine the detailed results of three individual runs
(shown in Fig.~\ref{fig:snapshots})
that serve as illustrative examples for
the three main (qualitatively different) types of disc evolution
seen in the simulations.
These runs use the same initial disc radius ($ r_0 = 1 $\,R$_\odot$),
but have different disc masses:
\begin{description}
  \item \textbf{\textit{Low-mass disc:}}
        The top-left panel
        shows
        a disc of initial mass
        $ M\sub{disc} = 10^{24}\,\mathrm{g} \sim 1$\,Ceres mass,
        whose initial surface mass density is below
        $ \varSigma\sub{crit} $.
        The evolution of this disc is governed almost entirely by PR drag
        and can be described well by the results of
        \citet[see Sect.~\ref{s:disc_evol_pr}]{2011ApJ...741...36B}.
  \item \textbf{\textit{Intermediate-mass disc:}}
        The top-right panel is for an initial mass
        $ M\sub{disc} = 10^{25}\,\mathrm{g} \sim 1$\,Pluto mass.
        The disc's initial surface density is above
        $ \varSigma\sub{crit} $,
        but its initial mass is still below $ M\sub{form} $.
        The disc initially
        spreads viscously
        (until about $ t = 1$\,Gyr),
        but before the outer edge has reached the Roche limit
        the surface density drops below the critical level
        and PR drag takes over the further evolution.
  \item \textbf{\textit{High-mass disc:}}
        The bottom panel is for a disc of initial mass
        $ M\sub{disc} = 10^{26}\,\mathrm{g} \sim 1$\,Io mass,
        which is above $ M\sub{form} $.
        Its evolution is dominated by viscous spreading
        for most of the simulation.
        At about $ t = 30 $\,Myr,
        the disc spreads beyond the Roche limit and new minor planets
        start being formed.
        These migrate outwards and merge with one another,
        roughly following the pyramidal-regime mass-distance relationship
        (Eq.~\eqref{eq:q_delta_pyra}).
        Eventually most of the minor planets coagulate into a
        Pluto-mass body orbiting at $ r_{2:1} $.
        Around $ t = 10$\,Gyr, the disc decouples from the Roche limit and
        the formation of new bodies stops.
        For some time after this point, however,
        the existing small bodies close to the Roche limit
        still migrate outwards through interaction with the retreating disc.
        This results in a deviation from the analytical mass-distance relation
        towards the end of the run
        (see also Sect.~\ref{s:num_res_planets}).
\end{description}

\subsubsection{Main results from the full set of simulations}
\label{s:num_res_full}

\begin{figure*}
  \includegraphics[width=8.25cm]{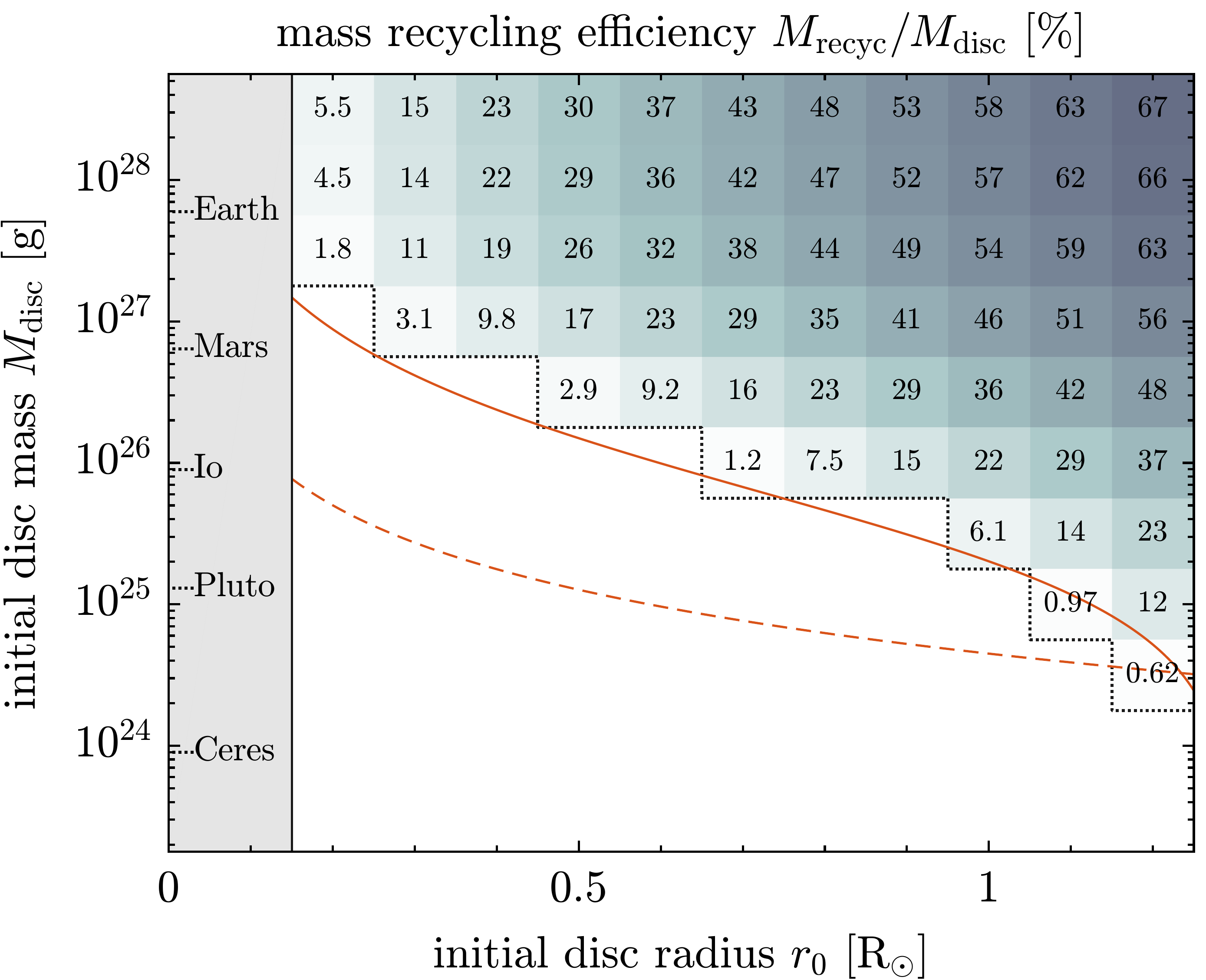}\hspace{26pt}
  \includegraphics[width=8.25cm]{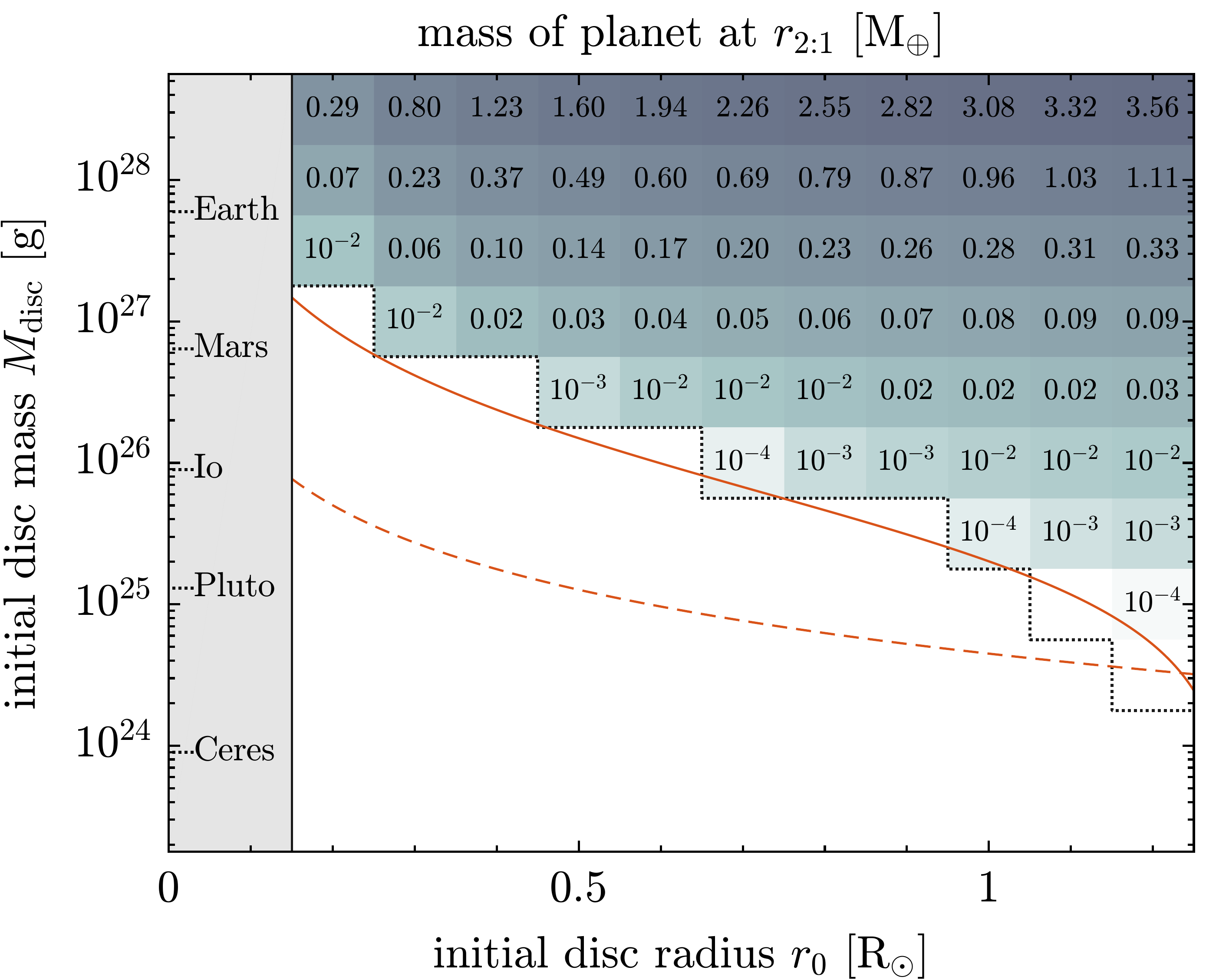}\vspace{7pt}
  \includegraphics[width=8.25cm]{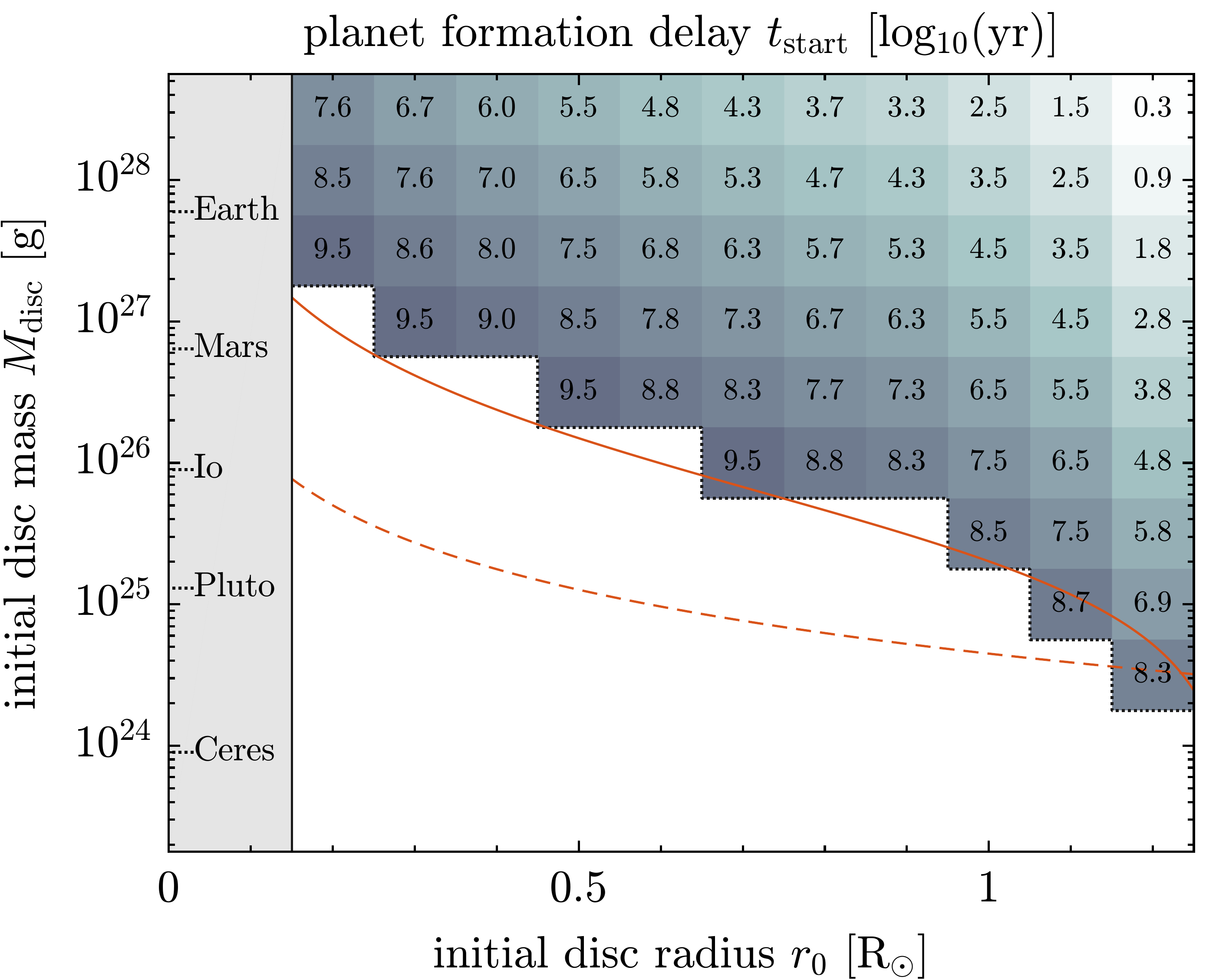}\hspace{26pt}
  \includegraphics[width=8.25cm]{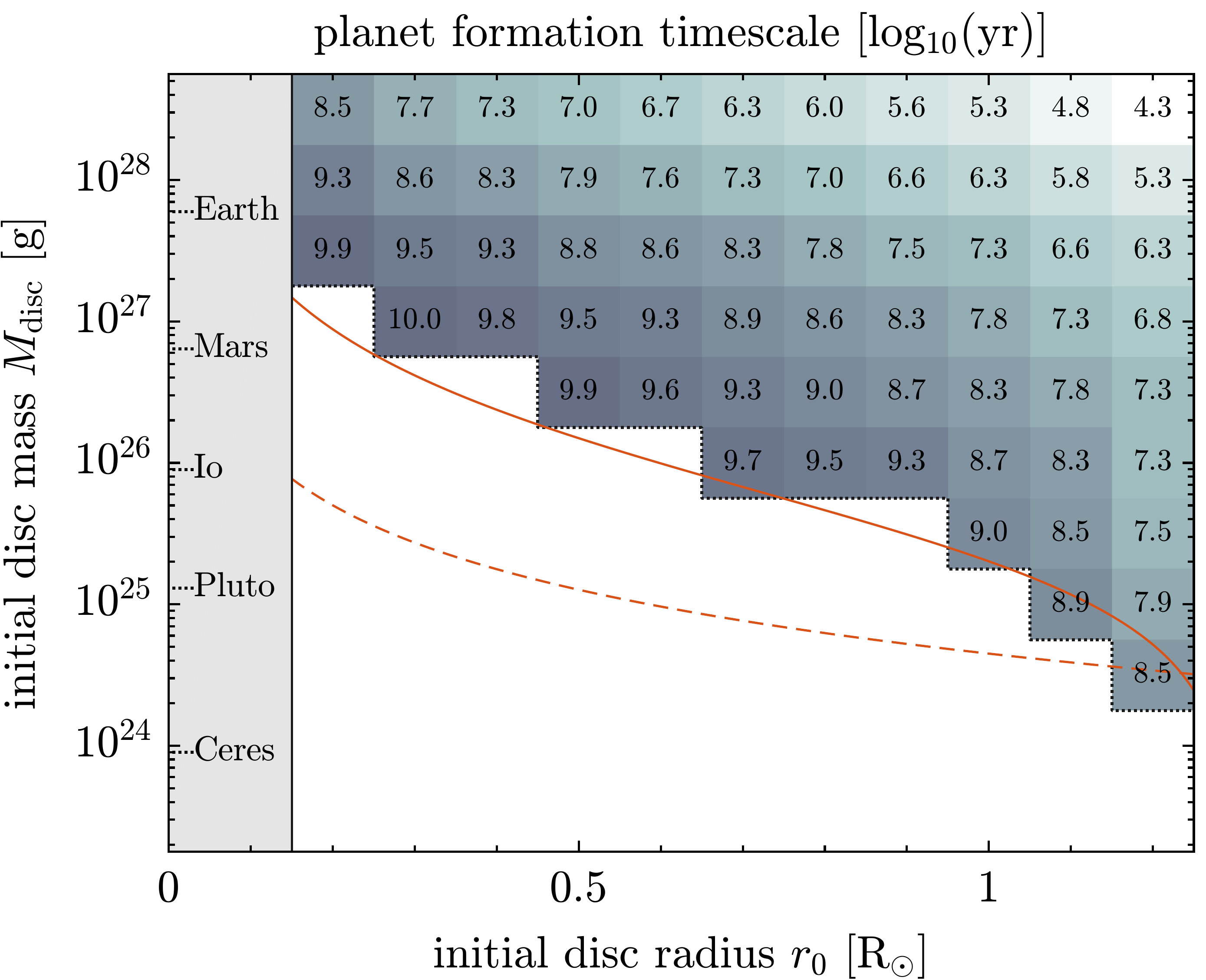}\vspace{7pt}
  \includegraphics[width=8.25cm]{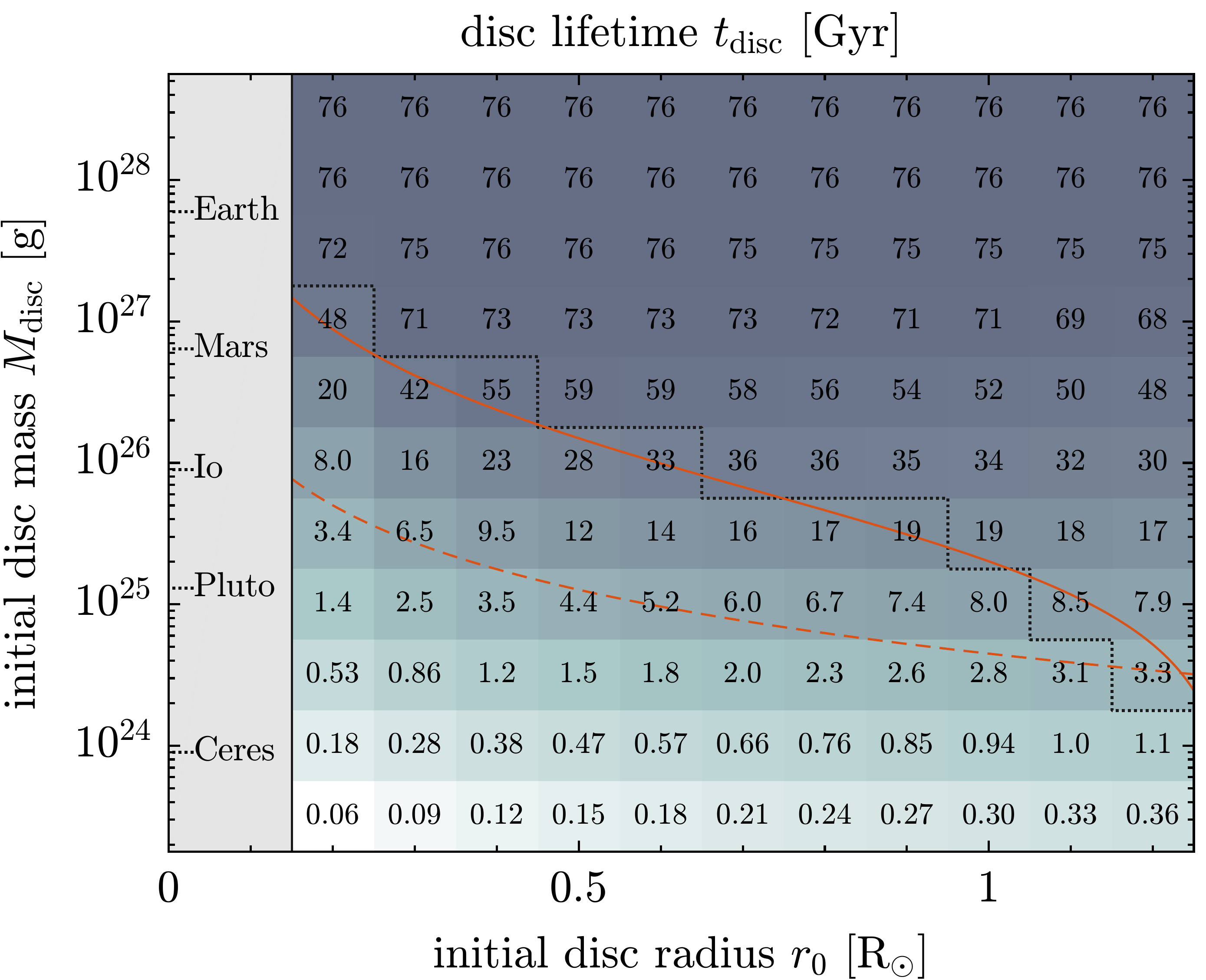}\hspace{26pt}
  \includegraphics[width=8.25cm]{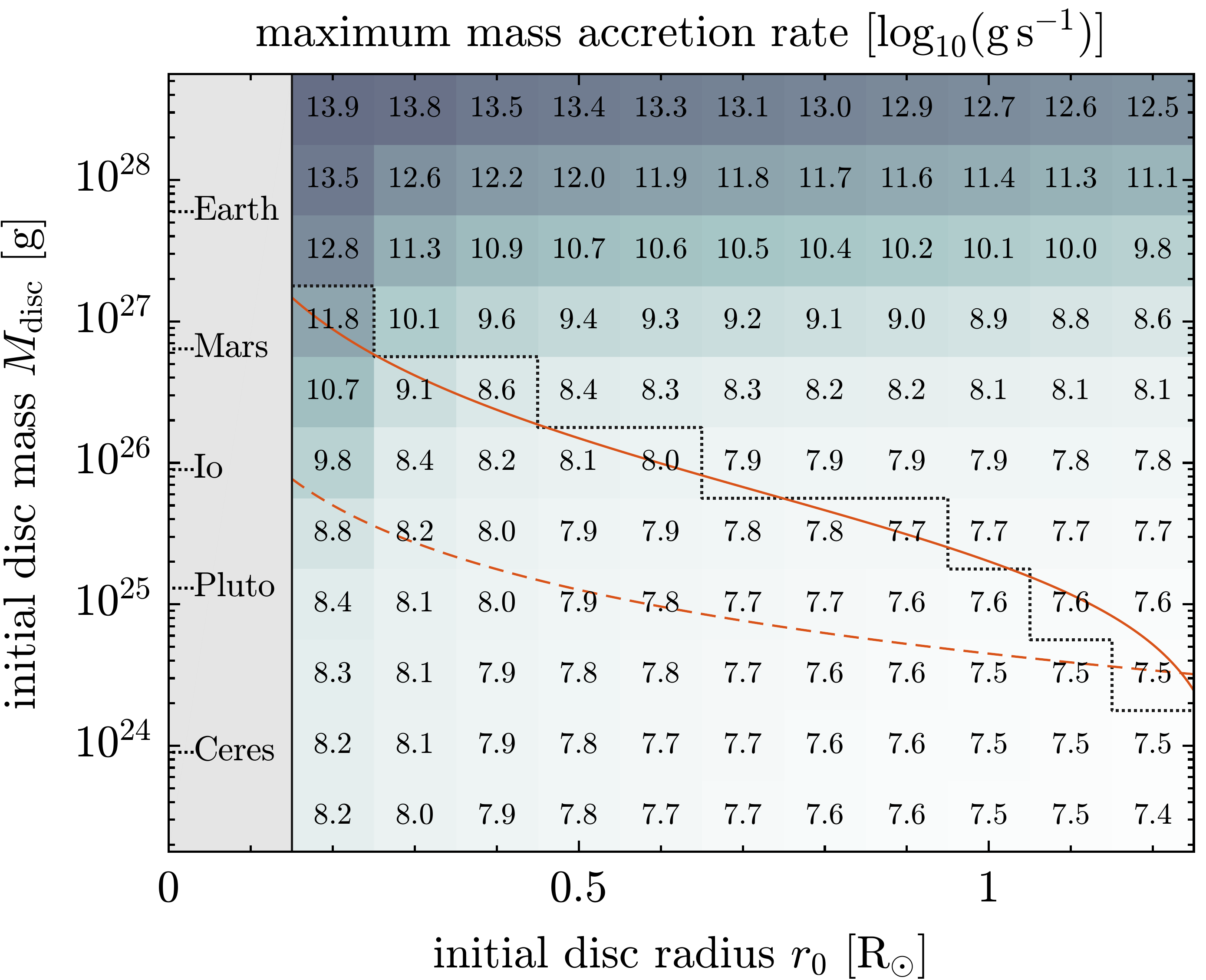}
  \caption{%
  Key results of the numerical simulations,
  shown as a function of the initial central radius of the disc $ r_0 $
  and its initial mass $ M\sub{disc} $.
  Each individual run is represented by a coloured cell with a number inside.
  The colouring is included to guide the eye;
  it contains the same information as the numbers.
  A dotted black line separates runs in which (minor) planets are formed
  from those in which this does not happen.
  The red lines
  are the same as those in Fig.~\ref{fig:mdisk_form}:
  they show the analytical approximations of
  the minimum disc mass for planet formation $ M\sub{form} $ (solid line) and
  the critical disc mass $ M\sub{crit} $ above which viscous spreading
  initially dominates
  the disc's evolution (dashed line).
  \textbf{Top-left panel:} Total mass in (minor) planets
  after 10\,Gyr,
  scaled to the initial disc mass.
  \textbf{Top-right panel:} Mass of the
  body at $ r_{2:1} $
  after 10\,Gyr,
  in Earth masses.
  \textbf{Middle-left panel:} Time at which the first planet is formed.
  \textbf{Middle-right panel:} Time at which half the final planet mass is formed.
  \textbf{Bottom-left panel:} Time until all of the disc's initial mass is
  converted into new (minor) planets and/or accreted onto the WD.
  \textbf{Bottom-right panel:} Highest absolute mass flow
  at the sublimation radius,
  averaged over 1\,Myr,
  recorded in the first 10\,Gyr of the simulation.}
  \label{fig:maps}
\end{figure*}

A broad overview of the simulation results is given by Fig.~\ref{fig:maps},
in which several key quantities extracted from the simulations are mapped
against the initial disc masses and radii of the runs.
Based on these and other results from the simulations,
we make the following observations and inferences:
\begin{enumerate}
  \item The analytical approximation for
        the minimum disc mass for planet formation $ M\sub{form} $
        (Eq.~\eqref{eq:mdisk_form}) closely matches
        the line separating systems with and without planets
        in the numerical results.
        The small discrepancy at $ r_0 > \mathrm{1\,R}\sub{\odot} $ can be explained by
        the non-zero initial width of the simulated discs:
        if the outer parts of a simulated disc start off close enough to the Roche limit,
        the disc can form (minor) planets despite having
        an initial mass below $ M\sub{form} $.
  \item A large fraction of the initial disc mass can be converted into planets.
        For the highest initial disc masses we consider
        ($ M\sub{disc} \gtrsim 10^{28} $\,g),
        this recycled mass fraction $ M\sub{recyc} / M\sub{disc} $
        roughly converges to the values
        predicted by Eq.~\eqref{eq:f_recyc_simple}.
        A small discrepancy between the analytical and numerical results is
        caused by the fact that some angular momentum is lost in the simulations
        by material flowing inside the sublimation radius.
  \item In almost all of the runs,
        most of the planet mass eventually accumulates in a single large body
        orbiting at $ r_{2:1} $.
        Only a few systems (all with initial disc masses close to $ M\sub{form} $)
        have a largest body containing less than 90\% of the total planet mass
        at the end of the run.
        In only two of these exceptions
        (the two empty cells in the top-right panel of Fig.~\ref{fig:maps}),
        the largest body does not reach $ r_{2:1} $.
        These two systems populate only part of the pyramidal regime
        before PR drag terminates their planet formation.
  \item The time at which the first bit of disc material reaches the Roche limit
        and planets start forming $ t\sub{start} $ varies dramatically, from
        a fraction of a year
        (for the highest mass disc starting closest to the Roche limit),
        to several Gyr (for initially compact discs with masses close to $ M\sub{form} $).
        Its scaling with initial disc mass and radius
        roughly resembles that of the viscous timescale
        (Fig.~\ref{fig:time_visc_pr}).
        The low values demonstrate that small bodies located just
        outside the Roche limit
        can form very quickly.
  \item The time at which
        the total mass in planets $ M\sub{recyc} $ has reached half its final value
        (referred to as the planet-formation timescale in Fig.~\ref{fig:maps})
        is relatively short in most runs
        (compared with, e.g.,
        the total length of time that planet formation can remain operating,
        which is often longer than the age of the universe;
        specifically, for massive discs it is roughly
        $ t\sub{crit}( r\sub{R} ) \approx 39\,$Gyr;
        see Sect.~\ref{s:disc_lifetime}).
        This shows that
        the final planet mass $ M\sub{recyc} $ is dominated by
        the early stages of planet formation.
        The reason for this is that
        the planet-forming mass flow through $ r\sub{R} $
        is highest at time $ t\sub{start} $
        and subsequently declines (because $ \varSigma\sub{R} $ goes down;
        see Sects.~\ref{s:sigma_roche} and \ref{s:largest_planet}).
  \item Disc lifetimes follow the predictions made in Sect.~\ref{s:disc_lifetime},
        converging to Eq.~\eqref{eq:t_disc_plan} for initial disc masses of
        $ M\sub{disc} \gtrsim 10^{28} $\,g,
        and following Eq.~\eqref{eq:t_disc_pr} in simulations far enough below the dashed $ M\sub{crit} $ line.
        Most planet-forming discs can survive viscous and PR-drag evolution
        for longer than the age of the universe.
  \item For low-mass systems, the inward mass flow rate at the sublimation radius
        (which we assume to be approximately equal to the mass accretion rate onto the WD)
        are consistent with the PR-drag-induced mass flow given by Eq.~\eqref{eq:mdot_pr_thick}
        and predicted in more detail by \citet{2011ApJ...741...36B}.
        In simulations with initial disc masses above $ M\sub{crit} $ (dashed line),
        viscous spreading moves the disc's inner edge inwards,
        which increases the mass flow due to PR drag slightly.
        A strong enhancement of the mass accretion rate due to viscous spreading
        (reaching $ \gtrsim $\,$ 10^9 $\,g\,s$^{-1}$)
        is only seen in discs with initial masses of $ M\sub{disc} \gtrsim 10^{27} $\,g.
        In these cases, the mass flow at $ r\sub{subl} $ is caused directly by viscous spreading.
        This confirms the conclusion of \citet[their Appendix~A]{2012MNRAS.423..505M} that,
        as an explanation for the highest accretion rates of polluted WDs,
        viscous spreading requires disc masses comparable to that of a terrestrial planet.
\end{enumerate}

\subsubsection{Minor planet properties}
\label{s:num_res_planets}

\begin{figure*}
  \includegraphics[width=\textwidth]{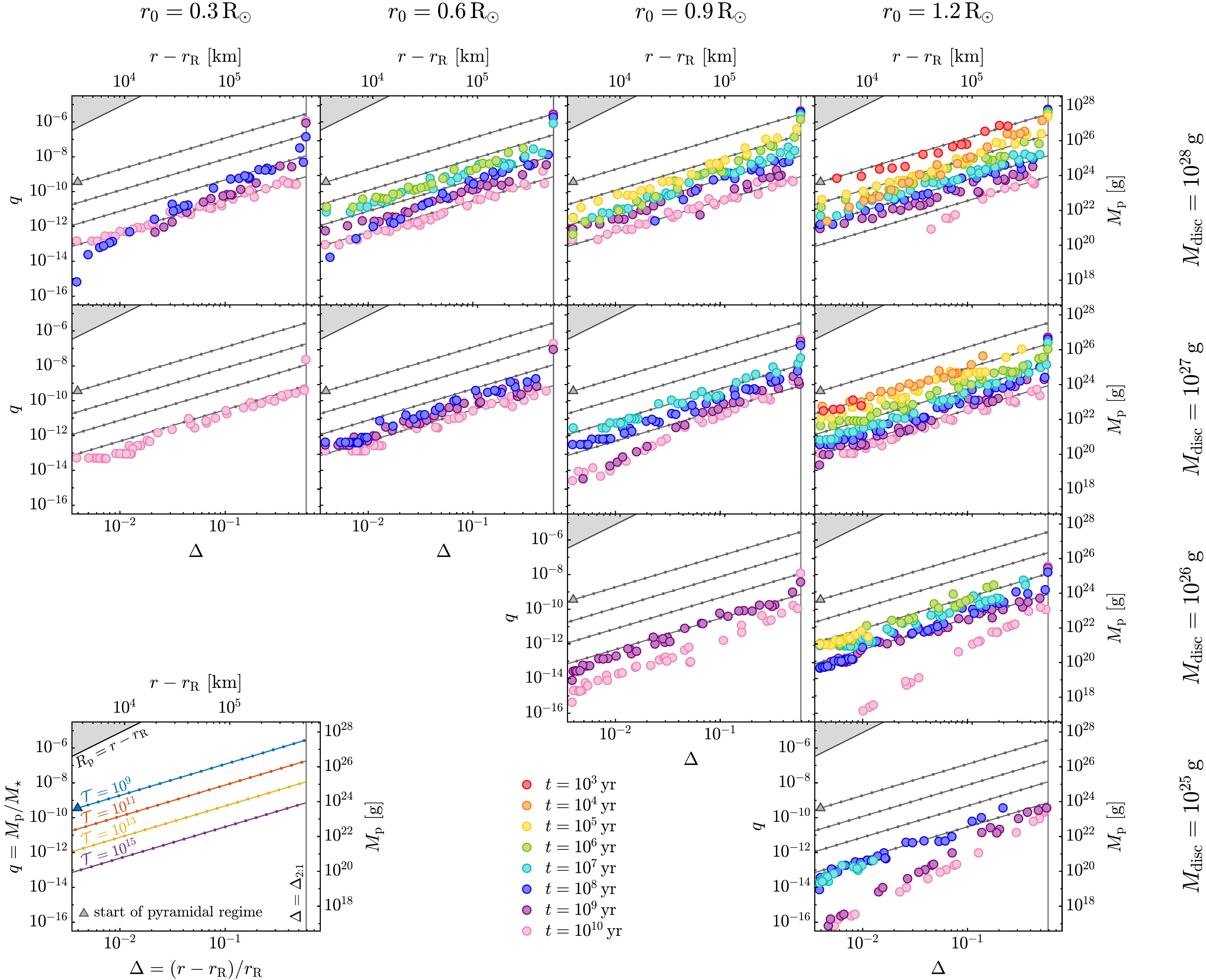}
  \caption{%
  Properties of the (minor) planets
  seen in a subset of the numerical simulations, overplotted on
  the analytical predictions
  found in Sect.~\ref{s:planets}.
  Each panel shows results from a single simulation,
  whose initial disc radius $ r_0 $ and disc mass $ M\sub{disc} $
  can be found
  along the top and right-hand side of the figure, respectively.
  The coloured circles represent individual (minor) planets,
  indicating their mass $ M\sub{p} $
  and distance from the Roche limit $ r - r\sub{R} $
  (or, in terms of dimensionless parameters, their mass ratio $ q = M\sub{p} / M_\star $
  and scaled distance $ \varDelta = ( r - r\sub{R} ) / r\sub{R} $).
  Different colours correspond to
  different times in the simulation,
  as labelled by the legend at the bottom-centre of the figure.
  In all panels, a vertical grey line marks
  the scaled distance $ \varDelta_{2:1} $
  at which a planet's 2:1 MMR coincides with the Roche limit.
  The grey, knotted lines
  in each panel
  are pyramidal-regime mass-distance relations
  for a range of plausible disc properties.
  They are the same for each panel, and a coloured, labelled version of these lines
  is reproduced in a separate panel in the bottom-left corner of the figure.
  This separate panel corresponds to a part of Fig.~\ref{fig:delta_q};
  see the caption of that figure for further information.
  The bottom-left section of the figure does not contain any panels with simulation results,
  because the runs with the corresponding initial parameters did not produce any minor planets.}
  \label{fig:num_planets}
\end{figure*}

Finally,
we review the detailed properties of the simulated minor-planet populations
and compare them to the predictions of the analytical model
presented in Sect.~\ref{s:planets}.
Figure~\ref{fig:num_planets} shows this comparison for a representative subset of the runs.
This shows that
the numerical results
generally
match the analytical predictions
for dimensionless disc evolution timescales of $ \mathcal{T} \sim 10^9 $ to $ 10^{15} $,
as expected.
In several instances, however, deviations occur
where the numerical results follow a steeper slope than the analytical predictions
and/or extend below the $ \mathcal{T} = 10^{15} $ line.
Further inspection of the simulation results reveals that this happens
when the outward mass flow is quenched,
which can occur for two reasons.
(1)~When a strong MMR of a massive body
is located in the outermost parts of the disc,
the back-reaction of this body on the disc
can push the outer edge
away from the Roche limit
(e.g., the orange circles for $ t = 10^4 $\,yr in the top-right panel).
(2)~At late times, PR drag starts to reverse the mass flow at the Roche limit
(e.g., the purple and pink circles
for $ t = 10^9 $ and $ 10^{10} $\,yr in the bottom-right panel).

\section{Discussion}
\label{s:discussion}

\subsection{The occurrence of massive WD debris discs}
\label{s:occur_disc}

In the preceding sections,
we found that viscously spreading WD debris discs
will produce new (minor) planets when they spread beyond the Roche limit,
but that this only happens for very massive discs
(roughly $ M\sub{disc} \gtrsim 10^{26}\,$g;
see Fig.~\ref{fig:mdisk_form}).
This brings up the question whether
discs that are sufficiently massive (to reach the Roche limit)
actually exist.
As mentioned in Sect.~\ref{s:intro},
WD debris discs are usually assumed to have masses comparable to
Solar-System asteroids ($ M\sub{disc} \lesssim 10^{23}\,$g),
but the atmospheric-pollution masses on which this number is based
only give lower limits on the disc mass.
There are several reasons to believe that there could also be
discs with masses in the range $ 10^{26}$ to $ 10^{28}\,$g
(i.e., the mass of a dwarf planet, major moon, or terrestrial planet):
\begin{enumerate}
 \item Some heavily polluted
WDs with helium-dominated atmospheres
also contain significant
amounts of hydrogen
(which does not sink down).
In the case of one of these, GD\;362, it has been shown that this pattern of pollution
can be explained by the accretion of a terrestrial-planet-mass object
containing some water \citep{2009ApJ...699.1473J}.
A similar scenario may account for other WDs exhibiting the combination
of heavy atmospheric metal and hydrogen pollution,
although the hydrogen may also have been accumulated
through the accretion of multiple smaller water-rich bodies
over an extended period of time
\citep[see also][]{2015MNRAS.450.2083R,2017MNRAS.468..971G}.
 \item Theoretical studies
of planetary-system evolution
show that in some cases
planets have pericentre excursions
below the Roche limit
\citep[$\sim$\,0.1 of simulated systems;][]{2014MNRAS.437.1404M,2016MNRAS.458.3942V}.
Furthermore, in unstable planetary systems,
massive moons can
be liberated from their host planet
\citep{2016MNRAS.457..217P}
and scattered into the Roche limit \citep{2017MNRAS.464.2557P},
although the efficiency of this process has yet to be quantified.
 \item For WDs with binary companions,
mechanisms have been proposed to drive small bodies
from an outer planetesimal belt down to the WD
\citep{2015MNRAS.454...53B,2016MNRAS.462L..84H,2017ApJ...844L..16S}.
These mechanisms would apply equally to planets.
 \item The tidal disruption of a planet-sized body around a WD
might be necessary to explain the properties of the interstellar asteroid
\mbox{1I/2017} `Oumuamua \citep{2018arXiv180102658R}.
In such a tidal disruption event,
only a minority ($ \lesssim $\,30\%) of the debris
is ejected to interstellar space,
while the rest remains bound to the WD.
\end{enumerate}

An important question following from these points
is whether the tidal disruption of a massive body
would result in a debris disc with the flat, compact geometry
that we have used as the starting point of our analysis.
Studies considering the formation of planetary rings
from the tidal disruption of a comet or Kuiper Belt object
argue that this is possible \citep{1991Icar...92..194D,2017Icar..282..195H},
but recent studies focussing on WDs have brought forward problems
with the flat disc geometry \citep{2017ApJ...844..116K,2017ApJ...850...50K}.
The problem of disc formation and geometry is
more general to the understanding of circumstellar dust around WDs
and extends beyond the scope of this paper.
We reiterate that
our study does not presuppose any particular formation scenario for the disc.
All conclusions regarding second-generation planets
apply equally
if the required discs form via a different pathway.

\subsection{Recycled planets with or without a remnant debris disc?}
\label{s:destroy_disc}

Given the long lifetime of
massive WD debris discs
(see Sect.~\ref{s:disc_lifetime}),
any second-generation exoplanets produced via the recycling mechanism
are expected to be accompanied
by (a remnant of) the disc that produced them.
For this reason,
WDs exhibiting IR excess may be
the most promising targets to search for recycled planets.
For the same reason,
the recycling mechanism may be expected to not occur frequently,
because the discs that produce planets would also give rise to prominent IR excesses,
and the fraction of WDs with a detectable IR excess is
fairly limited (1 to 3\% for WDs with ages $ t\sub{cool} \lesssim 0.5$\,Gyr;
\citealt{2009ApJ...694..805F}; see also \citealt{2017MNRAS.468..154B}).

On the other hand,
planets may occur without an accompanying remnant debris disc if,
after the planets have already been formed,
the disc is destroyed by some mechanism unaccounted for in our analysis.
Possible examples of such mechanisms are
the run-away accretion of the disc induced by gas drag
\citep{2011MNRAS.416L..55R,2012MNRAS.423..505M}
and a vaporising collision between the disc and a newly-inwards-scattered
body from the star's original planetary system \citep{2008AJ....135.1785J}.
In the case of disc destruction by run-away accretion,
second-generation planets may be expected to have formed before the disc is gone,
because the delay time for run-away accretion
is of the order of 0.1 to 10\,Myr
\citep[although this estimate is highly uncertain; see Eq.~(59) of][]{2012MNRAS.423..505M},
while the formation of planets starts earlier
for many planet-producing discs
(see Fig.~\ref{fig:maps}, middle-left panel).

If a massive disc is destroyed
after some second-generation planets have already been formed,
the planets
may act as a long-term (Gyr-timescale) reservoir of mass,
located just outside the Roche limit
and available for future accretion onto the WD.
Two possible Solar-System analogues of this situation are
the Uranian and Neptunian satellite systems,
both of which may have formed out of ancient massive planetary rings
that have since disappeared \citepalias{2012Sci...338.1196C}.
In a scenario in which
WD debris discs are generally only created
very early in a WD's life
(possibly as a result of some aspect of post-main-sequence evolution)
and then rapidly destroyed (e.g., by run-away accretion),
the long-term storage of mass in second-generation planets
could be the explanation for the occurrence of atmospheric metal pollution in older WDs.
A testable prediction for this scenario is that a large fraction of WDs
(at least as high as the fraction of WDs that show pollution; i.e., tens of per cent)
should be orbited by (minor) planets just outside the Roche limit.

\subsection{Planets in the habitable zone of WDs}
\label{s:hab_zone}

A WD's habitable zone
(in which
a terrestrial planet
can harbour liquid water on its surface)
is located at a few Solar radii from the star,
gradually moving inwards as the WD cools
\citep{2011ApJ...731L..31A,2013AsBio..13..279B}.
Since the region close to a WD ($ r \lesssim 1$\,AU),
including the habitable zone,
has been cleared during the star's post-main-sequence evolution,
no planets are generally expected here
\citep{2012ApJ...761..121M},
except maybe massive gas giants that have survived
a common-envelope event \citep{2013MNRAS.432..500N}.
If the recycling process studied in this paper occurs, however,
it predicts the existence of
low-mass ($ M\sub{p} \ll 1$\,M$\sub{Jup}$) planets
in compact ($ r \ll 1$\,AU) orbits around WDs.

Interestingly,
the orbital radius that our model predicts for the largest planet,
$ r_{2:1} $ (see Eq.~\eqref{eq:r_2to1}; Sect.~\ref{s:largest_planet}), is
squarely in the middle of
the WD `continuously habitable zone'
identified by \citet{2011ApJ...731L..31A}.
This is defined as the 
range of distances
that remain habitable
for an extended period of time (e.g., at least 3\,Gyr).
Specifically,
for a planet orbiting a $ 0.6\,\mathrm{M\sub{\odot}} $ WD
at $ r_{2:1} \approx 2.1\,\mathrm{R\sub{\odot}} $,
the habitable zone starts to coincide with the planet's location
when the WD has cooled to an effective temperature of about
$ T\sub{eff} \approx 6000 $\,K,
corresponding to a WD age of around 2\,Gyr,
and it
has moved too far inwards at $ T\sub{eff} \approx 4000 $\,K,
at an age of about 10\,Gyr
\citep{2011ApJ...731L..31A}.

This situation raises further questions regarding the potential habitability of
planets formed through the recycling mechanism:
Would any volatiles survive
the processing of material in the disc?
If not, could enough water be delivered onto the planet from outside reservoirs later on?
Although a detailed consideration of these issues is beyond the scope of this paper,
we now briefly discuss
the question of water delivery.

\citet{2014MNRAS.445.4175V} show that the trace amounts of hydrogen
that are observed in the atmospheres of helium-dominated WDs can be explained by
the accretion of water-rich comets from remnant Oort-cloud-type reservoirs.
Over a timespan of $\sim$\,10\,Gyr, this mechanism could provide
$ 10^{22} $ to $ 10^{25} $\,g in H onto the WD
\citep[see also Fig.~5 of][]{2015MNRAS.450.2083R},
corresponding to $ 10^{23} $ to $ 10^{26} $\,g in H$_2$O.
Assuming the WD's catchment area corresponds to the cross-section of its
tidal disruption sphere,
a planet can accrete a fraction on the order of
$ ( R\sub{p} / r\sub{R} )^2 $ of this mass, ignoring gravitational focussing.
For an Earth-sized body, this would yield about $ 10^{19} $ to $ 10^{22} $\,g in water
(equivalent to $ 10^{-5} $ to $ 10^{-2} $ times the mass of Earth's oceans),
some fraction of which will stay on the planet during the impact process,
depending on the size and velocity of the impactors
\citep[see, e.g.,][]{2009M&PS...44.1095S,2018SSRv..214...34S,2018arXiv180205034K}.
Whether this amount of water would be sufficient for the emergence of life
is another question that remains to be answered.

Finally,
if habitable planets around WDs exist
and they can be studied in transit,
they would make uniquely favourable targets
for attempts to detect biomarkers \citep{2013MNRAS.432L..11L}.
Our study contributes to this intriguing subject by showing that
the formation of planets in the habitable zone of WDs is possible,
indeed likely, for certain conditions.

\subsection{Observational constraints and prospects}
\label{s:obs}

The existence of
planets in compact orbits around WDs
is an interesting conjecture that warrants to be tested observationally.
Given the small orbital radii and large planet-to-star size ratio of the predicted planets,
transit surveys are likely to be a fruitful strategy for constraining
how frequently
they occur.
The planets
can cause very deep transits
or even complete occultations of their host star,
recurring at periods less than a day.
Thus,
even relatively low-precision surveys
(like some present-day ground-based efforts)
are capable of detecting them
\citep[see][]{2011MNRAS.410..899F,2011ApJ...731L..31A}.
Because the durations of WD planetary transits are relatively short
(typically 1 to 3\,min for the planets we consider;
e.g., \citealt{2011MNRAS.410..899F}),
the sensitivity of such surveys to small planets will depend critically
on the exposure time and/or observational cadence used.

\begin{table}
  \caption{Published results of transit surveys searching for WD planets}
  \label{tbl:obs_pub}
  \begin{center}
  \begin{tabular}{lcccc}
    \hline
  Ref. & Using          & WDs   & Cadence  & Min. detectable $ R\sub{p} $ \\
       &                &       & [min]    & [R$_\oplus$] \\
    \hline
  F11  & SuperWASP      & 194   & 8 to 10  & 0.4 to 1 \\
  F14  & Pan-STARRS     & 1718  & 4        & 1 to 2 \\
  S16  & HST/COS        & 7     & 0.25     & 0.08 to 0.7 \\
  B16  & DECam          & 111   & 1.5      & 0.3 to 1 \\
  V18  & K2             & 1148  & 1 or 30  & 0.125 to 1 \\
  W18  & ARCSAT         & 5     & 1        & 0.3 to 0.5 \\
    \hline
  \end{tabular}
  \end{center}
\textbf{References.}
B16:~\citet{2016MNRAS.462.2506B};
F11:~\citet{2011MNRAS.410..899F};
F14:~\citet{2014ApJ...796..114F};
S16:~\citet{2016ApJ...823...49S};
V18:~\citet{2018MNRAS.474.4603V};
W18:~\citet{2018arXiv180303584W}.
\end{table}

Several transit surveys have been conducted so far to search for planets around WDs
(see Table~\ref{tbl:obs_pub} for some of their specifics),
but no solid-body planetary companions have yet been found
(irregular transits were discovered for \mbox{WD~1145+017}, probably caused by clouds of dust;
see Sect.~\ref{s:wd1145}).
Overall,
the constraints on the occurrence rate of
small planets around WDs
are still relatively weak.
While the presence of Mars-to-Earth-sized bodies in \mbox{10-h} orbits
around half of all WDs or more can be ruled out
\citep{2018MNRAS.474.4603V},
a high-cadence or high-precision transit survey
with a larger sample size would be needed
to put stronger limits on the occurrence rate of
the smaller ($ R\sub{p} < 0.5$\,R$_\oplus$) bodies predicted in this paper.
Given that the geometric transit probability of these planets
is of the order of 1\% \citep[e.g.,][]{2011MNRAS.410..899F},
a detection would likely require a sample size of at least a few hundreds
divided by the occurrence rate \citep[see also][]{2011ApJ...731L..31A}.

To evaluate
the prospects of future transit surveys
to put meaningful constraints on the occurrence rate of
(or to detect) planets created via the recycling mechanism,
we now make estimates of the transit properties of these bodies.
For a planet of radius $ R\sub{p} $ on a circular orbit with radius $ r $,
the geometric transit probability is
\begin{equation}
  \label{eq:tr_prob}
  p\sub{tr} = \frac{ R\sub{p} + R_\star }{ r }.
\end{equation}
If, additionally, the planet's orbit is oriented edge-on (i.e., zero impact parameter),
the transit depth $ \delta\sub{tr} $ and full duration $ D\sub{tr} $ are given by
\citep[e.g.,][]{2011MNRAS.410..899F}
\begin{equation}
  \label{eq:tr_depth}
  \delta\sub{tr}
    =
      \begin{dcases}
        \, \left( \dfrac{ R\sub{p} }{ R_\star } \right)^2
          \approx 0.54 \;
            \Biggl( \dfrac{ R\sub{p} }{ \mathrm{1\,R_\oplus} } \Biggr)^2
            \Biggl( \dfrac{ R_\star }{ \mathrm{0.0125\,R_\odot} } \Biggr)^{-2}
          & \text{for} \;\, R\sub{p} < R_\star \\
        \, 1
          & \text{for} \;\, R\sub{p} \geq R_\star
      \end{dcases}
\end{equation}
\begin{align}
  \label{eq:tr_dur}
  D\sub{tr}
    & = 2 \sqrt{ \frac{ r }{ G M_\star } } \, \bigl( R\sub{p} + R_\star \bigr)
  \nonumber \\
    & \approx 2.1 \, \mathrm{min} \;
        \Biggl( \frac{ M_\star }{ \mathrm{0.6\,M\sub{\odot}} } \Biggr)^{-1/2}
        \Biggl( \frac{ r }{ \mathrm{2\,R_\odot} } \Biggr)^{1/2}
        \Biggl( \frac{ R\sub{p} + R_\star }
          { \mathrm{1\,R_\oplus} + \mathrm{0.0125\,R_\odot} } \Biggr).
\end{align}

The detectability of a planet using the transit method
depends on the signal-to-noise ratio \mbox{(S/N)} of the transit signal,
which can be estimated as
\citep[e.g.,][]{2014ApJ...784...45R}
\begin{equation}
  \label{eq:tr_snr}
  \mathrm{S/N} = \sqrt{ N\sub{in} } \, \frac{ \delta\sub{tr} }{ \sigma\sub{phot} }.
\end{equation}
Here, $ N\sub{in} $ is the total number of in-transit measurements
(i.e., summed over all observed transit events)
and $ \sigma\sub{phot} $ is the photometric precision on a single measurement.
The minimum \mbox{S/N} needed for reliable transit detection
depends on the desired detection efficiency and false-alarm probability.
Following \citet{2014ApJ...784...45R},
we adopt $ \mathrm{S/N} \geq 10 $.

\begin{figure}
  \includegraphics[width=\columnwidth]{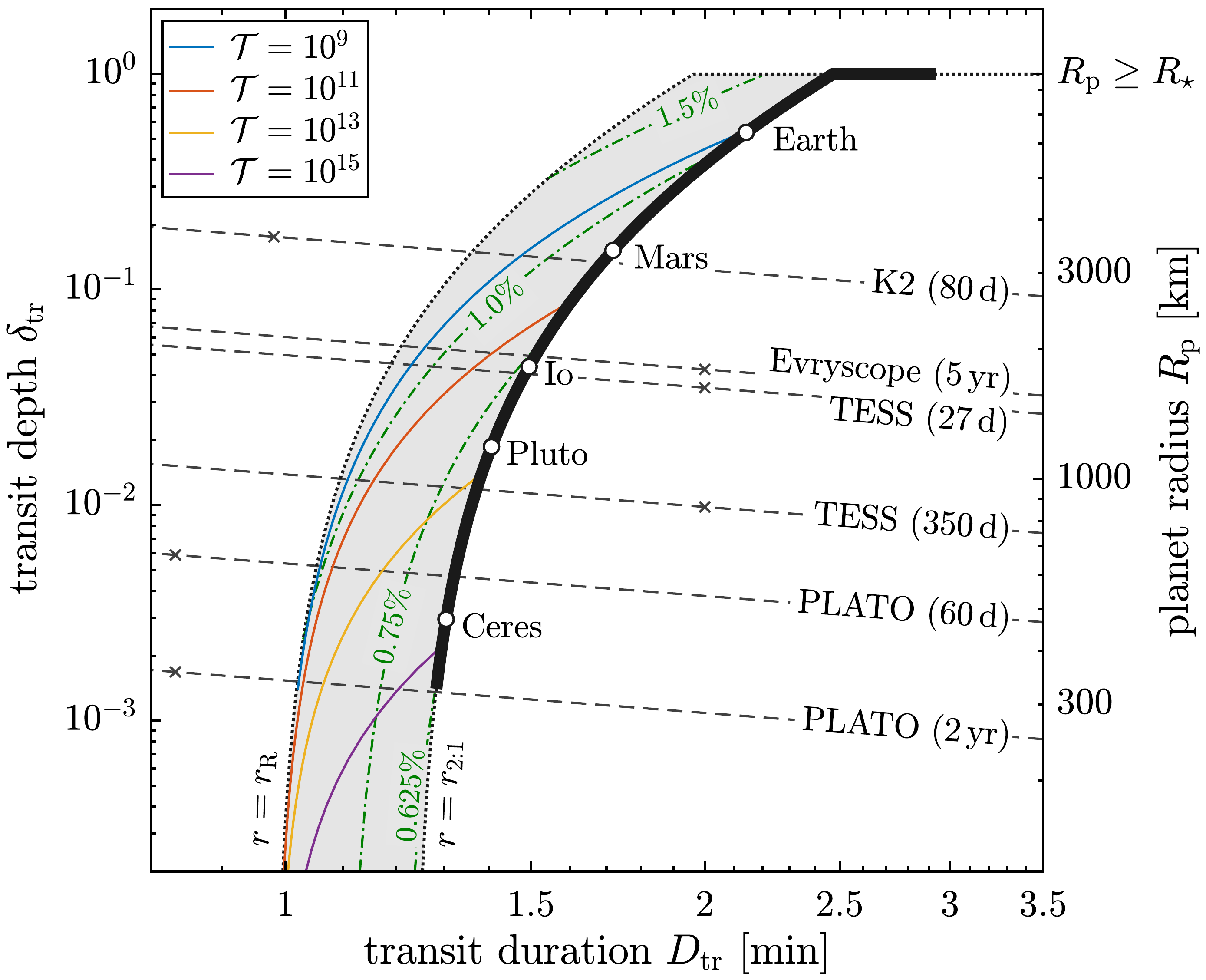}
  \caption{%
  Transit properties of recycled planets and their detectability.
  The shaded area marks the part of transit-depth-and-duration space
  in which transits by recycled planets can in principle occur.
  It is bounded by the dotted black curves,
  which correspond to bodies orbiting at radii $ r\sub{R} $ and $ r_{2:1} $,
  and extends out to longer transit durations along the horizontal dotted line
  at a transit depth of unity
  (for planets that are larger than the star).
  The thick black segment indicates the transit properties of
  the $ r_{2:1} $ bodies seen in the numerical simulations
  (see Fig.~\ref{fig:maps}, top-right panel).
  The four coloured solid curves correspond to pyramidal-regime bodies for different disc properties,
  parametrised by the disc's dimensionless Roche-limit viscous evolution timescale $ \mathcal{T} $
  (see Eq.~\eqref{eq:dimless_time} and Fig.~\ref{fig:delta_q}).
  Lines of constant geometric transit probability are shown
  by green dash-dotted curves (Eq.~\eqref{eq:tr_prob}). 
  Transit parameters were computed assuming
  a WD with mass $ M_\star = 0.6\,\mathrm{M\sub{\odot}} $
  and radius $ R_\star = 0.0125\,\mathrm{R\sub{\odot}} $,
  and planets on edge-on orbits (i.e., zero impact parameter).
  To convert planet masses to radii,
  we assume a mean density of 3\,g\,cm$^{-3}$ for all bodies.
  Several Solar-System bodies are indicated on the black curve,
  showing their transit properties
  if they orbited the WD at $ r_{2:1} $.
  The dashed grey lines show detection limits
  of some ongoing and future transit surveys
  to achieve a S/N of 10 for planets orbiting at $ r_{2:1} $
  (Eq.~\eqref{eq:tr_snr}; using the properties listed in Table~\ref{tbl:obs_fut}).
  For objects in more compact orbits, these lines overestimate
  the detection limits only slightly (see footnote \ref{fn:det_lim}).
  An \mbox{$\times$-sym}bol marks the (possible) cadence of each survey.
  For transits with durations below the cadence, the detection limits
  will be somewhat higher because the transit signal will be diluted.}
  \label{fig:obs_plan}
\end{figure}

\begin{table}
  \caption{Properties of some current and future transit surveys targeting WDs}\vspace{-6pt}
  \label{tbl:obs_fut}
  \begin{center}
  \begin{tabular}{lcccccc}
    \hline
  Survey                 & Cad.                   & Mag.                & Prec.                 & Basel.        & Sky cov.      & WDs    \\
                         & [min]                  &                     & [\%]                  & [d]           & [\%]          &        \\
    \hline
  K2                     & 1, 30                  & 19                  & 3                     & 80            & 5             & 2000   \\
  Evryscope              & 2                      & 16                  & 2                     & (note)        & 100           & 1000s  \\
  \multirow{2}{*}{TESS}  & \multirow{2}{*}{2, 30} & \multirow{2}{*}{15} & \multirow{2}{*}{0.5}  & 27            & 85            & 1000   \\
                         &                        &                     &                       & 350           & 2             & 10s    \\
  \multirow{2}{*}{PLATO} & \multirow{2}{*}{1, 10} & \multirow{2}{*}{16} & \multirow{2}{*}{0.08} & 60            & 25            & 100s?  \\
                         &                        &                     &                       & 730           & 5             & 10s?   \\
    \hline
  \end{tabular}
  \end{center}
\textbf{Columns.}
Cad.:~Possible observational cadences.
Mag.:~Typical magnitude of WDs that can be observed (in the instrument's native waveband).
Prec.:~Expected photometric precision in 1\,h of observations at the listed magnitude.
Basel.:~Temporal baseline over which targets are monitored.
Sky cov.:~Fraction of the sky that is monitored for the listed baseline or longer.
WDs:~Estimated number of WDs of the listed magnitude or brighter that will be monitored
(still very uncertain for PLATO).
\\
\textbf{References.}
K2:~\citet{2014PASP..126..398H}; \texttt{http://www.k2wd.org} (J.~J. Hermes 2017).
Evryscope:~\citet{2015PASP..127..234L}.
TESS:~\citet{2015JATIS...1a4003R}; \citet{2017MNRAS.472.4173R}; \citet{2017arXiv170600495S};
TASC target lists of \mbox{WG-8} Compact pulsators (\texttt{https://tasoc.dk/info/targetlist.php}).
PLATO:~\citet{2014ExA....38..249R}, PLATO Definition Study Report (Red Book).
\\
\textbf{Note.}
Being a ground-based survey, Evryscope does not have a pre-defined temporal baseline
in the `constant-staring' sense that applies to the other entries in the table.
On average it monitors targets continuously for about 6\,h per night.
Additionally assuming operations for 5\,yr with 35\% weather losses,
this yields a total monitoring time of about 300\,d,
which is the effective baseline used for
computing the detection limit in Fig.~\ref{fig:obs_plan}.
\end{table}

Figure~\ref{fig:obs_plan} shows
the transit properties of the bodies predicted by the recycling mechanism,
together with rough estimates of
the detection thresholds of several current and future
transit surveys that target WDs,
whose properties
are listed in Table~\ref{tbl:obs_fut}.\footnote{%
\label{fn:det_lim}
The detection limits shown in
Fig.~\ref{fig:obs_plan} 
are for planets orbiting at $ r_{2:1} $.
Planets orbiting at $ r\sub{R} $ transit twice more frequently
than those at $ r_{2:1} $,
but with transit durations that are a factor $ 2^{1/3} $ shorter.
Both these differences affect $ N\sub{in} $,
so the detection limits for bodies at $ r\sub{R} $ should be lower by
a factor $ \sqrt{ 2 / 2^{1/3} } = 2^{1/3} \approx 1.26 $
compared with the dashed lines in the figure.}
This reveals that the sensitivity to transits of small objects around individual WDs
will improve substantially in the near future.
However, the number of WDs that can/will be targeted by the upcoming missions
(rightmost column in Table~\ref{tbl:obs_fut})
is fairly limited compared with
the reciprocal of the typical transit probability
($ 1 / p\sub{tr} \sim 100 $).
Therefore, if small WD planets are rare,
their detection via transits will likely remain elusive,
even though the constraints on their occurrence rate will improve.
The low transit probability
can in principle
be countered by preferentially targeting
systems known to be oriented edge-on,
for example
from the detection of
circumstellar-gas absorption
\citep{2012MNRAS.424..333G,2016ApJ...816L..22X},
but
there are currently only
a couple of
WDs with such inclination constraints.

Other techniques of detecting exoplanets may be considered,
but most have some drawbacks compared with the transit method:
\begin{enumerate}
\item
Detecting low-mass planets around WDs using the radial-velocity method is generally not feasible,
because there are so few strong absorption lines in WD spectra and they are very broad.
\item 
Substellar companions such as brown dwarfs can be detected around WDs
from the contribution of their IR emission to the system's
spectral energy distribution
\citep[e.g.,][]{2008ApJ...681.1470F},
although it can be difficult to distinguish between a companion
and circumstellar dust
using the IR photometry alone.
\citet{2016ApJ...823...49S} have investigated the potential of this method
for discovering small, non-transiting planets, but find that
it is generally sensitive only to bodies
with radii larger than about 2\,R$_\oplus$.
\item
Smaller bodies (possibly down to \mbox{100-km-sized} asteroids)
may be detected by considering the IR variability
caused by their hot dayside coming in and out of view
\citep{2014ApJ...793L..43L}.
For cool ($ T\sub{eff} \lesssim 10{,}000 $\,K) WDs, however,
this technique likely requires lengthy space-based observations
(e.g., using JWST),
because of the low temperatures expected for rocky bodies
located just outside the Roche limit.
\item
Close-in planets around magnetic WDs may produce detectable electron-cyclotron maser emission
through their interaction with the WD's magnetic field
\citep{1998ApJ...503L.151L,2004MNRAS.348..285W,2005A&A...432.1091W}.\footnote{%
Incidentally,
the \mbox{10-h}-period terrestrial planet proposed by
\citet{1998ApJ...503L.151L}
to explain the peculiar emission lines of the highly magnetic WD GD\;356
\citep[see also][]{2010MNRAS.404.1984W}
could have formed via
the exoplanet-recycling scenario
presented in this paper.}
Magnetic WDs, however,
only comprise a small fraction of all isolated WDs
\citep[2 to 20\%;][]{2015SSRv..191..111F}.
\end{enumerate}

\subsection[Possible application to WD~1145+017]{Possible application to \mbox{WD~1145+017}}
\label{s:wd1145}

The discovery of transits in the light curve of \mbox{WD~1145+017}
\citep{2015Natur.526..546V}
has opened up a new window onto the circumstellar environment of polluted WDs.
A wealth of observational data is now available for this star,
but the comprehensive picture of what ultimately causes the transits remains unclear.
Here, we briefly review the relevant observations of \mbox{WD~1145+017} and
discuss whether its curious properties can be explained by
the model of second-generation planet formation
presented in this paper.

\subsubsection{Observational clues and their interpretation}

\mbox{WD~1145+017} is a heavily polluted helium-atmosphere WD
whose light curve shows a multitude of
recurring transit features \citep{2015Natur.526..546V}.
The dips are deep (up to 60\%), long ($\gtrsim$\,several minutes), irregular in shape,
and occur at multiple longitudes along the orbit,
sometimes collectively covering a large fraction of the orbital cycle.
Their recurrence periods fall between about 4.5 and 5\,h,
corresponding to
Keplerian orbital periods
at and/or near the Roche limit
(see Eq.~\eqref{eq:period_roche}).
The shapes and depths of the dips evolve on timescales of days
and individual transit features seem to gradually appear and disappear
over the course of several weeks or months
\citep{2016ApJ...818L...7G,2016MNRAS.458.3904R,2017MNRAS.465.3267G}.

In addition to its peculiar light curve,
\mbox{WD~1145+017} also exhibits excess IR emission from circumstellar dust
\citep{2015Natur.526..546V,2018MNRAS.474.4795X}
and evolving spectral absorption features due to metallic circumstellar gas
\citep{2016ApJ...816L..22X,2017ApJ...839...42R}.
The WD itself has an effective temperature of
$ 15{,}900 \pm 500$\,K,
corresponding to a cooling age of
$ 175 \pm 75 $\,Myr
\citep{2015Natur.526..546V}.
Its atmospheric metal pollution constitutes
about $ 6.6 \times 10^{23} $\,g of material
(almost the mass of Ceres),
yielding an estimated steady-state mass accretion rate
of $ 4.3 \times 10^{10} $\,g\,s$^{-1}$ over sinking timescales of 0.3 to 0.6\,Myr
\citep{2016ApJ...816L..22X}.

The properties of the dips in the light curve of \mbox{WD~1145+017}
are inconsistent with transits of solid-body planets.
Instead, the irregular, slowly evolving shapes
and the transient nature of the dips
suggest that they are caused by transiting clouds of dust
\citep{2015Natur.526..546V}.
However, while individual transit features come and go,
the most notable periodicity associated with the transits (about 4.5\,h)
has remained present in the light curve
over a timespan of several years
\citep[e.g.,][]{2017MNRAS.465.3267G}.
Furthermore,
the periods of some transit features were found to remain stable
to within 20\,ppm over several months \citep{2018MNRAS.474..933R}.
These properties seem to be incompatible with dust clouds,
which are expected to disperse
due to Keplerian shear within a few orbits.

The long-term stability of the transit recurrence periods
can be explained by a scenario in which
the dust that causes the dips is released quasi-continuously
from massive bodies
that
carry enough angular momentum to have highly stable orbital periods
\citep{2015Natur.526..546V}.\footnote{%
An alternative explanation for the period stability of the transits is that
the transiting clouds of dust
are trapped in the WD's magnetosphere
and therefore corotate with the WD \citep{2017MNRAS.471L.145F}.
The magnetic field strength required for this scenario, however, seems to be
far above the observational upper limits for \mbox{WD~1145+017}
\citep{2018MNRAS.474..947F,2018MNRAS.474..933R}.}
The release of short-lived dust from
these much-longer-lived objects
may happen
either through their direct disintegration
\citep{2015Natur.526..546V}
or via cratering erosion of larger bodies by somewhat smaller ones
\citep{2017MNRAS.465.3267G}.
Tidal-disruption simulations show that
an internally differentiated body orbiting at the Roche limit
may indeed be able to sustain
prolonged intermittent mass loss \citep{2017MNRAS.465.1008V}.

Assuming the dust that causes the dips is indeed released from massive bodies,
the observed pattern of transit periods
may provide insight into
the orbital configuration and nature of these objects.
In the original K2 light curve
(from which the peculiar nature of \mbox{WD~1145+017} was discovered),
six periodicities could be discerned,
ranging from 4.499 to 4.858\,h \citep{2015Natur.526..546V}.
The two most prominent of these,
at 4.499 and 4.605\,h,
have since been recovered in ground-based data that was taken several years later
\citep{2016MNRAS.458.3904R,2017MNRAS.465.3267G},
suggesting
that at least these two periods
correspond to massive objects
and lending some credibility to the other four.
Most of the dips
seen in the ground-based light curves, however,
are associated with short-lived features and repeat at
slightly (but statistically significantly) shorter periods,
mostly between 4.490 and 4.495\,h
\citep{
2016ApJ...818L...7G,2016MNRAS.458.3904R,2018MNRAS.474..933R,2017ApJ...836...82C,2017MNRAS.465.3267G}.

To explain
the observed pattern of transit periods,
\citet{2016MNRAS.458.3904R} suggest that
the \mbox{4.499-h} period corresponds to a long-lived, massive body
that fills its Roche lobe,
while
the 4.490-to-\mbox{4.495-h} periods correspond to
short-lived fragments of this body, released via its L1 Lagrange point
onto slightly smaller orbits.
In this scenario, the observed ratio of transit periods can be used to
estimate the mass of the parent body,
for which \citet{2016MNRAS.458.3904R} find
$ 10^{23} $\,g (about 0.1\,Ceres mass).
Also regarding this scenario specifically,
\citet{2017MNRAS.464..321G} find that
the parent body cannot be more massive than $ 10^{23} $\,g,
else
its perturbations of the fragments would cause period deviations
violating the observational upper limits on their period stability
\citep[see also][]{2016MNRAS.461.1413V}.

\subsubsection{The origin of massive bodies just outside the Roche limit}

It is clear that
many of the properties of \mbox{WD~1145+017}'s transits can be explained
if
the dust causing the dips is released from long-lived, massive objects.
An important unresolved issue
for this scenario, however, is
how these bodies
came to be in
4.5-to-\mbox{5-h}-period orbits,
which
lie
just outside the Roche limit,
far inside the region
that was cleared during earlier stages of
the star's evolution.
As an additional complication,
the putative bodies likely need to be on low-eccentricity
($\lesssim$\,0.1) orbits
to avoid immediate tidal disruption,
because the efficiency of this process
quickly increases with decreasing pericentre distance
\citep{2017MNRAS.465.1008V}.

\citet{2016ApJ...818L...7G} have suggested
that the close-in bodies may
be the result of
the tidal circularisation of a (minor) planet that was previously
on a highly eccentric orbit with its pericentre close to the Roche limit
\citep[see also Sect.~1.2 of][]{2015MNRAS.447.1049V}.
A potential problem with this scenario is that,
because of angular-momentum conservation,
the radius of the circularised orbit
would come to be a factor $ \! \sqrt{2} \,\! $ larger than
the original orbit's pericentre distance.
Since the latter would have to be outside the Roche limit
to avoid immediate tidal disruption,
it is difficult to intactly place a massive body
on an orbit very close to the Roche limit ($ r / r\sub{R} < \sqrt{2} $)
through tidal circularisation.
Radiative forces, such as PR drag or the Yarkovsky effect, can
remove angular momentum,
circularising orbits without increasing their pericentre distance,
but over the lifetime of a WD
this only works
for smaller objects
\citep[i.e., dust, pebbles, and boulders;][]{2015MNRAS.451.3453V}.
Alternatively,
the circumstellar gas disc detected around \mbox{WD~1145+017}
may have helped in the circularisation process
\citep{2016ApJ...818L...7G}.
It is still unclear, however, whether enough gas is available
around \mbox{WD~1145+017}
for this to work and, if so, whether it can dissipate energy fast enough.

In short,
it is challenging to explain
the proposed massive bodies
by invoking the migration of intact planetesimals
from high-eccentricity, AU-scale orbits
to low-eccentricity, 4.5-to-\mbox{5-h}-period ones.
In contrast,
the recycling mechanism studied in this paper
gives a natural explanation
for massive bodies on circular orbits close to the Roche limit.
In the region just outside the Roche limit, it predicts
a large number of small bodies that orbit close to one another,
which may explain the observed multitude of transit periods.
Furthermore, it allows the formation of minor planets as massive as 0.1\,Ceres mass
\citep[as suggested by][]{2016MNRAS.458.3904R} within the \mbox{175-Myr} age of the system
(see \mbox{Figs.~\ref{fig:snapshots}--\ref{fig:num_planets}}),
although the occurrence of such massive objects
very close to the Roche limit
requires a very massive disc
(e.g., the top-right panel of Fig.~\ref{fig:num_planets} shows
\mbox{$ 10^{23} $-g} bodies at $ \varDelta \lesssim 0.01 $),
unless the objects can somehow migrate inwards
after being assembled further out
(for instance through interaction with the gaseous disc).
A remnant of
the massive debris disc that produced the bodies
could either still be present, or it may
since have been destroyed
(see Sects.~\ref{s:destroy_disc} and \ref{s:wd1145_model_ir_mdot}).

\subsubsection{Ongoing planet formation in the \mbox{WD~1145+017} system?}
\label{s:wd1145_model_intro}

In the scenario sketched above,
the dust clouds that
cause the dips in the light curve of \mbox{WD~1145+017}
are emitted by pre-existing massive bodies that
are gradually destroyed.
Our study brings forward another possibility:
the dust clouds
may be related to
the ongoing formation of massive bodies
from a disc that is currently overflowing the Roche limit.
In this process of second-generation planet formation
(one of the main topics of this paper),
numerous collisions occur between rocky building blocks.
So far, we have
treated these collisions as idealised instantaneous perfect mergers,
but in reality many of them likely cause
at least some initial partial disruption of the bodies involved,
after which the resulting debris gradually reagglomerates
\citep[cf.][]{2015NatGe...8..686H,2017AJ....154...34H}.
The dust clouds transiting \mbox{WD~1145+017} may be manifestations of such disruptive collisions.

\begin{figure}
  \includegraphics[width=\columnwidth]{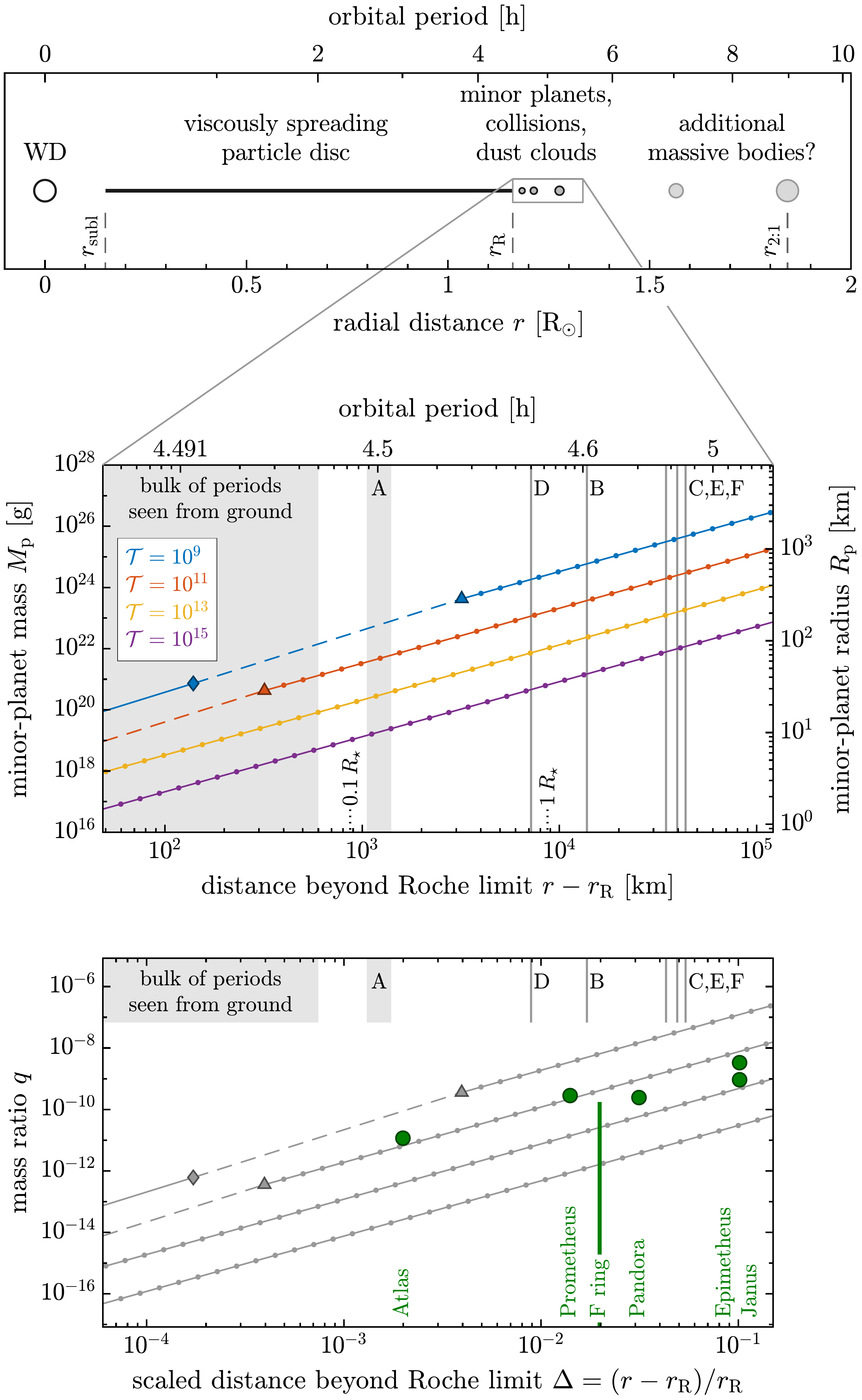}
  \caption{%
  Overview of a possible model for the \mbox{WD~1145+017} system,
  which assumes that
  the main transit period seen in the star's light curve
  corresponds to the outer edge of a viscously spreading debris disc,
  located at the Roche limit (i.e., $ P\sub{R} = 4.490 $\,h).
  In addition, we have assumed
  a WD mass of $ M_\star = 0.6\,\mathrm{M\sub{\odot}} $
  and a Roche prefactor of $ C\sub{R} = 2.0 $.
  \textbf{Top panel:} Cartoon showing a radial cross-section of the system.
  Key radial distances are marked:
  $ r\sub{subl} $, the sublimation radius;
  $ r\sub{R} $, the Roche limit;
  $ r_{2:1} $, the radius for which the 2:1 MMR
  coincides with the Roche limit.
  The WD and the planetary bodies are not drawn to scale.
  \textbf{Middle panel:} Log-scale zoom of
  the region just outside the Roche limit,
  showing the observed transit periods
  (vertical grey bands and lines)
  along with analytical minor-planet mass-distance relations
  (coloured lines and symbols).
  Labels `A' through `F' mark the periodicities seen in the K2 data,
  in order of their signal amplitude
  \citep{2015Natur.526..546V}.
  The various colours, line-styles, and symbols
  used in the analytical mass-distance relations
  have the same meaning as in Fig.~\ref{fig:delta_q}
  (see caption of that figure).
  Also indicated (along the bottom axis)
  are distances equalling the WD's radius and one tenth of that,
  below which the radial extent of the transiting dust clouds may
  be greater than the radial separation of adjacent orbits.
  \textbf{Bottom panel:} Same as the middle panel,
  but in dimensionless parameters.
  This allows a comparison with the Saturnian system,
  which is shown in green.
  The properties of Saturn's moons
  were taken from JPL's Solar System Dynamics website
  (\texttt{http://ssd.jpl.nasa.gov}).
  The location of Saturn's F ring comes from \citet{2002Icar..157...57B};
  its mass (though poorly constrained) is probably not greater than that of its
  two shepherding moons \citep[see][]{2015NatGe...8..686H}.
  We set the Roche limit for Saturn to be at $ r\sub{R} = 137{,}500 $\,km,
  (i.e.,
  in between the outer edge of Saturn's A ring
  and Atlas, the first satellite outside the main rings).}
  \label{fig:wd1145}
\end{figure}

In this scenario,
the most prominent cluster of transit periods seen in the light curve
(around 4.5\,h)
could be associated with
the outer edge of the disc (at the Roche limit)
and/or the region immediately beyond it,
where frequent collisions occur between newly formed bodies.
Dips that repeat at somewhat longer periods
could then be due to collisions
involving bodies that formed some time ago
and have since migrated some distance away from the disc.
Figure~\ref{fig:wd1145} gives an overview of the \mbox{WD~1145+017} system in this scenario,
specifically assuming that
the Roche limit corresponds to an orbital period of $ P\sub{R} = 4.490 $\,h
(i.e., the lower end of the observed range of transit periods).
In the following sections, we discuss this model in more detail.

\subsubsection{The transit periods around 4.5\,h}

With the Roche limit at
an orbital period of $ P\sub{R} = 4.490 $\,h,
as we have assumed for Fig.~\ref{fig:wd1145},
the transit periods close to 4.490\,h
may correspond to dust clouds generated in the continuous and/or discrete regimes
of planet formation (see Sect.~\ref{s:delta_q_cont_disc}).
If this is the case,
the days and weeks-to-months timescales
on which these dips seem to vary
\citep[see, e.g., Figs. 3 and 4 of][]{2018MNRAS.474..933R}
could be related to
the timescale on which new bodies gradually form
(see Eqs.~\eqref{eq:t_c} and \eqref{eq:t_d}).
Specifically regarding the formation of
discrete-regime bodies,
collisions (of continuous-regime building blocks onto a forming discrete-regime body)
are expected to gradually become less frequent
as the body
grows and moves away from the Roche limit,
because this allows
the continuous-regime building blocks to build up to a larger size
before they merge with the discrete-regime body
(see also Fig.~S1 of \citetalias{2012Sci...338.1196C}).
This cycle would then repeat once
the formation of one discrete-regime body is complete
and the production of the next one starts
-- i.e., on a timescale of $ t\sub{d} \approx 1.5$\,months.

On the other hand,
for plausible values of $ \mathcal{T} $ (see Eq.~\eqref{eq:dimless_time}),
the continuous and discrete regimes end very close to the Roche limit
($ \varDelta\sub{d} r\sub{R} \lesssim 100 $\,km
for $ \mathcal{T} \gtrsim 10^9 $).
At such small distance scales,
certain implicit assumptions of our model
may start to break down.
Firstly, our model assumes the Roche limit
is
a sharp boundary.
However, variations in material density $ \rho\sub{d} $
of the order of 1\%, for example,
will move its location by a few thousand km
(see Eq.~\eqref{eq:r_roche}).
Secondly,
the vertical size of the dust clouds orbiting \mbox{WD~1145+017}
must be comparable to the size of the WD
(in order to account for transit depths of 60\%).
If their radial extent is similar to this,
then clouds on orbits
with periods between 4.490 and 4.499\,h will overlap,
so they may interact in ways not accounted for.
Finally,
the dispersion of transit periods around 4.5\,h may also be due to
the effects of radiation pressure,
which can affect the orbital periods of (clouds of) dust grains significantly.\footnote{%
For example, spherical grey dust grains of size $ s = 3 $\,$\upmu$m
(well above the minimum grain size
derived from multiwavelength transit observations; \citealt{2018MNRAS.474.4795X})
and density $ \rho\sub{d} = 3 $\,g\,cm$^{-3}$
in orbit around \mbox{WD~1145+017}
(with parameters $ M_\star \approx 0.6 $\,M$_\odot$, $ L_\star \approx 0.009 $\,L$_\odot$;
\citealt{2015Natur.526..546V})
would have a radiation-pressure-to-gravity force ratio of
$ \beta \approx 0.001 $ \citep[e.g.,][]{1979Icar...40....1B}.
After being released from a massive ($ \beta \simeq 0 $) body
whose orbit is circular,
such grains would orbit the WD at a period that is a factor
$ ( 1 - \beta ) / ( 1 - 2 \beta )^{3/2} \simeq 1 + 2 \beta \approx 1.002 $
longer than that of their parent body
\citep[e.g., Eq. (B6) of][]{2014ApJ...784...40R},
which is comparable to the ratio of the observed periods
($ 4.499\,\mathrm{h} / 4.490\,\mathrm{h} \approx 1.002 $).}
This all goes to say that
our simple analytical model
(Sect.~\ref{s:planets})
may not include all the physics necessary
to fully understand the cluster of periodicities around 4.5\,h.

\subsubsection{Minor planets at longer periods}
\label{s:wd1145_model_longer}

The middle panel of Fig.~\ref{fig:wd1145} shows what
our model means
quantitatively
for the longer periodicities seen in the K2 data.
Because the disc needs to be younger than the age of the system
(175\,Myr),
rough lower limits can be put on the masses of the bodies
that presumably orbit at these periods.
Our numerical simulations show that
discs below this age
generally
do not produce minor planets with masses far below the purple $ \mathcal{T} = 10^{15} $ line
(see Fig.~\ref{fig:num_planets}).
In addition,
the analytical
mass-distance relations with a high normalisation (i.e., low $ \mathcal{T} $)
require a recently created and/or massive disc.
Given the low likelihood of witnessing an $ M\sub{disc} \gtrsim 10^{28} $\,g disc
within 1000\,yr of its formation,
minor planets with masses above the blue $ \mathcal{T} = 10^9 $ line
are improbable
(see Fig.~\ref{fig:num_planets}).

Combining the above two considerations yields rough constraints on the masses of the bodies
corresponding to the various observed transit periods.
The `B' period at 4.605\,h, for example, would likely correspond to a body
of mass $ 10^{21} \mathrm{\,g} \lesssim M\sub{p} \lesssim  10^{25} $\,g
(i.e., a radius between a few tens and several hundreds of km),
with masses towards the lower end of this range being more likely.
If the presence of massive bodies around \mbox{WD~1145+017} is confirmed,
this hypothesis of their origin could be tested by estimating
the masses of the bodies.
Most stringently,
their relative masses as a function of their separation from the Roche limit
should roughly follow the pyramidal-regime mass-distance relation
($ M\sub{p} \propto \varDelta^{9/5} $).

Additional massive bodies may be present around \mbox{WD~1145+017},
including more massive ones at longer orbital periods.
In particular, there may be a large planet orbiting at
$ r_{2:1} $ (i.e., with an orbital period of $ P_{2:1} \approx 9 $\,h),
as seen in many of the numerical simulations
(see Figs.~\ref{fig:maps}, top-right panel, and \ref{fig:num_planets}).
This planet could have evaded detection so far because it does not transit the WD
and does not emit dust like the bodies responsible for the observed transits.
Our model, however, does not require the presence of any additional bodies.
The disc's Roche-limit overflow may have started relatively recently,
such that the pyramidal regime has not yet been filled
(see, e.g., the cyan circles for $ t = 10^7 $\,yr in the bottom-right panel of
Fig.~\ref{fig:num_planets}).

Also shown in Fig.~\ref{fig:wd1145} (in the bottom panel)
is a comparison of \mbox{WD~1145+017} with the Saturnian system.
This shows that
(when expressed in dimensionless distance $ \varDelta $)
the positioning and spacing of \mbox{WD~1145+017}'s transit periods
bears some resemblance to that of
Saturn's small inner moons,
which are likely to have formed out of the planet's rings
(\citealt{2010Natur.465..752C,2011Icar..216..535C}; \citetalias{2012Sci...338.1196C}).
Saturn's rings are not currently overflowing the Roche limit
\citep[being held back temporarily by satellite MMR torques;][]{2017ApJS..232...28T},
explaining why this system at present
does not show the frequent, disruptive, dust-generating collisions
that we infer to be associated with the planet-formation process around \mbox{WD~1145+017}.
Nevertheless,
the presence of Saturn's F ring
\citep[likely the result of a relatively recent collision between two small moons;][]{2015NatGe...8..686H}
shows that
violent collisions between building blocks are
a regular occurrence in
these systems.
At times of active satellite formation,
collisions
are likely to be
much more frequent.

Figure~\ref{fig:wd1145} also shows that
the spacing of objects in both systems
is markedly wider than the predictions of the analytical model
(Eq.~\eqref{eq:dn_ddelta_pyra}; shown by the dots in the mass-distance relations).
For \mbox{WD~1145+017},
it could be that additional bodies are present
but remain undetected because they have not recently undergone dust-producing collisions.
If this is the case, continued monitoring of the WD should reveal
transits at periods between the observed ones.
Alternatively,
our analytical model may underestimate the spacing of the bodies,
because it does not take into account that
gravitational interactions between the massive bodies
may increase the average spacing at which mergers happen
(currently set to two mutual Hill radii in the model;
see Sect.~\ref{s:delta_q_pyra}).
Future modelling efforts
could include N-body simulations to resolve this issue.
Alternatively, the merger distance in the model
could simply be increased to match the Saturn data,
or it could be replaced with a prescription
derived from stability-analysis studies \citep[e.g.,][]{1996Icar..119..261C}.

The more general pyramidal-regime prediction of
$ \dif n / \dif \varDelta \propto 1 / \varDelta $
does seem to hold approximately for the Saturnian moons
(i.e., they are roughly equidistant in $ \log( \varDelta ) $,
especially when considering the Epimetheus-Janus pair as a single object;
see also Fig.~1b of \citetalias{2012Sci...338.1196C}).
For \mbox{WD~1145+017},
the distribution of transit periods found in the K2 data
may be
closer to equidistant in $ \varDelta $
\citep[see also Fig.~1a of ][]{2015Natur.526..546V},
but most of these periodicities still need to be confirmed.
Given the difficulty of
measuring the masses of very small bodies around a WD,
an accurate census of the periodicities in \mbox{WD~1145+017}'s light curve
may be the most accessible way of testing
our model.

\subsubsection{Implications for IR excess and mass accretion rate}
\label{s:wd1145_model_ir_mdot}

If \mbox{WD~1145+017} currently hosts a massive debris disc
that generates (minor) planets at its outer edge
(as postulated in Sect.~\ref{s:wd1145_model_intro}),
then this disc will cause IR emission and metal accretion onto the WD.
Here, we discuss these implications
in the context of observational constraints.

The observed IR excess of \mbox{WD~1145+017}
constitutes
a fractional luminosity of about 0.3 to 0.5\%
and requires a
projected area of dust
that is roughly 50 to 250 times the cross-section of the stellar disc
\citep{2015Natur.526..546V,2016MNRAS.463.4422Z,2018MNRAS.474.4795X}.
The postulated massive debris disc would produce a significant amount of IR emission
(by reprocessing stellar radiation),
but,
since the disc must be oriented close to edge-on
to produce bodies at (nearly) transiting inclinations,
we would only detect a small fraction of its emission.
\citet{2015Natur.526..546V} find that
a flat, optically thick disc with an inclination aligned with the transiting bodies
produces less IR emission than the observed excess.
Therefore,
the existing photometric data are consistent with the presence of an edge-on massive disc.
Fully explaining the observed excess, however,
also requires an additional dust population with a different configuration.
One possibility is that the IR excess is mostly due to
the dust released from the massive bodies.
The dust clouds seen in transit
already account for
some 10 to 20\% of the emitting surface area
inferred from the IR excess \citep{2018MNRAS.474.4795X}
and it is likely that there is more dust associated with these clouds
that passes above or below the stellar disc from our viewpoint.

The mass accretion rate of \mbox{WD~1145+017} inferred from its atmospheric pollution
\citep[$ 4.3 \times 10^{10} $\,g\,s$^{-1}$;][]{2016ApJ...816L..22X}
is relatively high
(although, unlike the transits, this is not unique to this WD;
see, e.g., \citealt{2012ApJ...749....6D}).
PR drag (on a physically thin, flat disc) can only explain
accretion rates up to $ \sim $\,$10^{8} $\,g\,s$^{-1}$
(see Eq.~\eqref{eq:mdot_pr_thick}; \citealt{2011ApJ...732L...3R}),
so an additional effect must be at play.
We find that
viscous spreading can yield the observed accretion rate
for discs with masses of a few times 10$^{27}$\,g
(see Fig.~\ref{fig:maps}, bottom-right panel).
Hence, if the debris disc that we propose for \mbox{WD~1145+017} is massive enough,
it can potentially explain
two uncommon properties of this WD:
transits with stable periods and a high accretion rate.

Finally,
the presence of a disc that overflows the Roche limit
relaxes the requirement that
the mass flow inferred from the transits
\citep[$ \sim $\,10$^{11}$\,g\,s$^{-1}$;][]{2016ApJ...818L...7G,2016MNRAS.458.3904R}
must equal
that
derived from the atmospheric pollution,
because the system is not in steady state in this case.
Instead, the disc constitutes a large mass reservoir,
from which mass flows both inwards
(explaining the mass accretion onto the WD)
and outwards
(producing the massive bodies associated with the transits).

\subsection{Exoplanet recycling around non-WD hosts}
\label{s:other_hosts}

While our study is primarily concerned with the formation of second-generation planets around WDs,
the recycling scenario could in principle also operate around other astrophysical objects,
as long as they are dense enough for the Roche limit (for appropriate material densities)
to lie far enough outside their surface (see Eq.~\eqref{eq:r_roche}).
Here, we briefly discuss a few interesting cases,
noting that further study would be needed to evaluate
the planet-forming potential of compact particle discs
in these different settings in more detail.

\subsubsection{Neutron stars}

Because neutron stars are even more compact than WDs,
the space between their surface and the Roche limit,
where a tidal disc of particles may be located, is substantial.
Incidentally,
low-mass ($ M\sub{p} \ll 1$\,M$\sub{Jup}$) planets can be detected efficiently around pulsars
(highly magnetised, rapidly rotating neutron stars that emit collimated beams of electromagnetic radiation)
by searching for planetary perturbations on the otherwise highly stable timing of their pulsed emission
\citep[e.g.,][]{1993ASPC...36...43C}.
Using this method, two super-Earth-mass planets and one Moon-mass planet
were inferred to orbit the pulsar \mbox{PSR~B1257+12}
on low-eccentricity orbits with periods between 25 and 100\,d
\citep{1992Natur.355..145W,1994Sci...264..538W,2003ApJ...591L.147K}.
The orbital periods of these planets are much longer than those predicted
for the recycled planets in this paper
(up to $\sim$\,10\,h; see Eq.~\eqref{eq:period_2to1}),
indicating that the recycling mechanism as presented here is not responsible for
them.
We note, however, that one of the suggested formation scenarios for the pulsar planets
involves the tidal disruption of a companion star
\citep[e.g.,][]{1993ASPC...36..371P,2007ApJ...666.1232C,2017MNRAS.465.2790M},
bearing some resemblance to the recycling scenario presented in this paper.

Despite the high sensitivity to planets
\citep[e.g.,][]{1992ApJ...387L..69T},
the trio of planets around \mbox{PSR~B1257+12} remains
the only confidently detected set of low-mass planets orbiting a neutron star.
Possible reasons why the recycling mechanism presented in this paper
does not (often) produce planets around neutron stars are as follows.
(1)~Neutron-star systems with planetary material available to form a debris disc may be rare,
because the planets that orbit a neutron-star progenitor are easily lost during the supernova
in which the neutron star is created.
Asymmetric supernovae may kick neutron stars away from their planetary systems
and, even in symmetric supernovae, the sudden mass loss involved can place planets on unbound hyperbolic trajectories
\citep{1993ApJ...419L..65T}.
(2)~The sublimation radius around a neutron star may lie outside the Roche limit,
precluding the formation of a rocky debris disc.
The location of the sublimation radius
depends on what fraction of the star's spin-down luminosity
is available for the heating of dust and debris,
which is not well understood.
Estimates by \citet{2006Natur.440..772W} and \citet{2013ApJ...766....5S}
place the sublimation radius for rocky material beyond the Roche limit.

\subsubsection{Hot subdwarfs}

Hot subdwarfs
are core-helium-burning stars at or beyond the blue end of the horizontal branch,
thought to originate from red-giant-branch stars that undergo enhanced mass loss
(possibly due to interaction with a close binary companion)
or from mergers of low-mass-WD pairs \citep{2009ARA&A..47..211H,2016PASP..128h2001H}.
With typical masses close to 0.5\,M$_\odot$ and radii of around 0.2\,R$_\odot$
\citep[e.g.,][]{2010A&A...524A..63V,2012A&A...539A..12F},
the Roche limit is well outside the stellar surface for these stars.
However, hot subdwarfs have effective temperatures exceeding 20{,}000\,K,
which may complicate the survival of rocky debris
inside the Roche limit against sublimation.

Two B-type hot subdwarfs have been reported to host
close-in, Earth-sized candidate exoplanets:
\mbox{Kepler-70}, a \mbox{2-planet} system \citep{2011Natur.480..496C},
and KIC~10001893, a \mbox{3-planet} system \citep{2014A&A...570A.130S}.\footnote{%
The evidence for the hot-subdwarf planets --
sinusoidal modulations in the light curves of the stars,
interpreted as the daysides of planets coming in and out of view
\citep{2011Natur.480..496C,2014A&A...570A.130S} --
is far from conclusive.
\citet{2015A&A...581A...7K}, for instance, has challenged the planetary origin of the light-curve modulations.}
In both cases,
the claimed planets have orbital periods
of 5 to 20\,h,
with period ratios close to MMRs.
The suggested explanation for the origin of such planets
is the engulfment of
some of a
star's original gas-giant planets during its red-giant-branch evolution,
after which the planets undergo rapid inward migration and extensive mass loss,
while the star is stripped of much of its envelop and becomes a hot subdwarf
(\citealt{2011Natur.480..496C}; see also \citealt{2012ApJ...759L..30P}).
In a variation of this formation scenario, only a single giant planet is engulfed
and the observed planets are remnants of its tidally disrupted core
\citep{2012ApJ...749L..14B}.

Could the recycling mechanism presented in this paper
play a role in the formation of the putative hot-subdwarf planets?
On the one hand, their orbital periods are broadly consistent with
the prediction of Eq.~\eqref{eq:period_2to1},
which argues in favour of the recycling scenario.
On the other hand,
both reported systems feature multiple Earth-sized candidate exoplanets,
contrary to the predicted architecture of the recycled planetary systems,
which typically consist of a single large planet plus a set of much smaller ones.
Our model, however, does not include resonant interactions between planets.
Satellite-formation studies in which such effects are taken into account
show that systems with two or three bodies of similar mass,
orbiting in or close to MMRs with one another,
can sometimes occur
\citep{2015ApJ...799...40H,2017ApJ...836..109S}.
Finally, as mentioned above,
the sublimation of rocky material inside the Roche limit
may be a problem for the recycling scenario.
However,
circumstances may exist
in which the debris is stable against sublimation
(e.g., a high ambient gas pressure can quench sublimation; see also \citealt{2012ApJ...760..123R}).
To sum up,
the hot-subdwarf planets (if real) have the need to be generated
after their host star's giant-branch phase
and roughly fit the expected orbital periods, which is intriguing,
but a more sophisticated model that can handle resonant interactions between planets
is needed to understand whether
they could have spawned from a massive tidal disc.

\subsubsection{Late-type main-sequence stars and substellar objects}

\begin{figure}
  \includegraphics[width=\columnwidth]{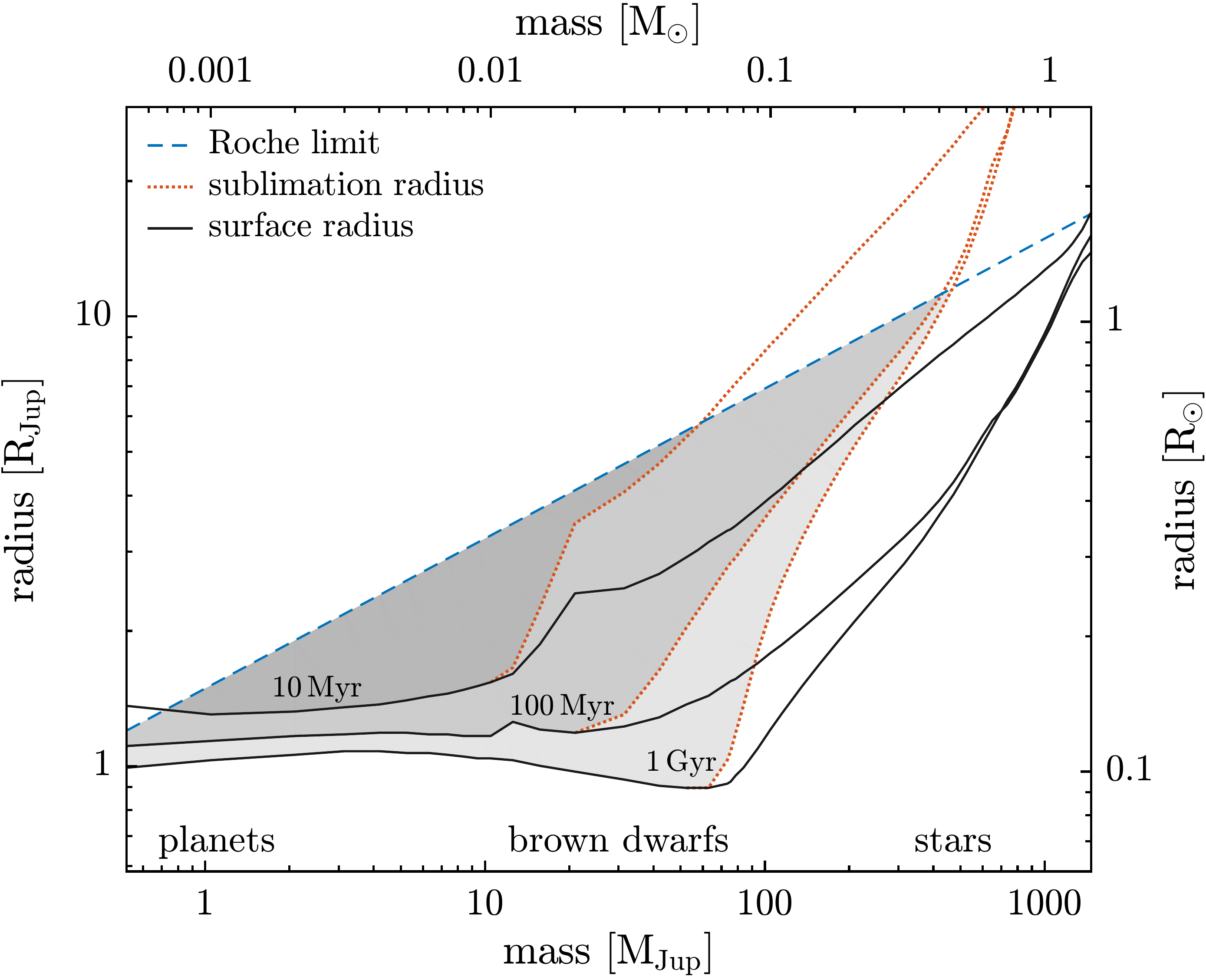}
  \caption{%
  Mass-radius relations for low-mass stars and substellar objects
  at three different ages (black solid lines),
  together with the associated sublimation distances (red dotted lines),
  compared with the Roche limit (blue dashed line).
  Shaded areas mark the region where a tidal disc made of rocky debris can exist.
  The mass-radius relations combine results from
  \citet[brown dwarfs and stars]{1998A&A...337..403B}
  and \citet[planets and brown dwarfs]{2003A&A...402..701B},
  with the transition between the two models at
  the (time-dependent) point where the central object has an effective temperature of 2600\,K.
  The Roche limit is calculated from Eq.~\eqref{eq:r_roche}, assuming
  a material density of $ \rho\sub{d} = 3\,\mathrm{g\,cm}^{-3} $
  and a Roche prefactor of $ C\sub{R} = 2.0 $.
  For the sublimation radius, we use
  the radial distance at which grey dust reaches a temperature of 1500\,K,
  using central-object temperatures given by the model isochrones.}
  \label{fig:mrr_roche}
\end{figure}

Most types of
main-sequence stars have a relatively low mean density and a high luminosity,
with the consequence that the Roche limit for such stars is inside the sublimation radius
(or even inside the stellar radius for early-type stars).
At the low-mass end of the main-sequence, however,
the mean density goes up steeply
(reaching a maximum of around $ \rho_\star \sim 10^2 $\,g\,cm$^{-3}$ at the star/brown-dwarf boundary),
while luminosity goes down.
As a result, late M-type main-sequence stars,
as well as substellar objects such as brown dwarfs and giant planets,
can host rocky debris discs
in the space between their surface or the sublimation radius and the Roche limit.
For most of these central objects, however,
this space is not available immediately
(i.e., while first-generation planets may be forming),
because they
take a long time to contract to their eventual compact size
(see Fig.~\ref{fig:mrr_roche}).
Hence,
any possible tidal discs around brown dwarfs and M dwarfs
would have to form later
(after several tens of Myr),
for instance when a late instability in the host's planetary system
sends a massive body into
the tidal disruption sphere.

If compact tidal discs do occur
around mature low-mass stars or substellar objects,
and if these discs are massive enough,
they may produce second-generation (minor) planets
via the recycling mechanism.
A complication may be that planet migration due to stellar tides
(which we neglected in our model for WDs)
may have to be taken into account for these hosts,
because they are not as compact as WDs.
If the central object's rotation period is longer than $ P_{2:1} $,
the tidal migration of planets at $ r \le r_{2:1} $ is directed inwards.
Mature brown dwarfs mostly have rotation periods of the order of a few hours
\citep[e.g.,][]{2014ApJ...793...75R,2015ApJ...799..154M},
so tidal inspiral is unlikely to pose a problem
for recycled planets around most of them.
The rotation periods of old M dwarfs, however,
often reach values of
the order of 1 to 100\,d
\citep[e.g.,][]{2011ApJ...727...56I}.
Therefore, if the tidal torques are strong enough,
they may prevent recycled planets around M dwarfs
from persisting
after the disc's mass decreases
\citep[see also][]{2016NatGe...9..581R}.

Do we know of any exoplanets
that could have formed in a recycling event?
Not many exoplanets are known that orbit late M dwarfs
and none are presently known around brown dwarfs
\citep{2017MNRAS.464.2687H}.
Most of the low-mass planets discovered around late M dwarfs
(e.g.,
GJ\;1132\;b, \citealt{2015Natur.527..204B};
Proxima Centauri\;b, \citealt{2016Natur.536..437A};
the \mbox{TRAPPIST-1} planets, \citealt{2016Natur.533..221G};
YZ\;Ceti\;bcd, \citealt{2017A&A...605L..11A};
LHS\;1140\;b, \citealt{2017Natur.544..333D})
have orbital periods that are
significantly
longer than the prediction of Eq.~\eqref{eq:period_2to1}.
An exception is \mbox{Kepler-42\;c}, which has an orbital period
of 10.8\,h \citep{2012ApJ...747..144M}.
The fact that this planet is situated
in a multi-planet system with similarly sized planets
further out,
however,
suggests a more conventional (first-generation) formation scenario.

In short, none of the currently known exoplanets
around low-mass stars
seem to have formed via the recycling scenario.
However, late M dwarfs and brown dwarfs have not yet
been surveyed for exoplanets very extensively.
It is possible that recycled exoplanets do occur around these objects,
but have not yet been
found.
On-going and
future
exoplanet-detection programs
specifically targeting late M dwarfs and/or brown dwarfs
(e.g., MEarth, \citealt{2008PASP..120..317N}; SPECULOOS, \citealt{2017arXiv171003775B})
may be able to determine whether recycled exoplanets occur around these hosts
and, if so, how common they are.

\section{Conclusions}
\label{s:conclusions}

This paper considered the evolution of massive debris discs around WDs,
using both analytical and numerical techniques.
Following previous work on WD debris discs, we start by assuming that these discs
are located inside the star's tidal disruption zone
and that they are geometrically and dynamically similar to Saturn's rings.
As a first conclusion, we reiterate the result known for planetary rings
\citep[e.g.,][]{2001Icar..154..296D}
that vertically optically thick discs
located in the outer parts of the tidal disruption zone are gravitationally unstable
(see Eqs.~\eqref{eq:tau_grav_stab} and \eqref{eq:tau_grav_stab2}; Fig.~\ref{fig:grav_stab}),
which means that they
have an enhanced effective viscosity due to self-gravity wakes.

Previous theoretical work on the evolution of WD debris discs has
focussed on disc evolution caused by PR drag
\citep{2011ApJ...732L...3R,2011ApJ...741...36B}.
Our findings for low-mass discs are consistent with
the results of these studies
(e.g., Fig.~\ref{fig:snapshots}, top-left panel;
Fig.~\ref{fig:maps}, bottom panels),
but we also quantify
the upper limit in surface mass density for which they are valid.
For values above 10$^{3}$ to 10$^{4}$\,g\,cm$^{-2}$
(corresponding to disc masses around the mass of Pluto; $\sim$\,10$^{25}$\,g),
viscous spreading becomes more important than PR drag for the disc's evolution
(see Eqs.~\eqref{eq:sigma_crit_time} and \eqref{eq:mdisk_crit}; Fig.~\ref{fig:time_visc_pr}).

The effective viscosity of a gravitationally unstable particle disc
increases rapidly with distance
(see Eq.~\eqref{eq:visc_sg}).
As a result,
discs whose evolution is dominated by viscous spreading have
an outer edge that moves outwards rapidly,
while a pile-up of material forms at their inner edge,
which only moves inwards slowly.
For intermediate-mass discs (with masses from 10$^{25}$ to 10$^{26}$\,g,
depending on their radial location), the viscous spreading can decrease
surface densities to the point where PR drag takes over and subsequent evolution
follows that known for low-mass discs (e.g., Fig.~\ref{fig:snapshots}, top-right panel).
For discs more massive than about 10$^{26}$\,g
(roughly the mass of Io),
however, the outer edge of the disc
has reached the Roche limit before this happens
(see Eq.~\eqref{eq:mdisk_form}; Fig.~\ref{fig:mdisk_form}).

When disc material spreads to beyond the Roche limit, it is no longer
prevented from coagulating by tidal forces and Keplerian shear,
so this material will gather into larger bodies.
Hence,
viscous-spreading-dominated discs
can produce
a set of second-generation (minor) planets just outside their Roche limit
(e.g., Fig.~\ref{fig:snapshots}, bottom panel)
in the same fashion that
small moonlets are being formed at the outer edge of Saturn's rings
\citep{2010Natur.465..752C}.
For massive discs, this process takes place on relatively short (sub-Myr) timescales
(see Fig.~\ref{fig:maps}, middle panels),
while the discs themselves can survive against viscous spreading and PR drag
for longer than the age of the universe
(see Eq.~\eqref{eq:t_disc_plan}; Fig.~\ref{fig:maps}, bottom-left panel).

The bodies that form at the disc's outer edge
migrate outwards by exchanging angular momentum with the disc via MMRs
and they grow through mutual mergers.
The interplay of these two effects results in a distinctive pattern
of minor planets whose masses and mutual spacing
increase with distance from the disc
(see Eqs.~\eqref{eq:q_delta_pyra} and \eqref{eq:dn_ddelta_pyra};
Fig.~\ref{fig:delta_q}),
as has previously been studied in detail for massive planetary-ring systems
\citepalias{2012Sci...338.1196C}.
At a given radial distance, minor-planet masses go down with time,
since mass flows through to larger bodies at greater radial distances,
while continued viscous spreading decreases the disc's surface density,
reducing the Roche-limit mass outflow
available for forming new bodies
(see Figs.~\ref{fig:snapshots}, bottom panel, and \ref{fig:num_planets}).

The outward migration of the bodies stops at $ r_{2:1} $,
the radius at which their 2:1 MMR coincides with the Roche limit
(because none of their first-order MMRs would overlap with the disc anymore
if they were to migrate beyond this point).
Only a limited amount of mass can be accommodated by minor planets
in between the Roche limit and $ r_{2:1} $ (see Eq.~\eqref{eq:mass_pyra}).
Once this space has been filled up
(which happens eventually for most planet-producing discs),
material starts to accumulate into a single massive body at $ r_{2:1} $.
This means that
massive WD debris discs
(i.e., with masses of about $ 10^{26} $\,g and higher)
generally give rise to
a single massive
planet orbiting at $ r_{2:1} $,
together with a set of smaller bodies
in smaller orbits
(see Figs.~\ref{fig:snapshots}, bottom panel, and \ref{fig:num_planets}).

The massive planet at $ r_{2:1} $
can take up a substantial fraction of the disc's original mass
(tens of per cent; see Eq.~\eqref{eq:f_recyc_simple}; Fig.~\ref{fig:maps}, top panels).
This means that
if a super Earth is tidally disrupted by a WD and forms a massive debris disc,
it can be partially recycled to form an Earth-mass second-generation planet.
Given its 2:1 MMR with the Roche limit,
the planet at $ r_{2:1} $ has
a clearly predictable orbital period
(around 10\,h; see Eq.~\eqref{eq:period_2to1}).
It is
located in the WD's habitable zone
for old, cool WDs (with cooling ages between about 2 and 10\,Gyr;
see Sect.~\ref{s:hab_zone}).
Constraints
on the occurrence rate of
this type of exoplanet
are still relatively weak, but they are likely to
be improved by several upcoming
large, high-cadence
transit surveys that include WD targets
(see Sect.~\ref{s:obs}; Fig.~\ref{fig:obs_plan}; Table~\ref{tbl:obs_fut}).
The best place to look for
such planets
may be
WDs that exhibit a strong IR excess
and/or circumstellar gas in absorption,
although it is also possible that a disc that gives rise to planets
is destroyed after producing them
(see Sect.~\ref{s:destroy_disc}).

As an application of our second-generation-planet-formation model,
we have investigated
whether
it
can shed light on
the curious
properties of \mbox{WD~1145+017} (Sect.~\ref{s:wd1145}).
The light curve of this WD
shows irregularly shaped
dips,
probably caused by
transiting
clouds of dust
orbiting close to the Roche limit.
These dips repeat with
remarkably stable periods,
which can be explained if the dust is released from long-lived, massive bodies.
The presence of such bodies
on low-eccentricity orbits
close to the Roche limit, however, is difficult to explain,
since this region
has been cleared
earlier in the star's evolution.
Second-generation planet formation from a massive debris disc, as presented in this paper,
provides a natural explanation
for the presence of massive bodies just outside the Roche limit.
Furthermore,
if \mbox{WD~1145+017} currently hosts a disc that is overflowing the Roche limit,
the transiting dust clouds may be
the result of collisions that are part of the planet-formation process.
In the context of this scenario,
we make rough predictions for the masses
of the putative bodies
associated with
the various observed transit periods
(see Fig.~\ref{fig:wd1145}).

While our study mainly focussed on what happens
around the outer edge
of massive WD debris discs,
we also
obtained results on the mass accretion rates of
WDs that host such discs.
Specifically,
we find that
in discs
more massive than about 10$^{27}$\,g
(i.e., the mass of a terrestrial planet)
viscous spreading can directly dictate
the inward mass flow at the sublimation radius,
causing
mass accretion rates of $ \gtrsim $\,$10^9 $\,g\,s$^{-1}$
(see Fig.~\ref{fig:maps}, bottom-right panel).
Thus,
the viscous spreading of very massive debris discs
is a possible
explanation for the
high
(super-PR-drag)
mass accretion rates
seen in
some heavily polluted WDs.
For \mbox{WD~1145+017}, in particular,
the presence of such a disc
may be able to simultaneously explain
the massive bodies related to its transits and
the high mass accretion rate inferred from its atmospheric pollution.

Finally,
we have explored the potential of other astrophysical objects
to host tidal discs that can spawn second-generation planets
(see Sect.~\ref{s:other_hosts}).
Objects with sufficient compactness
include neutron stars and hot subdwarf stars, but in both cases
sublimation may preclude the presence of rocky debris inside the Roche limit.
In the case of neutron stars,
the loss of planetary material during their birth supernovae
may be an additional obstruction.
A more likely class of hosts
consists of
late M-type main-sequence stars and substellar objects
like brown dwarfs and giant planets.
At a mature age, these objects are compact and cool enough
to allow rocky debris to persist inside the Roche limit
(see Fig.~\ref{fig:mrr_roche}).
Hence, if such discs are formed at the right time and if they are massive enough,
recycled planets may be produced around this class of hosts,
although stellar tides may impede their survival around M dwarfs.
We have examined whether any known (candidate) exoplanets
around the above types of objects may have formed via the recycling scenario,
but find that none conclusively show the expected characteristics.

\section*{Acknowledgements}

RvL, QK, MCW, and AS acknowledge support from the European Union through ERC grant number 279973.
RvL is also supported by the DISCSIM project, grant agreement 341137
funded by the European Research Council under ERC-2013-ADG.
QK acknowledges funding from STFC via the Institute of Astronomy, Cambridge Consolidated Grant.
AS is partially supported by funding from the Center for Exoplanets and Habitable Worlds.
The Center for Exoplanets and Habitable Worlds is supported by the Pennsylvania State University,
the Eberly College of Science, and the Pennsylvania Space Grant Consortium.

\bibliographystyle{mnras}
\bibliography{bib_ads}

\appendix

\section{Torque-density balance}
\label{app:visc_pr_torque}

Viscous spreading of a disc leads to the outward transport of angular momentum,
while PR drag acts to remove angular momentum.
Since angular-momentum change rates are equivalent to torques,
the conditions under which these two processes are in balance
can be found
by equating their radial torque densities:
\begin{equation}
  \label{eq:torque_dens_bal}
  \frac{ \dif \varGamma\sub{visc} }{ \dif r } = - \frac{ \dif \varGamma\sub{PR} }{ \dif r }.
\end{equation}
We now proceed to determine expressions for the individual torque densities
associated with viscosity and PR drag.

At any point in the disc, the viscous torque of the material inside radius $ r $
on material outside $ r $ can be written as \citep{1981ARA&A..19..137P}
\begin{equation}
  \label{eq:torque_visc}
  \varGamma\sub{visc} = 2 \uppi r \times \nu \varSigma A \times r,
  \qquad
  A = - r \frac{ \dif \varOmega\sub{K} }{ \dif r } = \frac{ 3 }{ 2 } \varOmega\sub{K},
\end{equation}
where
$ A $ is the local shearing rate.
Inserting the viscosity prescription for a gravitationally
unstable particle disc (Eq.~\eqref{eq:visc_sg})
and taking the radial derivative
(assuming the surface density is constant with radius)
gives the viscous torque density
\begin{equation}
  \label{eq:torque_dens_visc}
  \frac{ \dif \varGamma\sub{visc} }{ \dif r }
    \simeq 30 \uppi \varSigma \nu r \varOmega\sub{K}
    \simeq 780 \uppi r\sub{H}^{*\,5}
      \frac{ G^2 \varSigma^3 r }{ \varOmega\sub{K}^2 }.
\end{equation}

To find the PR-drag torque density, we consider the azimuthal force per unit area
exerted on the disc by the impinging stellar radiation through PR drag
\citep{2011ApJ...732L...3R}
\begin{equation}
  \label{eq:force_per_area_pr}
  f_{\varphi\mathrm{,PR}} = - \zeta \frac{ L_\star \phi_r }{ 4 \uppi r^2 c } \frac{ \varOmega\sub{K} r }{ c },
  \qquad
  \phi_r \simeq 1 - \exp( - \tau_\| ).
\end{equation}
Here,
$ \phi_r $ characterises the efficiency of radiative momentum absorption by the disc surface.
For a disc that is optically thick to the impinging radiation ($ \tau_\| \gg 1 $),
this factor simplifies to $ \phi_r \simeq 1 $.
We choose the negative sign because the azimuthal PR-drag force acts
in the direction opposite to the rotation of the disc.
The PR-drag torque on a narrow disc annulus of width $ \dif r $ is
$ 2 \uppi r \dif r \times f_{\varphi\mathrm{,PR}} \times r $,
and hence the radial density of PR-drag torque on the disc is
\begin{equation}
  \label{eq:torque_dens_pr}
  \frac{ \dif \varGamma\sub{PR} }{ \dif r }
    = 2 \uppi r \times f_{\varphi\mathrm{,PR}} \times r
    = - \frac{ \zeta \phi_r }{ 2 } \frac{ L_\star }{ c^2 } \varOmega\sub{K} r.
\end{equation}

Inserting expressions \eqref{eq:torque_dens_visc} and \eqref{eq:torque_dens_pr}
into Eq.~\eqref{eq:torque_dens_bal}, using $ \phi_r \simeq 1 $,
and solving for $ \varSigma $ gives the critical surface density
\begin{equation}
  \label{eq:sigma_crit_torque}
  \varSigma\sub{crit}
    \simeq
      \left( \frac{ \zeta L_\star }{ 1560 \uppi c^2 G^2 r\sub{H}^{*\,5} } \right)^{1/3}
      \varOmega\sub{K}.
\end{equation}
This result can be rewritten to produce
Eq.~\eqref{eq:sigma_crit_time} multiplied by a factor $ ( 2 / 5 )^{1/3} \approx 0.74 $.

\bsp  
\label{lastpage}
\end{document}